\pdfoutput=1
\documentclass[12pt,paper=letter,titlepage=true,twoside=true,open=right]{scrbook}
\typearea[current]{calc}
\areaset{6in}{9in}
\usepackage[T1]{fontenc}
\usepackage{ae,aecompl}
\usepackage{lmodern}
\usepackage{microtype}
\usepackage{ucs}
\usepackage[utf8x]{inputenc}
\usepackage{makeidx}
\makeindex
\usepackage{verse}
\usepackage[all,knot,poly,arc]{xy}
\usepackage[usenames]{color}
\usepackage{graphicx}
\usepackage{xspace}
\usepackage{setspace}
\usepackage[colorlinks=true,urlcolor=blue]{hyperref}
\hypersetup{
  pdftitle    = {Daniel's PhD Thesis},
  pdfauthor   = {Daniel Doro Ferrante <danieldf@het.brown.edu>},
  pdfsubject  = {High Energy Physics - Theory},
  pdfcreator  = {LaTeX2e with hyperref package},
  pdfproducer = {dvipdfm},
  pdfkeywords = {Quantum Field Theory, Lattice Quantum Field Theory, Higgs Bundles, Morse Theory, Symmetry Breaking,
                 Quantum Phase Transition, Classical Phase Transition, Modular Symmetry, Quantum Gravity},
  pdfview     = {FitH}
}   
\usepackage{amsmath,amsfonts,mathrsfs,dsfont,amstext,amssymb,amsbsy,amsopn,amsthm,eucal}
\usepackage[boxed]{algorithm}
\usepackage{algorithmic}
\DeclareMathAlphabet{\mathpzc}{T1}{pzc}{m}{it} 
\usepackage{mathrsfs}
\usepackage{wasysym}
\usepackage{empheq}
\usepackage{pifont}
\usepackage{textcomp}
\usepackage[Lenny]{fncychap}
\usepackage{fancyhdr}
\fancypagestyle{plain}{%
  \fancyhf{}
  \fancyfoot[C]{\bfseries\thepage}

}
\newcommand{\helv}{\fontfamily{phv}\fontseries{b}\fontsize{9}{11}\selectfont}
\makeatletter
\def\equalsfill{$\m@th\mathord=\mkern-7mu
  \cleaders\hbox{$\!\mathord=\!$}\hfill
  \mkern-7mu\mathord=$}
\makeatother
\def\hksqrt{\mathpalette\DHLhksqrt}
\def\DHLhksqrt#1#2{\setbox0=\hbox{$#1\sqrt{#2\,}$}\dimen0=\ht0
  \advance\dimen0-0.2\ht0
  \setbox2=\hbox{\vrule height\ht0 depth -\dimen0}%
{\box0\lower0.4pt\box2}}

\DeclareMathOperator{\tr}{tr}

\DeclareMathOperator{\T}{\mathds{T}}

\DeclareMathOperator{\ed}{d\!}
\DeclareMathOperator{\Ai}{Ai}
\DeclareMathOperator{\Bi}{Bi}

\DeclareMathOperator{\Res}{Res}

\DeclareMathOperator{\re}{\mathsf{Re}}

\DeclareMathOperator{\arcsinh}{arcsinh}

\newcommand{\ie}{\textrm{i.e.}\xspace}
\newcommand{\g}{\ensuremath{\boldsymbol{g}}\xspace}
\newcommand{\gE}{\ensuremath{\boldsymbol{\tilde{g}}_{E}}\xspace}
\newcommand{\M}{\ensuremath{\mathpzc{M}}\xspace}

\newcommand{\conv}[2]{\ensuremath{{#1}\ast{#2}}\xspace}
\newcommand{\hm}[2]{\ensuremath{\langle#1,#2\rangle}\xspace}

\newcommand{\cont}[1]{\ensuremath{{\mathscr{C}\/}^{#1}}\xspace}

\newcommand{\abs}[1]{\ensuremath{\left| #1\right|}\xspace}
\newtheorem{thm}{Theorem}[section]

\newtheorem*{deff}{Definition}
\newtheorem*{notat}{Notation}
\def\@spires#1{\href{http://www-spires.slac.stanford.edu/spires/find/hep/www?j=#1}} 
\catcode`\%=12
\catcode`\|=14
\newcommand\apa[3]    {\@spires{APASA
		{{\it Acta Phys.\ Austriaca }{\bf #1} (#2) #3}}
\newcommand\apas[3]    {\@spires{APAUA
		{{\it Acta Phys.\ Austriaca, Suppl.\ }{\bf #1} (#2) #3}}
\newcommand\appol[3] {\@spires{APPOA
		{{\it Acta Phys.\ Polon.\ }{\bf #1} (#2) #3}}
\newcommand\advm[3]  {\@spires{ADMTA
		{{\it Adv.\ Math.\ }{\bf #1} (#2) #3}}
\newcommand\adnp[3]   {\@spires{ANUPB
		{{\it Adv.\ Nucl.\ Phys.\ }{\bf #1} (#2) #3}}
\newcommand\adp[3]   {\@spires{ADPHA
		{{\it Adv.\ Phys.\ }{\bf #1} (#2) #3}}
\newcommand\atmp[3] {\@spires{00203
		{{\it Adv.\ Theor.\ Math.\ Phys.\ }{\bf #1} (#2) #3}}
\newcommand\am[3]    {\@spires{ANMAA
		{{\it Ann.\ Math.\ }{\bf #1} (#2) #3}}
\newcommand\ap[3]    {\@spires{APNYA
		{{\it Ann.\ Phys.\ (NY) }{\bf #1} (#2) #3}}
\newcommand\araa[3] {\@spires{ARAAA
		{{\it Ann.\ Rev.\ Astron.\ \& Astrophys.\ }{\bf #1} (#2) #3}}
\newcommand\arnps[3] {\@spires{ARNUA
		{{\it Ann.\ Rev.\ Nucl.\ Part.\ Sci.\ }{\bf #1} (#2) #3}}
\newcommand\asas[3]   {\@spires{AAEJA
		{{\it Astron.\ Astrophys.\ }{\bf #1} (#2) #3}}
\newcommand\asj[3]   {\@spires{ANJOA
		{{\it Astron.\ J.\ }{\bf #1} (#2) #3}}
\newcommand\app[3]   {\@spires{APHYE
		{{\it Astropart.\ Phys.\ }{\bf #1} (#2) #3}}
\newcommand\apj[3]    {\@spires{ASJOA
		{{\it Astrophys.\ J. }{\bf #1} (#2) #3}}
\newcommand\baas[3]   {\@spires{AASBA
		{{\it Bull.\ Am.\ Astron.\ Soc.\ }{\bf #1} (#2) #3}}
\newcommand\bams[3]   {\@spires{BAMOA
		{{\it Bull.\ Am.\ Math.\ Soc.\ }{\bf #1} (#2) #3}}
\newcommand\blms[3]   {\@spires{LMSBB
		{{\it Bull.\ London Math.\ Soc.\ }{\bf #1} (#2) #3}}
\newcommand\cjm[3]  {\@spires{CJMAA
		{{\it Can.\ J.\ Math.\ }{\bf #1} (#2) #3}}
\newcommand\cqg[3]   {\@spires{CQGRD
		{{\it Class.\ and Quant.\ Grav.\ }{\bf #1} (#2) #3}}
\newcommand\cmp[3]   {\@spires{CMPHA
		{{\it Commun.\ Math.\ Phys.\ }{\bf #1} (#2) #3}}
\newcommand\ctp[3]   {\@spires{CTPMD
		{{\it Commun.\ Theor.\ Phys.\ }{\bf #1} (#2) #3}}
\newcommand\cag[3]   {\@spires{00142
		{{\it Commun.\ Anal.\ Geom.\ }{\bf #1} (#2) #3}}
\newcommand\cpam[3]   {\@spires{CPAMA
		{{\it Commun.\ Pure Appl.\ Math.\ }{\bf #1} (#2) #3}}
\newcommand\cpc[3]   {\@spires{CPHCB
		{{\it Comput.\ Phys.\ Commun.\ }{\bf #1} (#2) #3}}
\newcommand\dmj[3]   {\@spires{DUMJA
		{{\it Duke Math.\ J. }{\bf #1} (#2) #3}}
\newcommand\epjc[3]  {\@spires{EPHJA
		{{\it Eur.\ Phys.\ J. }{\bf C #1} (#2) #3}}
\newcommand\epjd[3]  {\@spires{EPHJD
		{{\it Eur.\ Phys.\ J. Direct.\ }{\bf C #1} (#2) #3}}
\newcommand\epl[3]    {\@spires{EULEE
		{{\it Europhys.\ Lett. }{\bf #1} (#2) #3}}
\newcommand\forp[3]    {\@spires{FPYKA
		{{\it Fortschr.\ Phys.\ }{\bf #1} (#2) #3}}
\newcommand\faa[3]    {\@spires{FAAPB
		{{\it Funct.\ Anal.\ Appl.\ }{\bf #1} (#2) #3}}
\newcommand\grg[3]    {\@spires{GRGVA
		{{\it Gen.\ Rel.\ Grav.\ }{\bf #1} (#2) #3}}
\newcommand\hpa[3]   {\@spires{HPACA
		{{\it Helv.\ Phys.\ Acta }{\bf #1} (#2) #3}}
\newcommand\ijmpa[3] {\@spires{IMPAE
		{{\it Int.\ J.\ Mod.\ Phys.\ }{\bf A #1} (#2) #3}}
\newcommand\ijmpb[3] {\@spires{IMPAE
		{{\it Int.\ J.\ Mod.\ Phys.\ }{\bf B #1} (#2) #3}}
\newcommand\ijmpc[3] {\@spires{IMPAE
		{{\it Int.\ J.\ Mod.\ Phys.\ }{\bf C #1} (#2) #3}}
\newcommand\ijmpd[3] {\@spires{IMPAE
		{{\it Int.\ J.\ Mod.\ Phys.\ }{\bf D #1} (#2) #3}}
\newcommand\ijtp[3] {\@spires{IJTPB
		{{\it Int.\ J.\ Theor.\ Phys.\ }{\bf #1} (#2) #3}}
\newcommand\invm[3]  {\@spires{INVMB
		{{\it Invent.\ Math.\ }{\bf #1} (#2) #3}}
\newcommand\jag[3]   {\@spires{00124
		{{\it J.\ Alg.\ Geom.\ }{\bf #1} (#2) #3}}
\newcommand\jams[3]   {\@spires{00052
		{{\it J.\ Am.\ Math.\ Soc.\ }{\bf #1} (#2) #3}}
\newcommand\jap[3]   {\@spires{JAPIA
		{{\it J.\ Appl.\ Phys.\ }{\bf #1} (#2) #3}}
\newcommand\jdg[3]   {\@spires{JDGEA
		{{\it J.\ Diff.\ Geom.\ }{\bf #1} (#2) #3}}
\newcommand\jgp[3]   {\@spires{JGPHE
		{{\it J.\ Geom.\ Phys.\ }{\bf #1} (#2) #3}}
\newcommand\jhep[3]  {\href{http://jhep.sissa.it/stdsearch?paper=#1%28#2%29#3}
		{{\it J. High Energy Phys.\ }{\bf #1} (#2) #3}}
\newcommand\jmp[3]   {\@spires{JMAPA
		{{\it J.\ Math.\ Phys.\ }{\bf #1} (#2) #3}}
\newcommand\joth[3]  {\@spires{JOTHE
		{{\it J.\ Operator Theory }{\bf #1} (#2) #3}}
\newcommand\jpha[3]   {\@spires{JPAGB
		{{\it J. Phys.\ }{\bf A #1} (#2) #3}}
\newcommand\jphc[3]   {\@spires{JPAGB
		{{\it J. Phys.\ }{\bf C #1} (#2) #3}}
\newcommand\jphg[3]   {\@spires{JPAGB
		{{\it J. Phys.\ }{\bf G #1} (#2) #3}}
\newcommand\lmp[3]   {\@spires{LMPHD
		{{\it Lett.\ Math.\ Phys.\ }{\bf #1} (#2) #3}}
\newcommand\ncl[3]    {\@spires{NCLTA
		{{\it Lett.\ Nuovo Cim.\ }{\bf #1} (#2) #3}}
\newcommand\matan[3]  {\@spires{MAANA
		{{\it Math.\ Ann.\ }{\bf #1} (#2) #3}}
\newcommand\mussr[3]  {\@spires{MUSIA
		{{\it Math.\ USSR Izv.\ }{\bf #1} (#2) #3}}
\newcommand\mams[3]  {\@spires{MAMCA
		{{\it Mem.\ Am.\ Math.\ Soc.\ }{\bf #1} (#2) #3}}
\newcommand\mpla[3]  {\@spires{MPLAE
		{{\it Mod.\ Phys.\ Lett.\ }{\bf A #1} (#2) #3}}
\newcommand\mplb[3]  {\@spires{MPLAE
		{{\it Mod.\ Phys.\ Lett.\ }{\bf B #1} (#2) #3}}
\newcommand\nature[3]  {\@spires{NATUA
		{{\it Nature }{\bf #1} (#2) #3}}
\newcommand\nim[3]   {\@spires{NUIMA
		{{\it Nucl.\ Instrum.\ Meth.\ }{\bf #1} (#2) #3}}
\newcommand\npa[3]   {\@spires{NUPHA
		{{\it Nucl.\ Phys.\ }{\bf A #1} (#2) #3}}
\newcommand\npb[3]    {\@spires{NUPHA
		{{\it Nucl.\ Phys.\ }{\bf B #1} (#2) #3}}
\newcommand\npps[3]  {\@spires{NUPHZ
		{{\it Nucl.\ Phys.\ }{\bf #1} {\it(Proc.\ Suppl.)} (#2) #3}}
\newcommand\nc[3]    {\@spires{NUCIA
		{{\it Nuovo Cim.\ }{\bf #1} (#2) #3}}
\newcommand\ncs[3]  {\@spires{NUCUA
		{{\it Nuovo Cim.\ Suppl.\ }{\bf #1} (#2) #3}}
\newcommand\pan[3]  {\@spires{PANUE
		{{\it Phys.\ Atom.\ Nucl.\ }{\bf #1} (#2) #3}}
\newcommand\pla[3]   {\@spires{PHLTA
		{{\it Phys.\ Lett.\ }{\bf A #1} (#2) #3}}
\newcommand\plb[3]   {\@spires{PHLTA
		{{\it Phys.\ Lett.\ }{\bf B #1} (#2) #3}}
\newcommand\pr[3]    {\@spires{PHRVA
		{{\it Phys.\ Rev.\ }{\bf #1} (#2) #3}}
\newcommand\pra[3]   {\@spires{PHRVA
		{{\it Phys.\ Rev.\ }{\bf A #1} (#2) #3}}
\newcommand\prb[3]   {\@spires{PHRVA
		{{\it Phys.\ Rev.\ }{\bf B #1} (#2) #3}}
\newcommand\prc[3]   {\@spires{PHRVA
		{{\it Phys.\ Rev.\ }{\bf C #1} (#2) #3}}
\newcommand\prd[3]   {\@spires{PHRVA
		{{\it Phys.\ Rev.\ }{\bf D #1} (#2) #3}}
\newcommand\pre[3]   {\@spires{PHRVA
		{{\it Phys.\ Rev.\ }{\bf E #1} (#2) #3}}
\newcommand\prep[3]  {\@spires{PRPLC
		{{\it Phys.\ Rept.\ }{\bf #1} (#2) #3}}
\newcommand\prl[3]   {\@spires{PRLTA
		{{\it Phys.\ Rev.\ Lett.\ }{\bf #1} (#2) #3}}
\newcommand\phys[3]   {\@spires{PHYSA
		{{\it Physica }{\bf #1} (#2) #3}}
\newcommand\plms[3]   {\@spires{PHLTA
		{{\it Proc.\ London Math.\ Soc.\ }{\bf B #1} (#2) #3}}
\newcommand\pnas[3]  {\@spires{PNASA
		{{\it Proc.\ Nat.\ Acad.\ Sci.\ }{\bf #1} (#2) #3}}
\newcommand\ppnp[3]  {\@spires{PPNPD
		{{\it Prog.\ Part.\ Nucl.\ Phys.\ }{\bf #1} (#2) #3}}
\newcommand\ptp[3]   {\@spires{PTPKA
		{{\it Prog.\ Theor.\ Phys.\ }{\bf #1} (#2) #3}}
\newcommand\ptps[3]   {\@spires{PTPSA
		{{\it Prog.\ Theor.\ Phys.\ Suppl.\ }{\bf #1} (#2) #3}}
\newcommand\rmp[3]   {\@spires{RMPHA
		{{\it Rev.\ Mod.\ Phys.\ }{\bf #1} (#2) #3}}
\newcommand\sjnp[3]  {\@spires{SJNCA
		{{\it Sov.\ J.\ Nucl.\ Phys.\ }{\bf #1} (#2) #3}}
\newcommand\sjpn[3]  {\@spires{SJPNA
		{{\it Sov.\ J.\ Part.\ Nucl.\ }{\bf #1} (#2) #3}}
\newcommand\jetp[3]  {\@spires{SPHJA
		{{\it Sov.\ Phys.\ JETP\/ }{\bf #1} (#2) #3}}
\newcommand\jetpl[3]  {\@spires{JTPLA
		{{\it Sov.\ Phys.\ JETP Lett.\ }{\bf #1} (#2) #3}}
\newcommand\spu[3]  {\@spires{SOPUA
		{{\it Sov.\ Phys.\ Usp.\ }{\bf #1} (#2) #3}}
\newcommand\tmf[3]   {\@spires{TMFZA
		{{\it Teor.\ Mat.\ Fiz.\ }{\bf #1} (#2) #3}}
\newcommand\tmp[3]   {\@spires{TMPHA
		{{\it Theor.\ Math.\ Phys.\ }{\bf #1} (#2) #3}}
\newcommand\ufn[3]   {\@spires{UFNAA
		{{\it Usp.\ Fiz.\ Nauk.\ }{\bf #1} (#2) #3}}
| }}}}}}}}}}}}}}}}}}}}}} "|" is here a comment (catcode defined above) to
| }}}}}}}}}}}}}}}}}}}}}} include parenthesis for emacs to parse properly. 
\newcommand\ujp[3]   {\@spires{00267
		{{\it Ukr.\ J.\ Phys.\ }{\bf #1} (#2) #3}}
\newcommand\yf[3]    {\@spires{YAFIA
		{{\it Yad.\ Fiz.\ }{\bf #1} (#2) #3}}
\newcommand\zpc[3]   {\@spires{ZEPYA
		{{\it Z.\ Physik }{\bf C #1} (#2) #3}}
\newcommand\zetf[3]  {\@spires{ZETFA
		{{\it Zh.\ Eksp.\ Teor.\ Fiz.\ }{\bf #1} (#2) #3}}

\newcommand{\newjournal}[5]{\@spires{#2
		{{\it #1 }{\bf #3} (#4) #5}}

\newcommand\ibid[3]{{\it ibid.\ }{\bf #1} (#2) #3}
\catcode`\%=14
\catcode`\|=12
\newcommand{\hepth}[1]{\href{http://xxx.lanl.gov/abs/hep-th/#1}{\tt hep-th/#1}}
\newcommand{\hepph}[1]{\href{http://xxx.lanl.gov/abs/hep-ph/#1}{\tt hep-ph/#1}}
\newcommand{\heplat}[1]{\href{http://xxx.lanl.gov/abs/hep-lat/#1}{\tt hep-lat/#1}}
\newcommand{\hepex}[1]{\href{http://xxx.lanl.gov/abs/hep-ex/#1}{\tt hep-ex/#1}}
\newcommand{\nuclth}[1]{\href{http://xxx.lanl.gov/abs/nucl-th/#1}{\tt nucl-th/#1}}
\newcommand{\nuclex}[1]{\href{http://xxx.lanl.gov/abs/nucl-ex/#1}{\tt nucl-ex/#1}}
\newcommand{\grqc}[1]{\href{http://xxx.lanl.gov/abs/gr-qc/#1}{\tt gr-qc/#1}}
\newcommand{\qalg}[1]{\href{http://xxx.lanl.gov/abs/q-alg/#1}{\tt q-alg/#1}}
\newcommand{\accphys}[1]{\href{http://xxx.lanl.gov/abs/accphys/#1}{\tt accphys/#1}}
\newcommand{\alggeom}[1]{\href{http://xxx.lanl.gov/abs/alg-geom/#1}{\tt alg-geom/#1}}
\newcommand{\astroph}[1]{\href{http://xxx.lanl.gov/abs/astro-ph/#1}{\tt astro-ph/#1}}
\newcommand{\chaodyn}[1]{\href{http://xxx.lanl.gov/abs/chao-dyn/#1}{\tt chao-dyn/#1}}
\newcommand{\condmat}[1]{\href{http://xxx.lanl.gov/abs/cond-mat/#1}{\tt cond-mat/#1}}
\newcommand{\nlinsys}[1]{\href{http://xxx.lanl.gov/abs/nlin-sys/#1}{\tt nlin-sys/#1}}
\newcommand{\quantph}[1]{\href{http://xxx.lanl.gov/abs/quant-ph/#1}{\tt quant-ph/#1}}
\newcommand{\solvint}[1]{\href{http://xxx.lanl.gov/abs/solv-int/#1}{\tt solv-int/#1}}
\newcommand{\suprcon}[1]{\href{http://xxx.lanl.gov/abs/supr-con/#1}{\tt supr-con/#1}}
\newcommand{\Math}[2]{\href{http://xxx.lanl.gov/abs/math.#1/#2}{\tt math.#1/#2}}
\newcommand{\arXivid}[1]{\href{http://arxiv.org/abs/#1}{\tt arXiv:#1}}

\begin{document}
\frontmatter
\begin{titlepage}

  \vspace*{3cm}
  \begin{center}
    \begin{spacing}{2}
      {\LARGE \textbf{\textsf{Symmetry Breaking: A New Paradigm for
            Non-Perturbative QFT and Topological Transitions}}} \\
    \end{spacing}
   \vspace{5cm}
    \Large{\textsf{\href{http://creativecommons.org/licenses/by-sa/3.0/}{\textcopyleft}\, 2009 by \href{mailto:danieldf@het.brown.edu}{Daniel D. Ferrante}}} \\
    \vspace{3cm}

    \textsf{\href{http://www.physics.brown.edu/}{Physics Department}, \href{http://www.brown.edu/}{Brown University}} \\
    \textsf{Providence --- RI. 02912. USA.}\\

    \date{\today} 
 \end{center}
\end{titlepage}
\chapter*{Preface}

Lately, after $\sim\! 50$ years, there seems to be a convergence in the
languages used to describe Quantum Field Theory and String Theory, such that
it seems possible to relate objects from these two perspectives. Therefore, a
deeper investigation of the properties and features of QFT is a reasonable
thing to do: non-perturbative effects, dualities, emergent properties, non-commutative
structures, etc. This particular line of research uses Symmetry Breaking in
order to probe a few of the different topics mentioned above, i.e., Symmetry
Breaking is used as an ``underlying principle'', bringing different features
of QFT to the foreground. However, the understanding of Symmetry Breaking that
is used here is quite different from what is done in the mainstream: Symmetry
Breaking is understood as the solution set of a given QFT, its vacuum
manifold, or, more modernly, its Moduli Space. Distinct solutions correspond
to different sectors, phases, of the theory, which are nothing but distinct
foliations of the vacuum manifold, or points in the Moduli Space (for all
possible values of the parameters of the theory). Under this framework, three
different problems will be attacked: ``Mollifying QFT'', ``Topological
Transitions and Geometric Langlands Duality'' and ``Three-dimensional Gravity
and its Phase Transitions''. The first makes use of the Moduli Space of the
theory in order to construct an appropriate mollification of it, rendering it
viable to simulate a QFT in Lorentzian spaces, tackling the ``sign problem''
heads-on. The connections with Lee-Yang zeros and Stokes Phenomena will be
made clear. The second will show that each different phase has its own
topology which can be used as Superselection Rule;
moreover, the Euler Characteristic of each phase gives it quantization
condition. The mechanism via which several dualities work will also be
elucidated. The last one will generalize a 0-dimensional QFT, via dimensional
construction through its $D$-Module, and conjecture several connections
between the Lie-algebra-valued extension of the Airy function and the recent
Partition Function found for three-dimensional gravity with a negative
cosmological constant. These three problems, put together, should exhibit a
solid and robust framework for treating QFT under this new paradigm.
\chapter*{Acknowledgements}

The author would like to thank his advisor,
\href{http://en.wikipedia.org/wiki/Gerald_Guralnik}{G. Guralnik}, for invaluable
support, Physics- and other-wise.

Further thanks go to Professor
\href{http://www.brown.edu/Departments/Physics/people/facultypage.php?id=1143831933}{Michael
  Kosterlitz} and \href{http://www.linkedin.com/in/jameseodell}{James O'Dell},
for discussions about Physics (including phase transitions, non-equilibrium
statistical mechanics, symmetry breaking and renormalization methods),
Scientific and High Performance Computing.

Also, the author would like to thank Professors
\href{http://www.physics.brown.edu/people/detail.asp?id=14}{Herbert Fried},
\href{http://www.math.brown.edu/faculty/harris.html}{Bruno Harris}, \href{http://www.cs.brown.edu/~jes/}{John
  Savage},
\href{http://www.chem.brown.edu/people/facultypage.php?id=1106970114}{James
  Baird},
\href{http://www.brown.edu/Departments/Physics/people/facultypage.php?id=1106969931}{Ian
  Dell'Antonio},
\href{http://www.brown.edu/Departments/Physics/people/facultypage.php?id=1106969929}{Antal
  Jevicki}, and
\href{http://www.brown.edu/Departments/Physics/people/facultypage.php?id=1154701476}{Marcus
  Spradlin}, for all forms of discussions (academic or not) and various
``coffee and cookies'' moments.

Finally, \href{http://www.het.brown.edu/people/cengiz/}{C. Pehlevan} is
acknowledged for discussions and useful conversations. This work was supported
in part by funds provided by the US Department of Energy (\textsf{DOE}) under
contracts \textsf{DE-FG02-85ER40237} and \textsf{DE-FG02-91ER40688-TaskD}.
\tableofcontents

\mainmatter
\chapter{Motivations and Introductory Remarks}\label{chap:motives}
The paradigm of research in Physics and Mathematics around the 1960's was that
Physics dealt with local structures (QFT, ``Dual Models'', etc) while
Mathematics dealt with global tools (Morse Theory, Cohomologies, Index Theorems,
etc).

Nowadays, this situation is somewhat reversed, with physicists worrying about
questions of Yang-Mills over Riemann surfaces, Higgs Bundles, Topological
Quantization, etc; while mathematicians have gotten their hands full with
structures inspired by QFT and String Theory, e.g., $D$-Modules.

Thus, it seems to be that there is a convergence of these pictures into a single
language: after $\sim\! 50$ years it looks possible to translate between these
two pictures, i.e., it seems possible to relate objects from these two
perspectives.

Therefore, a deeper investigation into the properties and features of QFT seems
a reasonable thing to do. Studies in dualities (AdS/CFT, Langlands), emergent
properties (effective theories), non-commutative structures, non-perturbative
effects (branes), etc, seem like a rich venue to follow.

The present work started as a means to deepen the understanding between symmetry
breaking and the multiple solutions\footnote{Henceforth, when referring to QFT
  and its solutions, we shall use interchangeably the following terms:
  \emph{solution space}, \emph{configuration/phase space}, \emph{vacuum
  manifold} and \emph{moduli space}.} of a certain quantum field theory, in the
spirit already developed in \cite{garciaguralnik}.

To better understand this line of argument, let us focus on 0-dimensional
(i.e. ultralocal) models for the sake of simplicity. The idea is to compute the
Schwinger-Dyson Equations (SDEs) of the theory in question and solve it
analytically. Generically speaking, given that most theories are defined by a
Potential which is a polynomial in the fields, the SDEs are expected to have
more than one solution. Therefore, not all of these solutions will admit a
series expansion in terms of the coupling constants (which are normally called
``perturbative series''), implying that some solutions are non-perturbative,
even though they admit a series expansion in some other parameter. In this
sense, some of these solutions may keep the original symmetries (present in the
Action), but some may not. To fix the idea, let us look at two particular
examples.

\paragraph{Airy Potential} This theory is defined (in 0-dimensions) by $S(\phi)
= \phi^3/3 + J\, \phi$, and its Feynman Path Integral (FPI) representation is
given by,

\begin{equation*}
  \mathcal{Z}(J) = \frac{\displaystyle\int_{\mathds{C}} e^{i\, \frac{\phi^3}{3}
    + i\, J\, \phi}\, d\phi}{\displaystyle\int_{\mathds{C}} e^{i\,
    \frac{\phi^3}{3}}\, d\phi} < \infty \; ;
\end{equation*}
where $\mathds{C}$ is the contour of integration that renders this FPI
finite. On the other hand, its SDE is obtained in the following way:

\begin{equation*}
  (\phi^2 - J)\, \mathcal{Z}(J) = 0 \xrightarrow{\phi\, \mapsto
      -i\, \partial/\partial J} (\partial_J^2 - J)\, \mathcal{Z}(J) = 0 \; .
\end{equation*}

Now we make an important observation: the SDE above is second order and, thus,
must have two solutions. As it turns out, these solutions are given by the
$\Ai(J)$ and $\Bi(J)$ functions. At this point, it is not difficult to realize
that the FPI is nothing but the integral representation of the differential
problem posed above, which implies that,

\begin{equation*}
  \mathcal{Z}[J] = \frac{\displaystyle\int_{\mathds{C}_{1,2}} e^{i\,
    \phi^3/3 + i\, J\, \phi}\, d\phi}{\displaystyle\int_{\mathds{C}_{1,2}}
    e^{i\, \phi^3/3}\, d\phi} = \Ai[J] \text{ or } \Bi[J] \; ;
\end{equation*}
where the contours $\mathds{C}_{1,2}$ are such that the integral representation
above is finite.

The realization that should be accomplished at this stage is that $\phi$-space
had to be ``complexified'' so that we would be able to find both of these
solutions --- usually, when we think in terms of a FPI, it is \emph{presumed}
that the integration above should be done for $\phi \in \mathbb{R}$. However, as
we just showed above, in this very simple example we already see that if we do
not extend the FPI representation to $\phi \in \mathbb{C}$, we will not obtain
\emph{all} of the possible solutions to the problem at hand.

Having said that, it is important to realize that the partition function, in its
most general form, will be given by a linear combination of all possible
solutions, i.e.,

\begin{equation*}
  \mathcal{Z}(J) = \alpha\, \Ai(J) + \beta\, \Bi(J) \; ;
\end{equation*}
where $\alpha, \beta \in \mathbb{C}$ are scalars.

Further, the asymptotic series of $\Ai(J)$ and $\Bi(J)$ has a different form in
different quadrants of the complex plane, a fact known as the Stokes
phenomenon, implying that $\mathcal{Z}(J)$ will be a meromorphic
function. Therefore, resummations of these asymptotic series, if it has to be
done at all, must be tackled with the utmost care and attention, once mixing
different branches of these series renders all results obtained after that
completely useless.

Also, because $J\in \mathbb{C}$, we can study the partition function in terms of
its properties with respect to the M{\"o}bius group, which is the automorphism
group of the Riemann sphere. In particular, we may focus in one of its more
important subgroups, the Modular Group (SL$(2,\mathbb{Z})$, where we identify
the elements $A$ and $-A$), and study all of the connections with modular forms,
elliptic curves and many fractals --- some connections with String Theory can be
drawn once it is realized that the generators of this group are given by ``unit
translation to the right'' and ``inversion in the unit circle followed by
reflection about the line $\Re(z)=0$'', where this last generator is analogous
to T-duality (which can be understood in terms of Mirror Symmetry).

This is an interesting example because the cubic potential is usually said to
be unbounded from below and, therefore, yields no meaningful theory. However,
we clearly see above two solutions which are \emph{finite}. This happens because
we are explicitly looking for boundary conditions to solve the SDE, a fact that
is completely analogous to searching for contours that render the Path Integral
finite.

Furthermore, this example can be readily extended from a scalar-valued field
$\phi$, to more interesting cases such as that of a matrix-valued field
$\boldsymbol{\Phi}$, or of a Lie-algebra valued field $\boldsymbol{\varphi}$. As
expected, the only modification to the above equations is that the Airy
function is extended analogously to either a matrix-valued or
Lie-algebra-valued function --- in fact, these extensions are \emph{defined}
via the FPI representation above \cite{lieairy}.

Once these extensions are established, going from 0-dim to $d$-dim is
accomplished via the use of $D$-Modules: in this sense, the $D$-Modules are used
to compute the particular ``Thermodynamic Limit'' arising from the 0-dim
theory at hand. This will be important for future developments.

It is important to emphasize that the \emph{observables} of a theory should be
real, once they are the measurable quantities. But this says nothing about the
nature of the partition function nor of the fields themselves.

\paragraph{$\boldsymbol{\lambda\, \phi^4}$ Potential} This theory is defined by $S(\phi) = \mu\, \phi^2/2
  + \lambda\, \phi^4/4$. The SDE and FPI for this model are:

\begin{align*}
  &(\mu\, \phi + \lambda\, \phi^3 - J)\, \mathcal{Z}[J] = 0
  \xrightarrow[g=\mu/\lambda]{\phi\, \mapsto -i\, \partial/\partial J}
    (\partial_J^3 + g\, \partial_J - J)\, \mathcal{Z}[J] = 0 \; ; \\
  &\\
  \mathcal{Z}[J] &= \frac{\displaystyle\varint_{\mathds{C}_{1,2,3}} e^{i\,
    (\mu\, \phi^2/2 + \lambda\, \phi^4/4 - J\, \phi)}\,
    d\phi}{\displaystyle\varint_{\mathds{C}_{1,2,3}} e^{i\, (\mu\, \phi^2/2 +
    \lambda\, \phi^4/4)}\, d\phi} = U[g,J] \text{ or } V[g,J] \text{ or }
    W[g,J] \; ;
\end{align*}
where $g=\mu/\lambda$, and the three solutions for the SDE are given by the
Parabolic Cylinder Functions, $U[g,J]$, $V[g,J]$ and $W[g,J]$, which, in turn,
can be obtained with three different contours, $\mathds{C}_{1,2,3}$, in the FPI
representation (see Appendix \ref{sec:pcf}).

As mentioned earlier, each one of these parabolic cylinder functions
represents one particular solution: symmetric, broken-symmetric and solitonic.

Again, all of the considerations made above can be readily applied in this case:
complexification of $\phi$-space in order to obtain all possible solutions, the
partition function is a linear combination of these three solutions, Stokes
phenomenon is present in the asymptotic series for each one of these solutions,
the partition function is meromorphic and, finally, its properties with respect
to the Modular Group can be studied.
\section{The road ahead} \label{sec:roadahead}
We have just shown how extremely important boundary conditions (resp. contours)
are for properly defining a QFT and all of its solutions/sectors (resp. moduli
space); and this is the cornerstone of our work: from this point on, we can
attack several different problems in novel ways.

Here are the problems that we tackle in this thesis:

\begin{description}
\item[Mollifying QFT] We probe these multiple solutions using the help of a
  mollifier, which is essentially a band-pass filter that eliminates the
  meaningless highly-oscillatory parts of the Partition Function, but does
  retain its meaningful structure. This shows clearly the links with Lee-Yang
  Zeros (complex zeros of the Partition Function, seen as a polynomial in the
  coupling constants) and Stokes phenomena (the asymptotic expansion of each
  solution has a parameter-dependent limit, which means that there is a
  discontinuous change [in the Partition Function] when crossing Stokes lines
  separating distinct solutions/sectors of the theory). For more on this, please
  see \cite{mollifier}.
\item[Topological Transitions and Geometric Langlands Duality] We show that each
  different solution/sector of the theory has a distinct topology which can be used
  as a Superselection Rule (topological charge); moreover, the Euler
  Characteristic (second Chern class) of each phase gives its quantization
  condition. Also, these sectors can be labeled by the discriminant of a
  polynomial constructed from the potential, where the structure of this
  discriminant is related to singularities in the Partition Function, which, in
  turn, must be treated as a meromorphic function (which connects with the
  picture above about Lee-Yang zeros and Stokes Phenomena). However, the
  different solutions of the same theory are related by ``dualities'' in the
  values of the parameters of the theory: distinct boundary conditions establish
  different allowed ranges for these parameters. In this sense it is possible to
  construct several ``dualities'' that translate one sector into another, and
  this is how we make touch with the subject of Geometric Langlands
  Duality. Furthermore, using $D$-Modules, we can take the Thermodynamic Limit of
  each particular 0-dim theory into its full $d$-dim version, thus dimensionally
  constructing each different sector of the theory: each branch is ``grown''
  out of the ``natural differential operator'' defined by the specific potential
  and its parameters (determined by the appropriate boundary
  conditions). Therefore, these dualities are not only with respect to different
  sectors of the theory, but also with respect to each natural differential
  operator associated to each phase, thus establishing the possible allowed
  values for the geometric quantities (connection) associated to each point in
  the moduli space of the whole theory. For more on this, please refer to
  \cite{ferrante,glang}.
\item[Three-dim Gravity and Phase Transitions] This is where we make use of the
  extension to Lie-algebra-valued fields mentioned above when solving the Airy
  potential. Recently, Witten has shown that 3-dim gravity with negative
  cosmological constant has a dual 2-dim CFT, \cite{3dgravity}. In order to do
  so and obtain a holomorphically factorized Partition Function, he had to move
  from real saddle points (which do not account for the whole theory) to
  \emph{complex} ones. And, in doing so, he found that the Hawking-Page phase
  transition is analogous to the condensation of Lee-Yang zeros. The question we
  are tackling is that of extending the Airy function as previously discussed
  (to a Lie-algebra valued Airy function) and ``grow'' this 0-dim solution into
  a 3-dim one using its $D$-Module. Furthermore, there are some questions that
  arise from the connection with Braid and Knot theory, such as whether having
  two distinct Partition Functions (for $\Ai$ and $\Bi$) imply different
  polynomial invariants, or whether this implies that the structure constants of
  the Lie group in question can be dynamically determined (by the allowed
  contours), etc.
\end{description}

\part{Mollifying Quantum Field Theory}
%
%
\chapter{Motivation and Introduction} \label{sec:intro}
\section{Sign Problem}
One of the fundamental difficulties of Monte Carlo (MC) approaches is known as the ``sign
problem''. It is encountered when the functional integrals to be evaluated do not have a
positive definite measure. It is not related to any approximations or fundamental errors in
the MC scheme, but it describes the situation where the statistical error can become very
large. In general, any expectation value can be written as,

\begin{equation}
  \label{eq:genvev}
  \langle O \rangle = \frac{\displaystyle\varint O[\phi]\, \mu[\phi]}{\displaystyle\varint\mu[\phi]} \; ,
\end{equation}
where $\mu$ and $O$ are real-valued functions of the field variables and, in general, the
measure $\mu[\phi]$ need not be positive: in Lorentzian QFT this measure is
complex-valued and given by $e^{i\, S[\phi]}\, \mathcal{D}\phi$. However, if $\mu[\phi]$
changes sign, it cannot be considered a probability density (\textit{i.e.}, a measure).

The standard trick to avoid this problem is to modify the measure in the following way:
$\tilde\mu[\phi] = |\mu[\phi]|/\varint|\mu[\phi]|$. Then, one absorbs the sign of $\mu[\phi]$ in the
quantity to be measured:

\begin{equation}
  \label{eq:signtrick}
  \langle O \rangle = \frac{\displaystyle\varint O[\phi]\,\mathrm{sign}\bigl(\mu[\phi]\bigr)\,
    \tilde\mu[\phi]}{\displaystyle\varint\mathrm{sign}\bigl(\mu[\phi]\bigr)\,\tilde\mu[\phi]} \; .
\end{equation}
In some cases this may work. However, the random walk guided by $|\mu[\phi]|$ is very likely
to predominantly sample unimportant regions in phase space.

At this time no completely satisfactory solution to the sigh problem
exists, although there are some promising attempts including positive
projection \cite{pp1,pp2}, fractal decomposition scheme \cite{fd}, and
Berry's phase \cite{bp} and its Stiefel manifold \cite{sm}. We believe
that the techniques discussed in the present work lay the grounds
for another possible solution path. The general success of the methods
studied here will depend on the development of more computationally
efficient algorithms than the ones used to illustrate the simple
examples treated below.
\section{Smoothing out the Measure}
The idea behind the mollification technique is to use a convolution in order to smooth
out and filter the measure: the highly oscillatory measure given by $e^{i\,
  S[\phi]}\, \mathcal{D}\phi$ is convoluted with some suitable function (called
\emph{a mollifier}; see Appendix \ref{sec:background} for more details) and as a
result only an effective contribution is left (rapid oscillations of
the measure cancel out when integrated over the slowly varying mollifier).

A convolution is an integral that expresses the amount of overlap of one function $g$ as it
is shifted over another function $f$. It therefore ``blends'' one function with
another. In mathematics, mollifiers are smooth functions with special properties, used in
distribution theory (generalized functions) to create a sequence of smooth functions
approximating non-smooth functions, via a convolution.

In an  application of this idea to QFT we will replace the rapidly oscillating
measure in Minkowski space $(I = \exp\{i\, S[\phi]\})$ by smooth functions
$(I_{\epsilon} = \conv{\eta_{\epsilon}}{I})$ so that we can run numerical
simulations that are otherwise prohibitive (see \ref{sec:background}). At this
stage, the only requirement that is made is that: $\int \eta_{\epsilon}(x)\, dx
= 1$, where the integral is performed over the domain of $\eta_{\epsilon}$. The
parameter $\epsilon$ controls the approximation, the smoothness, being made, and
in the $\epsilon\rightarrow 0$ limit, the original expressions are
recovered. (The meaning of this $\epsilon$ parameter is that of dilating or
contracting the mollifier itself, thus controlling the range over which the
filtering is done. Note that the calculations should be done using a
non-vanishing value for $\epsilon$ in order to avoid the analytical expressions
which are hard to handle numerically.)
\chapter{Mollifying Quantum Field Theory} \label{sec:qft}
The technique described above (and detailed in \ref{sec:background})
can be applied to [Lattice] QFT in the following way:

\begin{equation}
  \label{eq:mollqft}
  \mathcal{Z}_{\epsilon}[J] \equiv \varint \biggl\{\varint \eta_{\epsilon}[\phi - \varphi]\,
    e^{i\, S[\varphi;J]}\, \mathcal{D}\varphi\biggr\}\, \mathcal{D}\phi \; .
\end{equation}
This form is useful to set up numerical computations. Performing the
mollification before performing the path integral can be very advantageous.

Mollifying the integrand (i.e., the complex exponential of the action)
smooths the highly oscillatory integral. The integrand changes its form, from
the canonical $I$ to the mollified $I_{\varepsilon}$, in the following fashion,

\begin{equation*}
  I[\phi;J] = e^{i\, S[\phi;J]} \; \longmapsto\; I_{\epsilon}[\varphi; J] =
    (\conv{\eta_{\epsilon}}{I})[\varphi; J] = \varint
    \eta_{\epsilon}[\varphi - \phi]\, e^{i\, S[\phi;J]}\, \mathcal{D}\phi \; ,
\end{equation*}
i.e., the convolution with $\eta_{\epsilon}$ changes variables: $\phi \mapsto
\varphi$. Taking [functional] derivatives of $\mathcal{Z}$ (which
yield Green's functions) is just the same as taking them with respect to
$\mathcal{Z}_{\epsilon}$, because the derivative operator commutes with the
mollification.

The [Feynman] Path Integral is constructed just as before:

\begin{align}
  \nonumber
  \mathcal{Z}[J] &= \varint I[\phi;J]\, \mathcal{D}\phi \; ; \\
  \nonumber
  \mathcal{Z}_{\epsilon}[J] &= \varint I_{\epsilon}[\varphi;J]\, \mathcal{D}\varphi\; .\\
  \intertext{Note that,}
  \nonumber
  \mathcal{Z}_{\epsilon}[J] &= \varint\Biggl\{ \underbrace{\varint \eta_{\epsilon}[\varphi
    - \phi]\, e^{i\, S[\phi;J]}\, \mathcal{D}\phi}_{I_{\epsilon}[\varphi;J]} \Biggr\}\,
    \mathcal{D}\varphi\; . \\
    \label{eq:zeqze}
  \therefore\; \mathcal{Z}_{\epsilon}[J] &\equiv \mathcal{Z}[J] \; ;
\end{align}
where standard properties of convolutions (see \cite{dtta}) have been used in the last
step.

Even though the above result is analytic, when one goes to the simulations, a
small dependence on $\epsilon$ shows up. (See Section \ref{subsec:tm} for a more detailed
discussion on this matter.)
\section{Importance Sampling}\label{subsec:isspm}
The next step consists of choosing an
appropriate sampling function (for the Monte Carlo simulation):

\begin{equation*}
  \mathcal{Z}_{\epsilon}[J] \equiv \varint \biggl\{\varint \eta_{\epsilon}[\phi - \varphi]\,
    \frac{e^{i\, S[\varphi;J]}}{W[\varphi]}\, \mathcal{D}W\biggr\}\, \mathcal{D}\phi \; ,
\end{equation*}
where $\mathcal{D}W = W[\varphi]\, \mathcal{D}\varphi$, i.e., just a change of
variables in order to make the computations more robust: it is possible to use the
integrand profile in order to speed up the calculation.

An appropriate choice for the importance sampling function, $W$, is usually
given by $W_{\epsilon}[\phi] = \big|I_{\epsilon}[\phi]\big|$, \cite{stcqp}.

A little digression is in order: If all we wanted to do was to simulate a
certain [Euclidean] QFT on a lattice, the above reasoning would be just as
valid, modulo the mollification, i.e., we would conclude that the proper
importance sampling function was $W[\phi] = \big|I[\phi]\big|$. The full-fledged
formula is useless for doing computations since it implies full knowledge of
the theory being calculated. A better choice is a simple approximation: we
choose a saddle-point approximation in order to make the expression more
manageable and still keep most of the characteristics of the integrand.

Thus, the importance sampling function ($W[\phi]$), which is used just to better
guide the simulation, would be nothing but the absolute value of the sum over
all saddle-points of the theory in question. Assuming we have only one such
saddle-point, the well known answer is given by:

\begin{align}
  \nonumber
  W[\phi] &\approx \Biggl| \frac{e^{i\, S[\phi_0]}}{\hksqrt{\det\bigl\{-\partial^2 + m^2 +
    \mathcal{V}''[\phi_0]\bigr\}\,{}}} \Biggr| \; ; \\
  \nonumber
  &= \Bigl| e^{i\, S[\phi_0]}\Bigr|\, \Bigl| e^{-\frac{1}{2}\,
    \mathrm{tr}\left\{\log(-\partial^2 + m^2 + \mathcal{V}''[\phi_0])\right\}} \Bigr|\;;\\
  \label{eq:sadw}
  &= e^{-\frac{1}{2}\, \mathrm{tr}\{\log(-\partial^2 + m^2 + \mathcal{V}''[\phi_0])\}}
    \equiv \Bigl(\det\bigl\{-\partial^2 + m^2 + \mathcal{V}''[\phi_0]\bigr\}\Bigr)^{-1/2}\;.
\end{align}

Using this to find an importance sampling function for a simple scalar QFT
on the lattice requires solving for the determinant above. In Lattice QCD, the 
problem of the \emph{fermion determinant} is a very time-consuming
operation that must be done for \emph{every} step of the MC calculation.

But this is not the only computational bottleneck. The more severe one comes from the so
called \emph{sign problem}: The exponent in $I[\phi]$ is not bounded from below,
therefore one cannot guarantee the ergodicity of the Markov Chain underlying the Monte
Carlo draws, i.e., a much bigger number of draws would be needed in order to
yield any meaningful answer. This is the reason why calculations are done in
Euclidean space rather than Lorentzian/Minkowski space. In Euclidean space
(Wick-rotating the integrand), the exponent in question becomes $I_E[\phi] =
e^{-S[\phi]}$: it is bounded from below and the Monte Carlo method works fine
(the Markov Chain behind it becomes ergodic).

It is to tackle this problem that the method of mollifiers comes in: mollifying $I[\phi]$
will yield a smooth function, $I_{\epsilon}[\phi]$, whose properties help the
convergence of the MC computation. Furthermore, if this can really be accomplished and
calculations with imaginary exponents become a reality, the next logical step is to
analyze the different phases of the given QFT. As shown in \cite{tvbcsde}, the different
phases of a theory can be picked out with a mere choice of boundary conditions, which is
the same as properly defining the measure of the [Feynman] Path Integral. The
upshot is that if we want to obtain answers other than the ones that can
be reached via perturbation theory, either the measure of the path integral or
the boundary conditions of the Schwinger-Dyson equations have to be modified
\cite{tvbcsde}.

Once there is nothing preventing the measure of the [Feynman] Path Integral,
from being (in the most general case) complex-valued --- in order to account for
the phase structure of the theory ---, the fact that its integrand is also
complex-valued has to be taken more seriously, afterall this is a
non-perturbative result: a perturbative series only works if we know, \emph{a
  priori}, in which phase we are working, so it can be tailored to that
particular sector of the theory.

Thus, the need to address the sign problem becomes even clearer: it is not
simply a problem of doing MC in Lorentzian/Minkowski space, it is mostly a
problem of being able to compute all possible solutions --- all the different
phases --- of a given QFT.
\section{Mollifying the Importance Sampling Function}\label{subsubsec:mis}
As outlined before, the idea is to use the mollification technique to handle
the highly oscillatory terms present in the importance sampling function.

There are \emph{three} possible choices for the sampling function $W$:

\begin{enumerate}
\item Taylor-expand and mollify the complex integrand: \textit{no} knowledge about the
  saddle-points is necessary, but it is very computer intensive, which is the
  reason we will not focus on it;
\item As done in \eqref{eq:fieldw}, where the integrand is saddle-point expanded and
  mollified, but the \textit{explicit} form of the mollifier is used in order to perform
  the remaining integral: not so heavy on the computer, but some knowledge about the
  saddle-points is needed;
\item As done in \eqref{eq:mollw}, where the integrand is saddle-point expanded and
  mollified, generalizing the standard saddle-point approximation: less intensive of all
  3, but even more knowledge about the saddle-points is needed.
\end{enumerate}

The difference between the second and third methods above is that the integral
in \eqref{eq:2ndordergradint} is explicitly carried out in the former, but not in
the latter. As explained below, the third method (listed above) consists of the
simple generalization of the well known saddle-point approximation; however, we
need to note that more knowledge about the particular saddle-points is needed in
order to perform the contour integration involved in this scheme (and this is
the difference with respect to the second method).

The calculation below shows the second method outlined above, known as
\emph{2\,\raisebox{0.9ex}{\tiny nd}-order gradient approximation}:

\begin{align}
  \nonumber
  W_{\epsilon}[\phi] &= \big|I_{\epsilon}[\phi]\big| = \left| \varint
    \eta_{\epsilon}[\phi - \varphi]\, e^{i\, S[\varphi]} \,
    \mathcal{D}\varphi\right| \; . \\
  \intertext{Now, a saddle-point expansion is performed on the action,}
  \nonumber
  W_{\epsilon}[\phi] &\approx \left| \varint \eta_{\epsilon}[\phi - \varphi]\,
    e^{i\, \left(S[\varphi_0] + \frac{1}{2}\, (\varphi - \varphi_0)^2\,
        S''[\varphi_0]\right)} \mathcal{D}\varphi \right| \; . \\
  \intertext{A mollifier needs to be chosen, and for the purposes of this
  calculation, a \emph{Gaussian} one is our choice:}
  \nonumber
  \eta_{\epsilon}[\phi - \varphi] &= e^{-\frac{1}{2}\, (\phi -
    \varphi)^2/\epsilon^2} \; ;\\
  \label{eq:2ndordergradint}
  W_{\epsilon}[\phi] &= \left| \varint_{-\infty}^{\infty} e^{-\frac{1}{2}\,
    (\phi - \varphi)^2/\epsilon^2}\, e^{i\, \left(S[\varphi_0] + \frac{1}{2}\,
      (\varphi - \varphi_0)^2\, S''[\varphi_0]\right)} \mathcal{D}\varphi
      \right| \; ; \\
  \label{eq:fieldw}
  \Rightarrow\; W_{\epsilon}[\phi] &= \left| \hksqrt{2\, \pi\, \epsilon^2\xspace}\,
    \frac{\exp\left\{i\, S[\varphi_0] + \frac{i}{2}\, \frac{(\phi - \varphi_0)\,
    S''[\varphi_0]\, (\phi - \varphi_0)}{1 - i\, \epsilon\, S''[\varphi_0]\,
    \epsilon}\right\}}{\hksqrt{\det\big\{1 - i\, \epsilon\, S''[\varphi_0]\,
    \epsilon\big\}\xspace\,}}  \right| \; .
\end{align}

The important point to note about the derivation above is that the \emph{explicit}
form of the mollifier, $\eta_{\epsilon}$, had to be used and, because of 
that and the nature of the MC simulation, we do not necessarily need a good knowledge
about $\varphi_0$. This means that we can compute the above importance sampling function,
\eqref{eq:fieldw}, and use a \emph{trial} $\varphi_0^{\text{trial}}$: the [MC] simulation
will do the job of moving towards the exact $\varphi_0$ and pick out the different
phases of the theory. As shown elsewhere, \cite{nspmc,rtnnm,mmc,stcqp}, this works quite
fine.

However, for more relevant cases (e.g. 4-dimensional QFTs) the time required for
simulations becomes unrealistic. The bottom-line is that a simple-minded
implementation of the method will have the code/com\-put\-er doing the work of
mapping the $(\phi,\varphi)$-space, while one could use that information
beforehand in order to speed things.

The way out of this is to generalize the following well known result:
$\mathcal{I}(s) = \int_{\mathds{C}} g(z)\, e^{s\, f(z)}\, dz = \hksqrt{2\,
  \pi\,{}}\, g(z_0)\, e^{s\, f(z_0)}\, e^{i\, \alpha}/\hksqrt{|s\, f''(z_0)|}\,
;\; s\in\mathbb{R}, \; z\in\mathbb{C}$; i.e.,

\begin{equation}
  \label{eq:mollw}
  W_{\epsilon}[\phi] = \left| \hksqrt{2\, \pi\,{}}\, \frac{\eta_{\epsilon}[\phi - \varphi_0]\,
    e^{i\, S[\varphi_0]}}{\hksqrt{\left|\det\bigl\{S''[\varphi_0]\bigr\}\right|\,{}}}\right| \; .
\end{equation}

At this point, the similarity between \eqref{eq:sadw} and
\eqref{eq:mollw} is clear, in fact, they are the same formula, except that, in
the former case there is no mollification while in the latter there is
mollification.
\chapter{Tuning the Mollification} \label{subsec:tm}
At this point we are almost done with the analysis of the mollification technique, the
only aspect remaining being the optimal choice of $\epsilon$, the parameter that regulates
the approximation.

From the discussion in appendix \ref{sec:background}, it is clear that in the
$\epsilon \rightarrow 0$ limit we recover the original theory. However,
numerically, this is not exactly what happens, because of numerical and
statistical fluctuations. Therefore, the choice of an optimal value for
$\epsilon$ is crucial.

Using an information theoretic viewpoint, we can choose $\epsilon$ to
optimally compress $W_{\epsilon}[\phi]$ around each stationary phase
point. Therefore, we can define an information entropy based on
$W_{\epsilon}[\phi]$ in the following way:

\begin{equation}
  \label{eq:infentropy}
  \mathcal{S}[\epsilon] = - \frac{\displaystyle \varint W_{\epsilon}[\phi]\, \log\bigl(
    W_{\epsilon}[\phi]\bigr) \, [d\phi]}{\biggl(\displaystyle\varint W_{\epsilon}[\phi]\, [d\phi]\biggr) +
    \log\biggl(\displaystyle\varint W_{\epsilon}[\phi] \, [d\phi]\biggr)} \; .
\end{equation}
The optimal $\epsilon$ is the one that minimizes $\mathcal{S}[\epsilon]$. This
value corresponds to maximal compression of the information in the importance
sampling function.

In order to illustrate the above, let us consider a very simple [0-dimensional] example,
given by $S[\phi] = m^2\, \phi^2/2$. In this case, the saddle-point is $\phi_0=0$ and the
importance sampling function and the information theoretic entropy are given by:

\begin{align*}
  W_{\epsilon}[\phi] &= \frac{e^{-\phi^2/2\epsilon^2}}{\hksqrt{2\,\pi\,\epsilon^2}}\,
    \frac{e^{i\, S[\phi_0]}}{\hksqrt{S''[\phi_0]}} \; ;\\
  &= \frac{e^{-\phi^2/2\epsilon^2}}{m\, \epsilon\, \hksqrt{2\, \pi}} \; ;\\
  \mathcal{S}[\epsilon] &= \frac{\log\bigl(m\, \epsilon\,
    \hksqrt{2\, \pi}\bigr) - \epsilon^2}{1 - m\, \log(m)} \; .
\end{align*}

Plotted below are the graphs of the importance sampling function (also showing how it
varies with $\epsilon$) and of the information theoretic entropy (both using $m = 1$):
\begin{center}
  \hfill
  \includegraphics[scale=0.6]{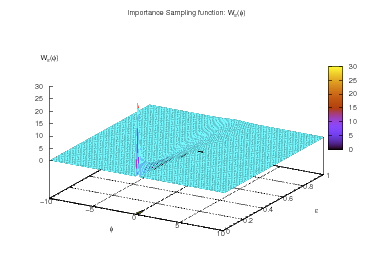}
  \hfill
  \includegraphics[scale=0.47]{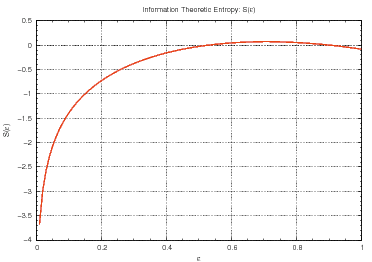}
  \bigskip \bigskip

  {\footnotesize \noindent\textbf{Figure 1:} Importance Sampling function
    (leftmost plot, $m = 1$, $W_{\epsilon}[\phi] = e^{-\phi^2/2\epsilon^2}/\epsilon\,
    \hksqrt{2\, \pi}$) and information theoretic entropy (rightmost plot, $m =
  1$, $\mathcal{S}[\epsilon] = \log\bigl(\epsilon\, \hksqrt{2\, \pi}\bigr) -
  \epsilon^2$) for the action $S[\phi] = m^2\, \phi^2/2$.}
\end{center}

We can clearly see the importance function peaking around $\phi=0$, which is the expected
behavior (\textit{i.e.}, it happens around the saddle-point $\phi_0=0$), while the
information theoretic entropy shows that the best parameter for the
mollification is $\epsilon \rightarrow 0$: this is not desirable, once
$\lim_{\epsilon \rightarrow 0}\mathcal{S}(\epsilon) \rightarrow -\infty$, but it 
simply shows that we should use a small non-vanishing value for $\epsilon$, as
mentioned before in Section \ref{sec:intro}.
\chapter{Simulation Details}\label{subsec:std}
In order to perform the simulation as outlined in \ref{subsubsec:mmls}, some technical
details had to be taken into more serious account. Among them, the important ones are:
random-number generation and non-local algorithms. Those will be the bottle-necks, as
noted in \cite{selcrngws}.
\section{Random Number Generators}
Even though some consider the random-number question a solved one, this is not
always the case \cite{selcrngws}. For this reason, the choice of which
random-number generator (RNG) to use is still a critical one, not only for speed reasons
but because of systematic errors as well. In order to address this issue the choice made
was for the Mersenne Twister RNG \cite{mtrng} (period of $2^{19937} - 1$).
\section{Non-local Algorithms}
As for non-local algorithms, the choice made was for a Genetic Algorithm (GA)
\cite{ugosumcmci}. The ``population'', in the Lattice QFT sense, consists of the lattice
(\textit{i.e.}\/ spacetime) points. For a given ``recombination rate'', an initial
population gets \emph{randomly} divided into pairs of [lattice] points and
``crossing-over'' (or ``genetic'') operators are applied to these pairs in order to
generate different sets of pairs. In this context, the crossing-over operators exchange
the coordinates of given points, thus there are 4 of those operators, namely:
$T_t,\, T_x,\, T_y\,, T_z$. That is, if the crossing-over operator $T_x$ is
applied to the points $\phi = (\phi_t, \phi_x, \phi_y, \phi_z)$ and $\phi' =
(\phi'_t, \phi'_x, \phi'_y, \phi'_z)$ the outcome will be: $T_x [\phi,\phi'] =
\{(\phi_t, \phi'_x, \phi_y, \phi_z),\, (\phi'_t, \phi_x, \phi'_y,
\phi'_z)\}$. Analogous definitions are valid for the other crossing-over
operators. The key features for this choice were: which $T_i$ to use is random;
whether or not the change $\phi \mapsto \phi'$ is made depends on a given
probability and the recombination rate can be arbitrarily chosen (although
keeping it below 1\% showed to be a good tune). Moreover, the genetic operators
are unitary, i.e., $T_i = T_i^{-1}$. This implies that $T_i^2 = \mathds{1}$,
which guarantees the so-called \emph{detailed balance} of the MC simulation
(this is the \emph{ergodicity} of the algorithm). At this point, the only step
remaining is the explanation of the \emph{probability profile} used in algorithm
\ref{alg:gl} below.

\begin{algorithm}
  \caption{Genetic Algorithm}
  \label{alg:ga}
  \begin{algorithmic}[1]
    \STATE Choose the recombination rate such that: $0 \leqslant r\leqslant 1$ \COMMENT{usually
      between $0.5\%$ and $1.0\%$}
    \STATE Draw a random number $\beta\in [0,1]$ and compare with $r$
    \IF{$\beta \geqslant r$}
      \STATE Metropolis Monte Carlo loop \COMMENT{using the \emph{random walk} technique}
    \ELSE
      \STATE Genetic loop \COMMENT{see algorithm \ref{alg:gl} below}
    \ENDIF
  \end{algorithmic}
\end{algorithm}
\begin{algorithm}
  \caption{Genetic Loop}
  \label{alg:gl}
  \begin{algorithmic}[1]
    \STATE Choose a random pair: $\{\phi_1, \phi_2\} = \{\phi(t_1,x_1,y_1,z_1),\,
      \phi(t_2,x_2,y_2,z_2)\}$
    \STATE Generate the pair $\{\phi'_1, \phi'_2\} = \{\phi(t'_1,x'_1,y'_1,z'_1),\,
      \phi(t'_2,x'_2,y'_2,z'_2)\} = T_i(\phi_1, \phi_2)$
    \STATE Draw a random number $c\in [0,1]$
    \IF{$c > P(\phi'_1, \phi'_2)/P(\phi_1, \phi_2)$}
      \STATE do \textbf{nothing}
    \ELSE
      \STATE perform the exchange $(\phi_1, \phi_2) \mapsto (\phi'_1, \phi'_2)$
    \ENDIF
  \end{algorithmic}
\end{algorithm}

In the notation used below, $P$ is such probability profile
and $P(\phi'_i,\phi'_j)$ means that this profile is calculated using the [GA generated]
points $\phi'_i$ and $\phi'_j$. It proved useful to chose the mollified
importance sampling function as this profile $P$. Also, note that $\phi_i$ and
$\phi_j$, just like $\phi'_i$ and $\phi'_j$, are just particular values of the
field $\phi$ at the sites $i$ and $j$. A comparison between these two profiles
is then performed.
\section{Euclidean Lattice Simulations}\label{subsubsec:els}
In order to be able to understand better what will happen in the
Lorentzian/Minkowski case, a quick and dirty reminder of the Euclidean one is
presented below.

In what follows, $\Phi$ is the set of all possible field configurations and $||\Phi||$ is its
\emph{cardinality} (i.e., the number of elements in the set). For each element of $\Phi$
(i.e., for each field configuration) one has to average the observable over all
the lattice points (which are computed via Metropolis). Each element in the Markov Chain
(generated by the Metropolis MC algorithm) is denoted by $\phi^{[i]}$. To obtain the final
result, an average over all field configurations is made. (To make clear that the
final goal is the actual lattice computation, the measure is denoted as $[d\phi]$.)

\begin{align*}
  \langle O \rangle &= \frac{\displaystyle\varint O[\phi]\, \exp\bigl\{-S[\phi]\bigr\} \,
    [d\phi]}{\displaystyle\varint \exp\bigl\{-S[\phi]\bigr\} \, [d\phi]} \; , \\
  &= \frac{1}{||\Phi||}\, \sum_{\phi\in\Phi}
    \Biggl\{\frac{\Bigl[\frac{\text{Volume}}{N^d}\Bigr]\cdot \sum_{i=1}^{N^d}
    O[\phi^{[i]}]}{\Bigl[\frac{\text{Volume}}{N^d}\Bigr]\cdot N^d}\Biggr\} \; , \\
  \therefore\; \langle O \rangle &= \frac{1}{||\Phi||}\, \sum_{\phi\in\Phi} \Biggl\{
  \frac{1}{N^d}\, \sum_{i=1}^{N^d} O[\phi^{[i]}]\Biggr\} \; .
\end{align*}
\section{Mollified Minkowski Lattice Simulations}\label{subsubsec:mmls}
In the Minkowski version of the above, the Wick rotation is not performed, therefore we
are left with the original form of the exponent,

\begin{equation*}
   \langle O \rangle = \frac{\displaystyle\varint O[\phi]\, \exp\bigl\{i\, S[\phi]\bigr\}
    \, [d\phi]}{\displaystyle\varint \exp\bigl\{i\, S[\phi]\bigr\} \, [d\phi]} \; .
\end{equation*}

The above functional will be mollified in order to yield more tractable
expressions. The fact that there will be two functional integrations (rather than just
one, like above) should not bring many problems. The real question here stems from the
fact that the mollifier, $\eta$, \emph{mixes} the two of them.

\begin{align}
  \nonumber
  \langle O \rangle &= \frac{\displaystyle\varint\Bigl\{\varint \eta_{\epsilon}[\phi
    - \varphi]\, O[\varphi]\, \exp\bigl\{i\, S[\varphi]\bigr\} \, [d\varphi]\Bigr\}
    [d\phi]}{\displaystyle\varint\Bigl\{\displaystyle\varint \eta_{\epsilon}[\phi - \varphi]\,
    \exp\bigl\{i\, S[\varphi]\bigr\} \, [d\varphi]\Bigr\} [d\phi]} \; ; \\
  \label{eq:mollobs}
  \therefore\; \langle O \rangle &= \frac{1}{||\Phi||}\, \frac{1}{||F||}\,
  \sum_{\substack{\phi\in\Phi \\ \varphi\in F}}\, \Biggl\{\frac{\sum_{i,j=1}^{N^d}\;
    \eta_{\epsilon}[\phi^{[j]} - \varphi^{[i]}]\, O[\varphi^{[i]}]\, \exp\bigl\{i\,
    S[\varphi^{[i]}]\bigr\} / W_{\epsilon}^{\Phi}[\varphi^{[i]}]}{\sum_{i,j=1}^{N^d}\; \eta_{\epsilon}[\phi^{[j]}
    - \varphi^{[i]}]\, \exp\bigl\{i\, S[\varphi^{[i]}]\bigr\} /
    W_{\epsilon}^{\Phi}[\varphi^{[i]}]}\Biggr\} \; ;
\end{align}
where it is understood that the configurations in $\Phi$ are chosen with respect
to the importance sampling function, $W_{\epsilon}^{\Phi}[\phi]$, and the
configurations in $F$ are chosen with respect to a \emph{uniform}
distribution. This happens because the mollification process (namely the
\emph{convolution}) mixes the variables from $\Phi$ and $F$ together. Therefore,
the only way to implement this ``mixing'' is by having a uniform distribution
for $F$ and implementing the ``interaction'' with the variables in $\Phi$ ---
via the mollification --- explicitly, using $\eta$. Note that $||\Phi|| = ||F||$.
\chapter{Spontaneous Symmetry Breaking}\label{subsec:symb}
We have been discussing lattice QFT in Lorentzian/Minkowski spacetime because one
of the main objectives of this paper is to examine a numerical
approach to directly calculate Green's functions in different phases
in QFT. This necessitates being able to evaluate path integrals of
complex exponentials. The arguments for this have been given elsewhere \cite{tvbcsde}
and will be illustrated in a particular example in the following section,
but it is fairly easy to understand why this is the case. The
traditional [Feynman] Path Integral formulation of a QFT involves integration of
the exponential of the action over every field variable at every space
time point. Traditionally, these integrations (assume the action is
written in terms of self adjoint fields) range from negative to
positive infinity along the real axis. In the limit of small
couplings, this form of the [Feynman] Path Integral generates perturbation
theory and the results appear (for finite number of expansion terms) to
be regular at vanishing coupling. Thus, all expansions that are not
regular at vanishing coupling, such as the traditional symmetry
breaking expansion of quartic scalar field couplings, are excluded. In
order to avoid this restriction and produce all possible solutions of
the QFT it is necessary to extend the [Feynman] Path Integral integrations to
complex values of the fields in a way consistent with the field
equations and reality properties of the theory.

Indeed, as shown in \cite{tvbcsde} and section \ref{subsec:r}, the
different phases of the theory emerge through the varied boundary
conditions consistent with the equations of motion. Equivalently,
rather than varying the boundary conditions of equations of motion,
the measure of the Path Integral can be changed. In general the number
of choice of paths of integration for the path integral correspond to
the number of independent solutions of the Schwinger-Dyson differential equations.

If we define a QFT via its [Feynman] Path Integral, all we need to know is the action,
$S[\phi, J]$ (for a field $\phi$ whose source is $J$), of a given model, for then we can
write the generating functional as:

\begin{equation*}
  \mathcal{Z}[J] = \mathcal{N}\, \varint\exp\bigl\{i\, S[\phi; J]\bigr\}\,
    \mathcal{D}\phi\; ,
\end{equation*}
where $\mathcal{N}$ is a normalization constant such that $\mathcal{Z}[J=0] =
1$. The crucial question that remains unanswered in this approach is: \emph{``How does one
  properly define the measure $\mathcal{D}\phi$?''}

The best answer so far (for 4-dimensional systems) says that this can only be
done for \textit{free} QFTs, via the use of cylindrical functions
\cite{Baez,Ashtekar}. (Note that this only happens in the
continuum. In its Lattice formulation, QFT is free from such peculiarities
because of the lattice regularization.)

Analogously, defining a QFT via its Schwinger-Dyson equation,

\begin{equation*}
  \frac{\delta S[-i\, \tfrac{\delta}{\delta J}]}{\delta\phi}\mathcal{Z}[J] - J(x)\,
    \mathcal{Z}[J] = 0 \; ;
\end{equation*}
is equivalent to substituting $\phi \mapsto -i\, \tfrac{\delta}{\delta J}$ in the
action, requiring the equations of motion to be the solutions that extremize it. Note
that the Schwinger-Dyson equations are a system of infinitely many [partial] differential
equations, one per each point of spacetime.

Therefore, when thinking in terms of differential equations, the boundary conditions are
responsible for the phase structure of the theory. Intuitively, the picture that comes to
mind is that of a portion of space divided into as many subsets as there are solutions to
our equations of motion, such that in each of those regions, the equations of motion
satisfy appropriate boundary conditions.

On the other hand, when thinking about the [Feynman] Path Integral, usually it does not
seem bothersome that the measure is not properly well-defined. In fact, determining the
boundary conditions for the Schwinger-Dyson equations is analogous to determining the
measure for the Path Integral. Thus, just like the boundary conditions, the measure is
responsible for the phase structure of the theory (in the integral representation of the
problem).

For completeness sakes, this is how a QFT in different spacetime dimensions, $d$, behaves:

\begin{description}
 \item[$\boldsymbol{d = 0}$] For 0-dimensional QFTs --- i.e., QFT on a point (the
   [Feynman] Path Integral degenerates into a simple integral) --- there is no
   such thing as phase transition since there is no such thing as
   dynamics. However, the phase structure of the theory survives, given by the
   different boundary conditions (resp. measure) needed in order to determine all
   the solutions to the equations of motion.
 \item[$\boldsymbol{d = 1}$] For 1-dimensional QFTs --- i.e., Quantum
   Mechanics ---, again, there is no such thing as phase transitions, for if the
   theory has 2 different vacua we could take a linear combination of them to be
   the ``real'' vacuum state, given that they would be related by tunneling. (For
   the analogous case in Condensed Matter physics, please refer to \cite{llsp1}.)
 \item[$\boldsymbol{d \geqslant 2}$] In this case, phase structure and transition
   exist; this is the complete scenario. (Note that there is no operator that
   relates 2 inequivalent $\theta$-vacua of the theory, therefore they belong to
   different algebras.)
\end{description}

The real lesson to be learned from all of this is that Spontaneous Symmetry Breaking is a
phenomena generated by boundary conditions; whether you use them to define the measure of
the Path Integral or to define the Schwinger-Dyson equation is just a matter of personal
preference. It is not said anywhere that the limits of the Path Integral have to be real;
what we need to have are real observables. In fact, as shown below on section
\ref{subsec:r}, it turns out that in order to have symmetry breaking we need a
measure that is not necessarily real: this will enable the computation of all the
solutions of a given QFT \cite{tvbcsde}.
\chapter{Results}\label{subsec:r}
In what follows, the results obtained thus far are presented. In short, they are in 0,
1 (time) and 4 spacetime dimensions.
\section{Lower Dimensional}\label{subsubsec:ld}
Let us start by addressing the 0- and 1-dimensional results. Plainly and simply put, this
means that we are solving a simple integral in 0 spacetime dimensions since the
[Feynman] Path Integral degenerates into a standard integral and, in the 1-dimensional
(time) case, one will be doing Quantum Mechanics.

For the 1-dimensional results, refer to \cite{stcqp}. There, the Quantum
Chemistry of the problem is fully treated and addressed. Note, however, that the
importance sampling function chosen in \cite{stcqp} is different than the one
used in this work: the quantum chemistry was done using \eqref{eq:fieldw} while we use
\eqref{eq:mollw}.

Below, the 0-dimensional results are summarized and, in order to illustrate the features
of the mollifier technique, different properties of those models are made explicit.
\subsection{Airy Function}
The action for this model is given by: $S(x) = x^3/3 + J\, x$. This is an interesting
model because we can explicitly calculate the partition function and compare it with the
results coming from the Mollified Monte Carlo procedure. Moreover, this model has 2
stationary phase points and the results displayed are from the one in the complex plane
which is not accessible with normal Monte Carlo. The quantity of interest is:
\begin{center}
  \includegraphics[scale=0.45]{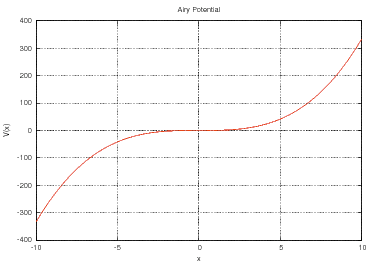}

  \medskip

  {\footnotesize\noindent\textbf{Figure 2:} Plot of the Airy potential, $V(x) = x^3/3$,
    highlighting the fact that it is not bounded from below.}
\end{center}

\begin{equation*}
  \mathcal{Z}[J] = \frac{\displaystyle\int_{-\infty}^{\infty}\, \exp\Bigl\{i\,
    \frac{x^3}{3} + i\, J\, x\Bigr\}\, dx}{\displaystyle\int_{-\infty}^{\infty}\,
    \exp\Bigl\{i\, \frac{x^3}{3}\Bigr\}\, dx} \equiv \frac{\Ai(J)}{\Ai(0)} \; .  
\end{equation*}

The [first two] graphs below show the highly oscillatory behavior of the
integrand in the partition function: the left one is its the real part, while the
right one is its the imaginary part.

\begin{center}
  \hspace{\fill}
  \includegraphics[scale=0.65]{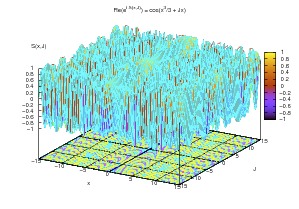}
  \hspace{\fill}
  \includegraphics[scale=0.65]{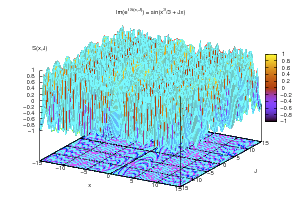}
  \hspace{\fill}

  \smallskip

  {\footnotesize\noindent\textbf{Figure 3:} Real and Imaginary parts of $e^{i\,
      x^3/3 + i\, J\, x}$ showing the highly oscillatory nature of the problem.}
\end{center}

Below we see the graphs of a particular 2-dimensional slice of the above pair,
where $J = -1$: the real (left) and imaginary (right) part of the mollified
Airy-integrand (in red, $\epsilon=0.1$) is in contrast to the non-mollified
integrand (green).

\newpage

\begin{center}
  \includegraphics[scale=0.5]{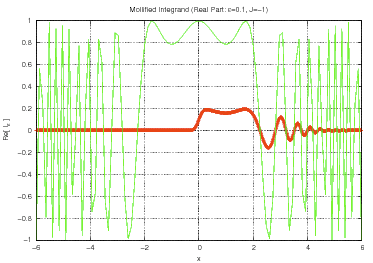}
  \hspace{\fill}
  \includegraphics[scale=0.5]{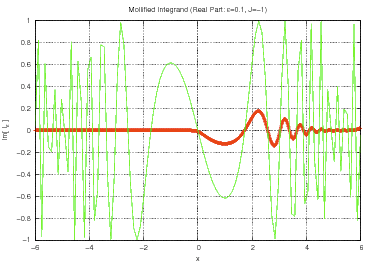}

  \bigskip 

  {\footnotesize\noindent\textbf{Figure 4:} Superimposed plots of mollified
    (red) and non-mollified (green) real and imaginary parts of the integrand
    $e^{i\, x^3/3 - i\, x}$, showing the smoothness achieved on the $J=-1$
    two-dimensional slice of the previous graphs.}
\end{center}

Below we see the graphs of the real (left) and imaginary (right) part of the mollified
Airy-integrand for different values of the field $x$ and of the mollification
parameter $\epsilon$: analogous to Figure 3 but for the mollified version of the
integrand (which corresponds to the red lined plots of Figure 4).

\begin{center}
  \includegraphics[scale=0.5]{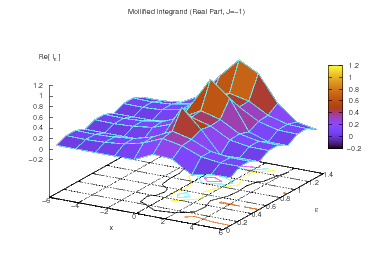}
  \hspace{\fill}
  \includegraphics[scale=0.5]{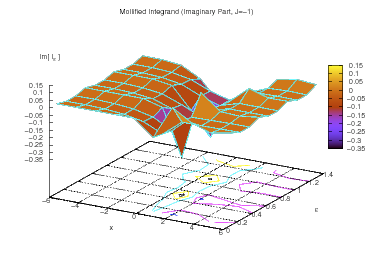}

  \bigskip

  {\footnotesize\noindent\textbf{Figure 5:} Plots of the real and imaginary
    parts of the mollified integrand for various values of $x$ and $\epsilon$.}
\end{center}

The 2 stationary phase points are given by $x_0 \in
\bigl\{\pm\hksqrt{-J}\bigr\}$, where $x_0$ is the solution of $S'(x_0) = 0$, and the
[properly mollified] importance sampling functions are:

\begin{align}
  \label{eq:mollisairypos}
  W_{\epsilon}^{J \geqslant 0}[x] &= \exp\biggl\{-\frac{1}{2}\, \frac{x^2 - J}{\epsilon^2} \biggr\} +
    \exp\biggl\{-\frac{1}{2}\, \frac{x^2 + J}{\epsilon^2} \biggr\}
    \; ;\quad J \geqslant 0 \; ; \\
  \label{eq:mollisairyneg}
  W_{\epsilon}^{J < 0}[x] &= \exp\Biggl\{-\frac{1}{2}\, \biggl(\frac{x -
    \hksqrt{-J\,{}}}{\epsilon}\biggl)^2 \Biggr\} + \exp\Biggl\{-\frac{1}{2}\, \biggl(\frac{x +
    \hksqrt{-J\,{}}}{\epsilon}\biggl)^2 \Biggr\}\; ; \quad J < 0 \; ;
\end{align}
It is customary to see the source term above
called $t$ and identified with time but this is not the approach taken here.

Below we see the graphs of $W_{\epsilon}$ with respect to the $\epsilon$
parameter and $x$, for fixed values of the source: on the left $J=4$, and on the right
$J=-16$,

\begin{center}
  \hspace{\fill}
  \includegraphics[scale=0.55]{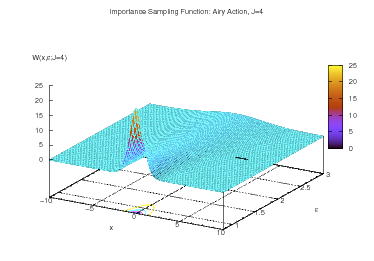}
  \hspace{\fill}
  \includegraphics[scale=0.55]{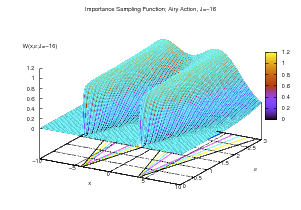}
  \hspace{\fill}

  \bigskip

  {\footnotesize\noindent\textbf{Figure 6:} Plots of \eqref{eq:mollisairypos},
    for $J=4$, and \eqref{eq:mollisairyneg}, for $J=-16$.}
  \bigskip \bigskip

  \includegraphics[scale=0.45]{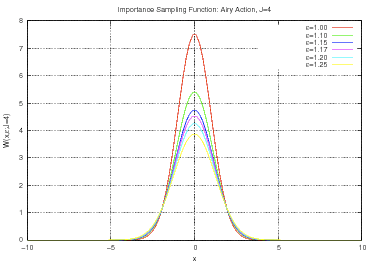}
  \hspace{\fill}
  \includegraphics[scale=0.45]{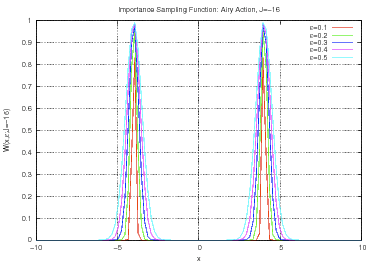}
  \hspace{\fill}

  \bigskip

  {\footnotesize\noindent\textbf{Figure 7:} Plots of 2-dimensional slices of
    \eqref{eq:mollisairypos} and \eqref{eq:mollisairyneg} for various values of
    $\epsilon$.}
\end{center}
\bigskip

It is not difficult to see that there is an optimal value for the parameter
$\epsilon$, as shown below in the graph for the information theoretic entropy (see
Appendix \ref{sec:entropy} for further details): we have the information theoretic entropy
for $J\geqslant 0$ on the left, and $J<0$ on the right,

\begin{center}
  \includegraphics[scale=0.55]{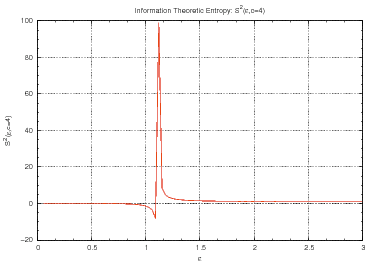}
  \hspace{\fill}
  \includegraphics[scale=0.55]{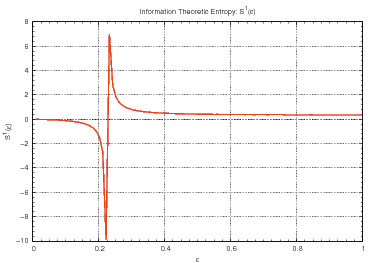}

  \bigskip

  {\footnotesize\noindent\textbf{Figure 8:} Information theoretical entropy for
    finding the optimal value of the mollification parameter: on the left we
    have $J \geqslant 0$, and on the right we have $J < 0$.}
\end{center}
\bigskip

Furthermore, it is easy to perform the analytical calculations outlined above (using the
Airy functions on the Generating Functional, rather than its mollification) to get the
pure evaluation of this model. The results show the mollified answers (given a
proper choice of $\epsilon$) are sensationally accurate and show, already in this
simple example, that a region not allowable in ordinary Monte Carlo approaches can be
computed. In fact, due to the simplicity of this example, modern computers can
calculate it straightforwardly, without the need of special tricks to handle the highly
oscillatory integrand. The point of this example is twofold: show the precision and the
accuracy that the mollifier method can achieve and also to illustrate how this smoothing
procedure works.
\subsection{0-dimensional $\phi^4$ Theory}
Also known as ultra-local $\phi^4$, its action is given by: $S[\phi] = \mu\, \phi^2/2 +
g\, \phi^4/4$. As before, the saddle-point $\phi_0$ is such that $S'[\phi_0] = 0$. The
graphs for the Action above (positive mass on the left, negative mass on the right) are
given by:

\begin{center}
  \hfill
  \includegraphics[scale=0.5]{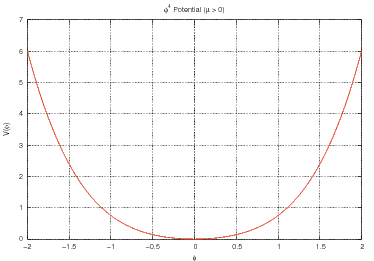}
  \hfill
  \includegraphics[scale=0.5]{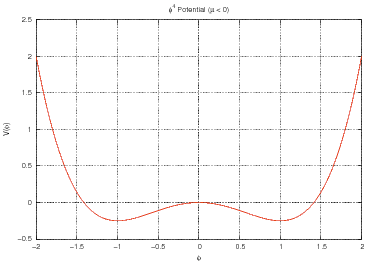}

  \bigskip

  {\footnotesize\noindent\textbf{Figure 9:} $\phi^4$ potential for positive and
    negative values of $\mu$.}
\end{center}

Below, we show the highly oscillatory behavior of the integrand of the Partition Function
given by the above action, i.e., $I = e^{i\, (\mu\, \phi^2/2 + g\, \phi^4/4)}$, for
positive and negative values of the $\mu$ parameter:

\begin{center}
  \hspace{\fill}
  \includegraphics[scale=0.5]{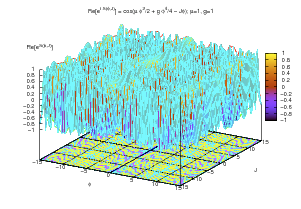}
  \hspace{\fill}
  \includegraphics[scale=0.5]{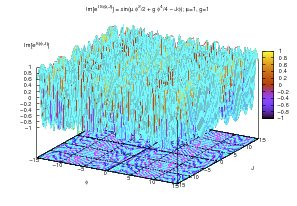}
  \hfill

  \hfill
  \includegraphics[scale=0.5]{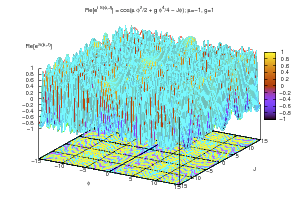}
  \hspace{\fill}
  \includegraphics[scale=0.5]{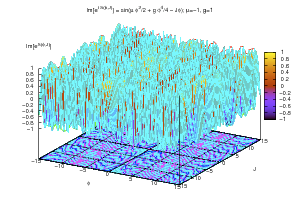}
  \hspace{\fill}

  \bigskip \bigskip

  {\footnotesize\noindent\textbf{Figure 10:} Plots of the Real and Imaginary
    parts of the integrand for various values of $J$ and $\mu = 1,\, g = 1$ and
    $\mu = -1,\, g = 1$.}
\end{center}

Below, we see the real (left) and imaginary (right) parts of the mollified
$\phi^4$-integrand (red) in comparison with their non-mollified counterparts (green):

\begin{center}
  \hfill
  \includegraphics[scale=0.5]{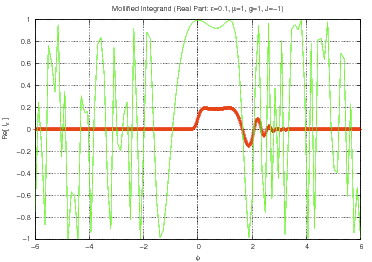}
  \hfill
  \includegraphics[scale=0.5]{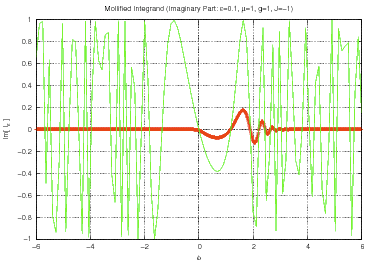}
  \hfill

  \bigskip \bigskip

  {\footnotesize\noindent\textbf{Figure 11:} Superimposed mollified (red) and
    non-mollified (green) real and imaginary parts of the integrand for $\mu
    =1,\, g = 1,\, J = -1,\, \epsilon = 0.1$.}
\end{center}

Below, we see the real (left) and imaginary (right) parts of the mollified
$\phi^4$-integrand for different values of the field $\phi$ and of the mollification
parameter $\epsilon$:

\begin{center}
  \hfill
  \includegraphics[scale=0.5]{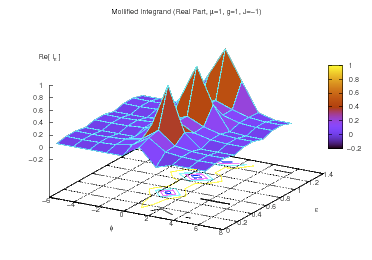}
  \hfill
  \includegraphics[scale=0.5]{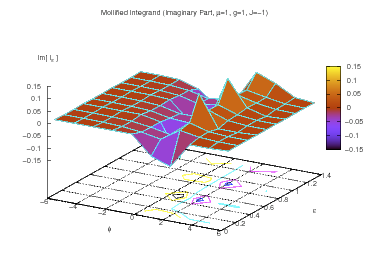}
  \hfill

  \bigskip \bigskip

  {\footnotesize\noindent\textbf{Figure 12:} Plots of the real and imaginary
    parts of the mollified integrand for various values of the field and the
    parameter $\epsilon$.}
\end{center}

As explained in \cite{tvbcsde,rtnnm}, the three different solutions/phases of this model
can be selected via an appropriate choice of boundary conditions: $\Gamma^0 = \mathbb{R}$,
$\Gamma^{+} = (-\infty,0) \cup (0, i\, \infty)$ and $\Gamma^{-} = (-\infty,0) \cup (0,
-i\, \infty)$. The leftmost figure below shows those boundaries in the Argand Plane:
$\Gamma^0 = \mathbb{R}$ is the green line, $\Gamma^{+} = (-\infty,0) \cup (0, i\, \infty)$
is the red line while $\Gamma^{-} = (-\infty,0) \cup (0, -i\, \infty)$ is the blue
line. The rightmost figure below shows the regions (shaded) of the complex $\phi$-plane,
$(\phi = \rho\, e^{i\, \theta})$, defined by $\cos(4\, \theta) \geqslant 0$. Any contour,
starting and ending at infinity, within one of these four domains corresponds to a
particular solution of the 0-dimensional $\phi^4$.

\begin{center}
  \begin{minipage}[b]{0.4\linewidth}
    \includegraphics{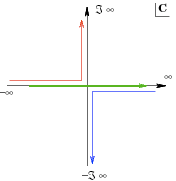}
  \end{minipage}
  \hspace{0.1\linewidth}
  \begin{minipage}[t]{0.4\linewidth}
    \includegraphics{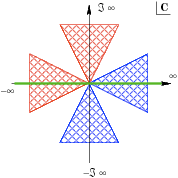}
  \end{minipage}
  \bigskip \bigskip

  {\footnotesize\noindent\textbf{Figure 13:} \textsc{Left:} Three different
    contours that render the Path Integral finite and depict the three distinct
    solutions obtained in this model. \textsc{Right:} Any contour that starts
    and finishes inside the same-color shaded regions (following the general
    directions shown on the left) corresponds to a particular finite solution.}
\end{center}

Using the three different contours above we are able to find the three different solutions
to the equation of motion of this 0-dimensional QFT. The Schwinger-Dyson equation is given by,

\begin{equation*}
  -i\, \biggl(\mu\, \frac{\delta}{\delta J} + g\, \frac{\delta^3}{\delta J^3}\biggr)\,
     \mathcal{Z}[J] = J \, \mathcal{Z}[J] \; ,
\end{equation*}
where we can easily see (again) that there has to be three solutions. These are
called \emph{Parabolic Cylinder Functions} and are denoted $U(\mu/g,J),\,
V(\mu/g,J),\, W(\mu/g,J)$. Those behave in the following manner (see Appendix
\ref{sec:pcf}):

\begin{description}
\item[Regular at $\boldsymbol{g \rightarrow 0}$] Consistent with perturbation theory,
\item[Singular $\boldsymbol{(\simeq\hksqrt{g})$ at $g \rightarrow 0}$] Symmetry Breaking,
\item[Singular $\boldsymbol{(\simeq\exp\{\mu/4\, g\})$ at $g \rightarrow 0}$] Instanton.
\end{description}

Thus,
\begin{equation*}
  \langle\mathcal{O}\rangle_{\epsilon}[J] = \frac{\displaystyle\varint_{-\infty}^{\infty}\,
    \displaystyle\varint_{\Gamma^{0,\pm}}\, \eta_{\epsilon}(\phi - \varphi)\, \mathcal{O}[\varphi]\,
    \exp\Bigl\{i\, \frac{\mu}{2}\, \varphi^2 + i\, \frac{g}{4}\, \varphi^4 - i\, J\,
    \varphi\Bigr\}\, d\varphi\, d\phi}{\displaystyle\varint_{-\infty}^{\infty}\,
    \displaystyle\varint_{\Gamma^{0,\pm}}\, \eta_{\epsilon}(\phi - \varphi)\, \exp\Bigl\{i\,
    \frac{\mu}{2}\, \varphi^2 + i\, \frac{g}{4}\, \varphi^4\Bigr\}\, d\varphi\, d\phi} \; .
\end{equation*}

It should be noted, though, that depending on the chosen contour $(\Gamma^0, \Gamma^+,
\Gamma^-)$, the above integral will resemble a Fresnel one and, as such, its computation
is quite simplified.

For the graphs below, we used the following values for the saddle-points,

\begin{equation*}
    \phi_0 \in \Bigl\{0, \pm\hksqrt{-\mu/2\,g}\Bigr\} \; ,
\end{equation*}
where $S'[\phi_0] = 0$. Therefore, the different importance sampling functions are:

\begin{description}
\item[Symmetric Phase] Given by the $\Gamma^0$ contour, this is the one which is regular
  in the limit $g\rightarrow 0$ and accessible via perturbation theory,
  \begin{equation*}
    W_{\epsilon}^{0}[\phi] = \exp\biggl\{-\frac{1}{2}\,
      \Bigl(\frac{\phi}{\epsilon}\Bigr)^2\biggr\} \; ; \quad \phi_0 = 0 \; ;
  \end{equation*}
\item[Solitonic Phase] This phase is given by the linear combination of the
  contours $\Gamma^{+}$ and $\Gamma^{-}$ such that $\mu \geqslant 0$, i.e., this
  represents the soliton solution,
  \begin{equation*}
    \hspace{-5em} W_{\epsilon}^{\mu\geqslant 0}[\phi] = \exp\biggl\{-\frac{1}{2}\,
      \Bigl(\frac{\phi^2 + (\phi_0^{+})^2}{\epsilon^2}\Bigr)\biggr\} + \exp\biggl\{-\frac{1}{2}\,
      \Bigl(\frac{\phi^2 - (\phi_0^{-})^2}{\epsilon^2}\Bigr)\biggr\} \; ;\quad \phi_0^{\pm} = \pm
      i\, \hksqrt{\mu/2\,g \,{}\xspace} \; ; \quad (\mu \geqslant 0)\; ;    
  \end{equation*}
\item[Broken-Symmetric Phase] This one is given by the linear combination of the
  contours $\Gamma^{+}$ and $\Gamma^{-}$ such that $\mu < 0$, i.e., this
  represents the solution usually referred to as broken-symmetric,
  \begin{equation*}
    \hspace{-5em} W_{\epsilon}^{\mu < 0}[\phi] = \exp\biggl\{-\frac{1}{2}\,
      \Bigl(\frac{\phi + \phi_0^{+}}{\epsilon}\Bigr)^2\biggr\} + \exp\biggl\{-\frac{1}{2}\,
      \Bigl(\frac{\phi - \phi_0^{-}}{\epsilon}\Bigr)^2\biggr\} \; ;\quad \phi_0^{\pm} = \pm
      \hksqrt{-\mu/2\,g \,{}\xspace} \; ; \quad (\mu < 0)\; .    
  \end{equation*}
\end{description}

Below, we have the graphs of these importance sampling functions:
\vspace{-2em}
\begin{center}
  \includegraphics[scale=0.55]{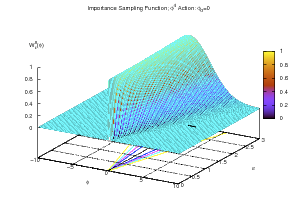}
  \includegraphics[scale=0.55]{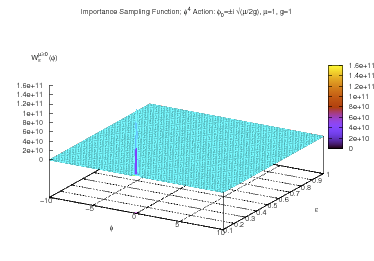}
  \includegraphics[scale=0.55]{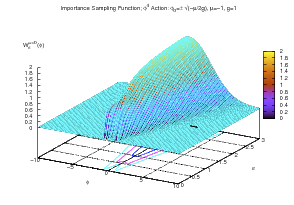}

  \bigskip \bigskip

  {\footnotesize\noindent\textbf{Figure 14:} Plots of the mollified importance
    sampling functions for various values of the field and $\epsilon$: symmetric
  ($\phi_0 = 0$), solitonic ($\mu = 1,\, g = 1$) and broken-symmetric ($\mu =
  -1,\, g = 1$).}
\end{center}

The graphs below show the entropy for the above importance sampling functions, as
discussed in Appendix \ref{sec:entropy}:

\begin{center}
  \hspace{\fill}
  \includegraphics[scale=0.35]{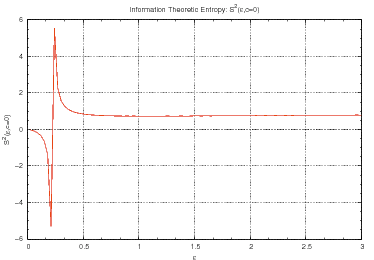}
  \hspace{\fill}
  \includegraphics[scale=0.35]{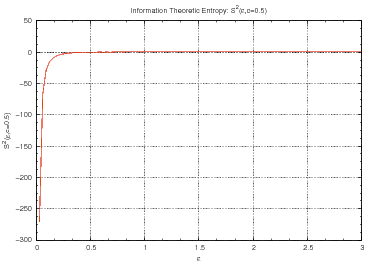}
  \hspace{\fill}
  \includegraphics[scale=0.35]{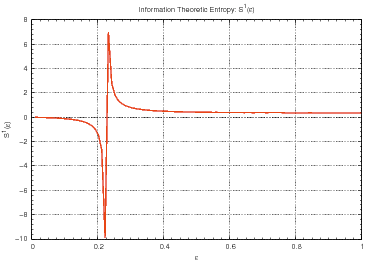}
  \hspace{\fill}

  \bigskip

  {\footnotesize\noindent\textbf{Figure 15:} Information theoretical entropy for
  finding the optimal $\epsilon$ in the symmetric, solitonic and
  broken-symmetric cases.}
\end{center}
\section{Higher Dimensional}\label{subsubsec:hd}
As for the 4-dimensional results [for a Free Scalar QFT] the news is not
encouraging. As mentioned in section \ref{subsubsec:mis}, there are severe 
efficiency constraints in this implementation of the mollification; in section
\ref{subsubsec:mis} three possible choices of solutions to this technical issue
were given.

It turns out that, from those three solutions, one is non-efficient while the other
two produce analogous results: only the imaginary parts of physically interesting
quantities (generating functional and Green's functions) can be tamed. The real parts
continue to show an oscillatory pattern.

Although this is at least disappointing, the solutions obtained are good enough
to control the imaginary part's oscillations.

It is our hope that changes in algorithms will allow some version of
mollification to work in real world problems.

\part{Solution Space and Moduli Space Topology}
\chapter{Introduction}\label{sec:introduction}
Classical gauge theories are well described through differential geometry, where
a gauge field is represented by a connection on a principal fibre bundle $P$ for
which the structure group is the symmetry group of the theory; see, for example,
\cite{frankel,nakahara,darling,ladies,kerbrat,fulpnorris,mayer,sardanashvily,simpson}
and references therein.

The phenomenon of symmetry breaking also has its own geometric
formulation \cite{ladies,kerbrat,fulpnorris,mayer,sardanashvily,simpson,neeman}: the
reduction of the principal bundle $P$. To this classical picture, the question
that arises is that of how does a quantum field theory select its vacua given by
the different possible reductions of the principal bundle $P$. The canonical
answer to this question is that radiative corrections (beyond the tree-level
approximation) cause the QFT to ``jump'' from the symmetric phase to the
broken-symmetric one(s) \cite{weinbergcoleman}.

However, \cite{weinbergcoleman} already expresses concerns about
some issues (e.g., the last paragraph on page 1894 and its continuation on page
1895) that are more explicitly treated in \cite{garciaguralnik}. For more
examples about these issues, see \cite{borcherds} (pages 27 and 28 discuss
the asymptotic nature of the series expansion, their maximum accuracy and issues
regarding Borel summation) and \cite{fredenhagen}.

Loosely speaking, what \cite{garciaguralnik} does, is to compute the
Schwinger-Dyson equations (for a given QFT) and note that these differential
equations have as many solutions as its order indicates. Moreover, the boundary
conditions that determine each of these solutions yield different possible
values for the parameters of the theory (mass, coupling constants, etc). Thus,
each one of these solutions has its own series expansion (which may or may not
be equivalent to the perturbative series) and particular behavior.

The analysis of the boundary conditions of the Schwinger-Dyson equation being
responsible for the different solutions of the theory done in \cite{garciaguralnik}
is more in the spirit of the self-adjoint extension of the associated
Hamiltonian, as done in \cite{reedsimon,dunfordschwartz}. Although it may not
seem so at first, \cite{garciaguralnik} is equivalent to a latticized approach to
QFT and, as such, requires 2 types of infinite limits in order to give rise
to its continuum version: the thermodynamic (number of particles in the box) and
the volume (size of the box) limit. These limits do not commute and fiddling
with them (as done in \cite{garciaguralnik}) is analogous to resumming the
perturbative series (as done in \cite{weinbergcoleman}).

The present work has the goal of developing a geometrical method that brings to
the foreground these issues of vacuum structure in QFT, its multiple solutions,
its moduli space, clearly showing that these different solutions are topologically
inequivalent. The fact that different configuration spaces for distinct
solutions of the equation of motion of a given Lagrangian have different
topologies shows that expansions must be performed separately for each
solution of a theory, \ie, each phase has to be regarded and treated as a
separate theory.

In this sense, this work streamlines how the parameters of a theory (mass,
coupling constants, etc) determine the topology of the vacuum manifold (moduli
space) which, combined with the picture presented in \cite{garciaguralnik},
gives a prescription of how to examine the solution space of a field theory: The
boundary conditions of the Schwinger-Dyson equations determine the parameters of
the theory which, in turn, determine its topology.
\chapter{Classical Field Theory and the Jacobi Metric} \label{sec:qftjm}
A typical action for a scalar field has the form,
\begin{align*}
  S[\phi] &= \int K(\pi,\pi) - V_{\tau}(\phi) \, d^{n}x \; ;\\
  K(\pi,\pi) &= \frac{1}{2}\, \g(\pi,\pi) = \frac{1}{2}\, g_{\mu\,\nu}\,
    \pi^{\mu}\,\pi^{\nu} \; .
\end{align*}
where $K$ is the kinetic quadratic form (bilinear and, in case $\pi\in\mathbb{C}$,
hermitian; defining the inner product \g) and $V_{\tau}$ is the potential, where
the index $\tau$ collectively denotes coupling constants, mass terms, etc.

We want to reparameterize our field using the arc-length parameterization such
that $\boldsymbol{\tilde{\g}}(\tilde{\pi},\tilde{\pi}) = \tilde{g}_{\mu\,
  \nu}\tilde{\pi}^{\mu}\, \tilde{\pi}^{\nu} = 1$, where
$\boldsymbol{\tilde{\g}}$ is the new metric and $\tilde{\pi}^{\mu}$ is the
momentum field redefined in terms of this new parameterization (assuming
$V_{\tau}(\phi)$ contains no derivative couplings). That is, we want to scale our
coordinate system in order to obtain a conformal transformation of the metric
that normalizes our momentum field.

In fact, there are three possible normalizations, depending on the nature
of $\tilde{\pi}$:
\begin{equation*}
  \boldsymbol{\tilde{\g}}(\tilde{\pi},\tilde{\pi}) = 
  \begin{cases}
     1, & \text{spacelike;} \\
     0, & \text{lightlike;} \\
    -1, & \text{timelike.}
  \end{cases}
\end{equation*}

However, this will not concern us here, once our main focus will
be to find the geodesics of $\boldsymbol{\tilde{\g}}$. Therefore, we have a
clear distinction among the three foliations of Lorentzian spaces:

\begin{description}
\item[$\boldsymbol{\tilde{\g}(\tilde{\pi},\tilde{\pi}) > 0}$] represents the
  spacelike-leaf and contains only non-physical objects --- after the arc-length
  reparameterization we will have $\tilde{\g}(\tilde{\pi},\tilde{\pi}) = 1$;
\item[$\boldsymbol{\tilde{\g}(\tilde{\pi},\tilde{\pi}) = 0}$] clearly singular,
  representing the null-leaf where there is no concept of distance, given that
  all objects here are massless;
\item[$\boldsymbol{\tilde{\g}(\tilde{\pi},\tilde{\pi}) < 0}$] represents the
  timelike-leaf and contains the physical objects --- after the arc-length
  reparameterization we will have $\tilde{\g}(\tilde{\pi},\tilde{\pi}) = -1$.
\end{description}

In order to find the arc-length parameterization, we will borrow an idea from
classical mechanics called \emph{Jacobi's metric} \cite{drg}. In the Newtonian
setting, Jacobi's metric gives an intrinsic geometry for the configuration
space (resp. phase space), where dynamical orbits become geodesics, i.e., it
maps every Hamiltonian flow into a geodesic one; therefore, solving the
equations of motion implies finding the geodesics of Jacobi's metric and
vice-versa. This can be done for any closed non-dissipative system (with total
energy $E$), regardless of the number of degrees of freedom.

Before we proceed any further, let us take a look at a couple of simple examples
in order to motivate our coming definition of Jacobi's metric.

\section{Harmonic Oscillator} \label{par:sho}
Let us consider a model similar to the harmonic oscillator (in
$(1+0)$-dimensions): without further considerations, we simply allow for the
analytic continuation of the frequency: $\mu = +\omega^2$ or $\mu =
-\omega^2$. Its Lagrangian is given by $L = \tfrac{1}{2}\, \dot{q}^2 \mp
\tfrac{\mu}{2}\, q^2$, where $\mu > 0$, from which we conclude that for a fixed
$E$ such that $E = \tfrac{1}{2}\, (\dot{q}^2 \pm \mu\,q^2)$, we have $\dot{q} =
\tfrac{dq}{dt} = \hksqrt{2\, (E \mp \tfrac{\mu}{2}\, q^2)}$.

So, in the spirit of what was said above, we want to find a reparameterization
of the time coordinate in order to have the normalization $\dot{q} = 1$, where
the dot represents a derivative with respect to this new time variable.

Thus,

\begin{align*}
  \frac{dq}{dt} &= \frac{dq}{ds}\, \frac{ds}{dt} = \hksqrt{2\, \biggl(E \mp
    \frac{\mu}{2}\, q^2\biggr)} \; ;\\
  \text{if}\quad \frac{ds}{dt} &= \hksqrt{2\, \biggl(E \mp \frac{\mu}{2}\,
    q^2\biggr)} \Rightarrow \frac{dq}{ds} \equiv 1\;; \\
  \therefore\; ds &= \hksqrt{2\, \biggl(E \mp \frac{\mu}{2}\, q^2\biggr)}\, dt\; ; \\
  \Rightarrow\; \gE &= 2\, \biggl(E \mp \frac{\mu}{2}\, q^2\biggr)\, \g \; .
\end{align*}

As expected, the new metric, \gE, is a conformal transformation of the original
one, \g; and under this new metric, our original Lagrangian is simply written
as,

\begin{align*}
  L(q,\dot{q}) &= \frac{1}{2}\, \dot{q}^2 \mp \frac{\mu}{2}\, q^2 \; ;\\
  &= \frac{1}{2}\, g_{i\, j} \dot{q}^i\,\dot{q}^j \mp\frac{\mu}{2}\, q^2\; ;\\
  &= \frac{1}{2}\, \g(\dot{q},\dot{q}) \mp\frac{\mu}{2}\, q^2\; ;\\
  \therefore\; L(q,\dot{q}) &= \gE(\dot{q},\dot{q}) \; .
\end{align*}

Now, as we said above \cite{drg}, the Euler-Lagrange equations for this
conformally transformed Lagrangian are simply the geodesics of the metric
\gE. The task of finding the geodesics $\gamma(s)$ of the \gE metric can be more
easily accomplished with the help of the normalization condition,
$\gE\bigl(\tfrac{d\gamma}{ds},\tfrac{d\gamma}{ds}\bigr) = 1$, and the initial
condition $\gamma(s=0) = 0$, given that, in this fashion, we only need to solve
a first order differential equation:

\begin{align*}
  \gE\biggl(\frac{d\gamma}{ds}, \frac{d\gamma}{ds}\biggr) &= 1 \; ;\\
  \Rightarrow\; 2\, \biggl(E \mp \frac{\mu}{2}\, \gamma^2\biggr)\,
    \g\biggl(\frac{d\gamma}{ds}, \frac{d\gamma}{ds}\biggr) &= 1 \; ;\\
  \therefore\; \gamma'(s) = \frac{d\gamma}{ds} &= \hksqrt{\frac{1}{2\,\bigl(E
      \mp \frac{\mu}{2}\, \gamma^2\bigr)}} \; ;\\
  \text{with the initial condition:}\quad \gamma(0) &= 0 \; .
\end{align*}

Here we should note that solving for $\gE(\gamma',\gamma') = 1$ (spacelike-leaf)
is analogous to solving $\gE(\gamma',\gamma') = -1$ (timelike-leaf), once the
two cases are symmetrical about the origin.

Analytically solving the equation above yields 2 possible answers, depending on
the particular form of the potential ($\mu > 0$ in both cases):

\begin{enumerate}
\item $V_{+} = +\mu\, \gamma^{2}/2$: For the case of a positive pre-factor, we
  get that $\gamma\, \hksqrt{\mu\, (2\, E - \mu\, \gamma^{2})} + 2\, E\,
  \arcsin\bigl(\gamma\, \hksqrt{\mu/2E}\bigr) - 2\, s\, \hksqrt{\mu} = 0$; and
\item $V_{-} = -\mu\, \gamma^{2}/2$:  For the case of a negative pre-factor, we
  find that $\gamma\, \hksqrt{\mu\, (2\, E + \mu\, \gamma^{2})} + 2\, E\,
  \arcsinh\bigl(\gamma\, \hksqrt{\mu/2E}\bigr) - 2\, s\, \hksqrt{\mu} = 0$.
\end{enumerate}

It is clear from the expression for $V_{+}$ that $\gamma^{2} \leqslant 2E/\mu$,
i.e., the length of the [classical] geodesic is bounded; this does not happen
with $V_{-}$.

These geodesics, $\gamma_{\pm}$, clearly depend on the parameters $E$ and $\mu$;
therefore, in order to plot $\gamma(s)$, we have to make 2 distinct choices:
$E/\mu = 1$ (left plot) and $E/\mu = -1$ (right plot). The last plot
comparatively depicts both geodesics.

\begin{center}
  \includegraphics[scale=0.5]{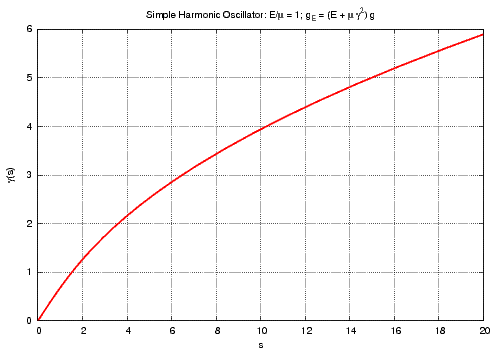} \hfill
  \includegraphics[scale=0.5]{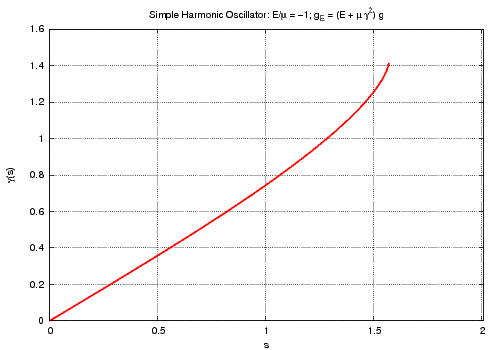} \hfill\\
  \includegraphics[scale=0.5]{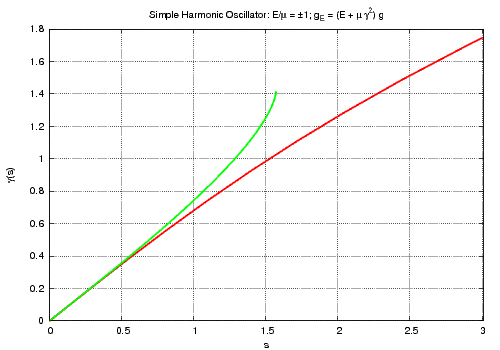}
  \bigskip

  {\footnotesize \noindent\textbf{Figure 1:} The top-left-corner plot shows the
    geodesic for $E/\mu = 1$, while the top-right-corner one depicts it for
    $E/\mu = -1$. The bottom-center plot superimposes both of them.}
\end{center}

This ``harmonic oscillator'' model already portrays the features that we want to
identify in forthcoming applications of this technique: changing the values of
the parameters of the potential we can identify 2 distinct types of geodesics
which will be related to different solutions on the coming examples.

Note, however, that in general the parameters will not be able to vary freely:
they will belong to some specified set; and they will be related to each other,
i.e., we will be able to find a function of the parameters that constrains their
behavior --- this will establish ``dualities'' among the parameters of a given
theory, as will be shown below.

\section{Cubic Potential} \label{par:airy}
For completeness sakes, let us try another example, that of a cubic potential
given by: $L = \tfrac{1}{2}\, \g(\dot{q},\dot{q}) + q^3/3$. It's Jacobi Metric
is given by $\gE = 2\, (E + q^3/3)\, \g$ and the equation we need to solve in
order to find the geodesics, $\gamma(s)$, of this metric is,

\begin{align*}
  \gE(\gamma',\gamma') = 2\, (E + q^3/3)\, \g(\gamma',\gamma') &= 1 \; ;\\
  2\, (E + q^3/3)\, \bigl(\gamma'(s)\bigr)^2 &= 1 \; ; \\
  \text{with the initial condition:}\quad \gamma(0) &= 0 \; .
\end{align*}

Once again, we should note that solving for $\gE(\gamma',\gamma') = -1$
(timelike-leaf) is analogous to what is being done above: the graphs are
reflected with respect to each other.

This example is slightly different from the previous one for the following
reason: before, it was the relative values of $E$ and $\mu$ that determined our
two solutions, i.e., one for $E/\mu \geqslant 0$ and one for $E/\mu < 0$. Now,
this cubic theory has \emph{no} free parameters in its potential, therefore the
geodesics are labeled by the arbitrary parameter $E$.

Just as before, the metric will be [artificially] degenerate when $E =
V(\gamma)$, which tells us that $\gamma \geqslant (-3\, E)^{1/3}$: when $E
\geqslant 0$ the geodesic is allowed to have any length, but when $E < 0$ the
geodesic vanishes for some time, after which it starts to grow.

This is very interesting because it essentially says that there is a ``lag
time'' before this solution comes alive: this leaf is non-existent for some
``proper time'' and then it springs into being quite abruptly.

\begin{center}
  \includegraphics[scale=0.5]{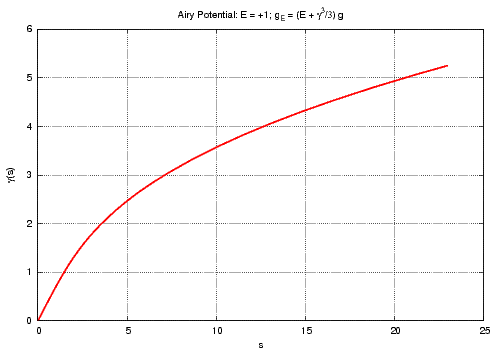} \hfill
  \includegraphics[scale=0.5]{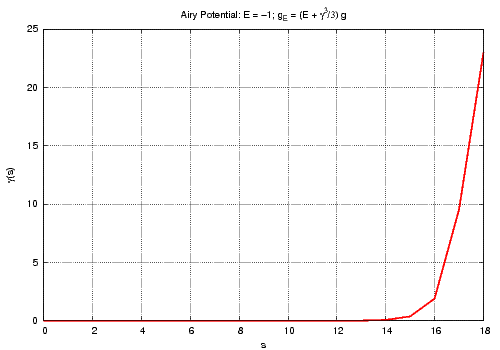} \hfill\\
  \includegraphics[scale=0.5]{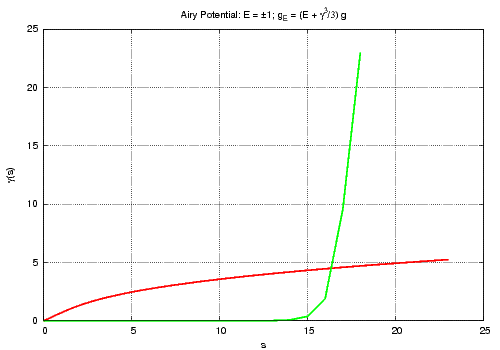}
  \bigskip

  {\footnotesize \noindent\textbf{Figure 2:} The top-left-corner plot shows the
    geodesic for $E = 1$, while the top-right-corner one depicts it for
    $E = -1$. The bottom-center plot superimposes both of them.}
\end{center}

Now that we are done with our examples, we are ready to find the generalization
of Jacobi's metric to the field theoretical setting can be accomplished via the
definition of the following conformally transformed metric:

\begin{align}
  \nonumber
  L &= \frac{1}{2}\, \g(\pi,\pi) - V_{\tau}(\phi) \; ; \\
  \nonumber
  &\equiv \gE^{\tau}(\tilde{\pi},\tilde{\pi}) \; ;\\
  \intertext{where,}
  \label{eq:jacobimetric}
  \gE^{\tau} &= 2\, \bigl(E - V_{\tau}(\phi)\bigr)\, \g \; ;
\end{align}
and $E$ (an arbitrary parameter) is the total energy of the system.
\chapter{Applications in Quantum Field Theory} \label{sec:appsqft}
We will use the approach in terms of Feynman Path Integrals in order to make
things more straightforward. However, we could as well talk in terms of the
phase space of a given QFT and its vacuum manifold, i.e., its moduli space.

Starting from the partition function, we have the following ($\hbar = 1$):

\begin{align*}
  \mathcal{Z}[J] &= \mathcal{N}\, \varint e^{i\, S[\phi] + i\int J(x)\,
    \phi(x)\, d^dx}\, \mathcal{D}\phi \; ;\\
  &= \mathcal{N}\, \varint e^{i\, \int \gE^{\tau}(\tilde{\pi},\tilde{\pi}) + J(x)\,
    \phi(x)\, d^dx}\, \mathcal{D}\phi \; ;
\end{align*}
where $\mathcal{N}$ is a normalization constant such that $\mathcal{Z}[J=0] =
1$, and we have already written the Action in terms of Jacobi's metric.

Now, let us expand the partition function above in terms of its classical part
and its quantum fluctuations, i.e., $\phi = \phi_{\text{cl}} + \delta\phi$,
and the classical Action is given by $S_{\text{cl}} = S[\phi_{\text{cl}}] =
\int_{\mathpzc{M}} \gE^{\tau}(\tilde{\pi}_{\text{cl}},\tilde{\pi}_{\text{cl}})$, where
$\mathpzc{M}$ is the particular region of spacetime where the integration is
performed and $\pi = \ed\phi$ --- for this calculation we are assuming the
fluctuations vanish at the boundary, $\delta\phi|_{\partial\mathpzc{M}} =
0$. Thus, we have that,

\begin{align}
  \nonumber
  \mathcal{Z} &= \varint \exp\bigl\{i\, S[\phi_{\text{cl}} +
    \delta\phi]\bigr\}\, \mathcal{D}\phi \; ; \\
  \label{eq:clfldecomp1}
  &= \varint \exp\biggl\{i\, \int_{\mathpzc{M}}
    \gE^{\tau}(\tilde{\pi}_{\text{cl}},\tilde{\pi}_{\text{cl}}) +
    \gE^{\tau}(\delta\tilde{\pi},\delta\tilde{\pi})\biggr\} \, \mathcal{D}\phi \; ; \\
  \label{eq:clfldecomp2}
  &= e^{i\, S_{\text{cl}}}\, F_{\tau}[\partial\mathpzc{M}] \; ; \\
  \intertext{where,}
  \label{eq:quantumfl}
  F_{\tau}[\partial\mathpzc{M}] = \varint e^{i\, S[\delta\phi]}\,
    \mathcal{D}(\delta\phi) &= \varint e^{i\, \int_{\mathpzc{M}}
      \gE^{\tau}(\delta\tilde{\pi},\delta\tilde{\pi})}\,
    \mathcal{D}(\delta\phi) \approx \frac{1}{\hksqrt{\det\bigl(-D^2_{\tau}\bigr)}} \; ;
\end{align}
where $F_{\tau}[\partial\mathpzc{M}]$ is the [quantum] fluctuation part of the
partition function, the index $\tau$ denoting collectively the parameters of the
potential (mass, coupling constants, etc), and $D^2_{\tau} = \nabla^{\mu}\,
\nabla_{\mu}$ is the Laplace-Beltrami operator constructed from the covariant
derivative $(\nabla_{\mu})$ associated to the Levi-Civita connection of
$\gE^{\tau}$. There are two important things to note from \eqref{eq:quantumfl}:
the fluctuation term can only depend on the parameters of the potential and on
the boundary $\partial\mathpzc{M}$, i.e., surface terms (that we assumed
vanishing); and its structure is analogous to that of the classical part, in
that $\gE^{\tau}$ is the same on both.

However, the above presumes that there is only one solution to the theory in
question. But, the method we are developing is exactly to use Jacobi's metric in
order to find and classify \emph{all} of the solutions of the theory given,
i.e., we want to be able to use this tool to study the moduli space of the
problem at hand. Therefore, we need to generalize the situation above for the
case of \emph{many} solutions, which is not a difficult task:

\begin{align}
  \label{eq:multiclfldecomp1}
  \mathcal{Z} &= \prod_{\nu=1}^{N} e^{i\, S_{\text{cl}}^{\nu}}\,
    F_{\tau}^{\nu}[\partial\mathpzc{M}] \; ;\\
  \label{eq:multiclfldecomp2}
  &\approx \prod_{\nu=1}^{N} \frac{e^{i\,
      S_{\text{cl}}^{\nu}}}{\hksqrt{\det\bigl(-D^2_{\tau}\bigr)}} \; ;
\end{align}
where $\nu$ counts the number of different solutions (i.e., the number of
critical points of $S[\phi]$, including its multiplicity) denoted by
$\phi^{\nu}_{\text{cl}}$, and $S_{\text{cl}}^{\nu} =
S[\phi^{\nu}_{\text{cl}}]$. Note that, in order to find all of the possible
critical points of $S[\phi]$, we need to take into account its
$\tau$-dependence, and in doing so we are implicitly assuming that $\tau \in
\mathbb{C}$, ensuring we are able to find them all.

There are two major observations to be done at this point:
\begin{enumerate}
\item There are two kinds of discontinuities present in the above construction,
  \cite{berrystokes}: first, critical \emph{points} $\phi^{\nu}_{\text{cl}}$ can
  coalesce, which happens in the complexified catastrophe set in parameter
  space; second, critical \emph{values} $\Im(i\, S[\phi^{\nu}_{\text{cl}}]) =
  S^{\nu}_{\text{cl}}$ can coalesce, which happens on the Stokes set in
  parameter space and corresponds to the appearance or disappearance of a
  subdominant exponential in a ``non-local bifurcation''. These correspond,
  respectively, to realizing that the partition function is to be taken over
  \emph{complex} fields (rather than real ones, as is customary), where the contour of
  integration (rendering the partition function finite, \cite{garciaguralnik}) will ultimately
  determine the parameter space (these are the Lee-Yang zeros of our theory);
  and the coalescing of $S^{\nu}_{\text{cl}}$ represents the Stokes phenomena of our
  theory. Both of these will determine the phase structure of our problem (see
  \cite{garciaguralnik}). Note that it is the contour of integration that
  connects both of these, once the appropriately chosen range of $\phi$ will
  render $e^{i\, S[\phi]}$ convergent.
\item The asymptotic behavior of the partition function depends \emph{only} on
  the critical points of $S$, i.e., on $\phi^{\nu}_{\text{cl}}$. Thus, when
  $S[\phi]$ (resp. $V[\phi]$) is a Morse function, i.e, smooth with no degenerate
  critical points, we can use Morse's lemma to show that the critical points are
  isolated, keeping in mind that the number of isolated points is a topological
  invariant. (It is worth noting this can be generalized in the thermodynamical
  limit via the Morse-Palais lemma, just as we can relax the condition of
  non-degeneracy of the critical points via Morse-Bott theory.) Using this, it
  can be shown that $\phi$-space (resp. Phase Space) is a CW-complex with a $\nu$-cell
  for each critical point of index $\nu$: the fluctuation term contains the
  Maslov-Morse index (in $\gE^{\tau}$ by means of its dependence on
  $V_{\tau}[\phi]$) that accounts for the discretization of Path Space and
  corrects for the thermodynamic limit of the particular solution in question,
  i.e., every time the denominator in \eqref{eq:multiclfldecomp2} vanishes,
  $F_{\tau}^{\nu}[\partial\mathpzc{M}]$ passes a singularity in such a way as to
  ensure the proper phase; the phase factor that arises in this way is nothing
  but $e^{i\, \alpha\, \nu}$, where $\alpha$ is some angle and $\nu$ counts the
  number of zeros (with multiplicity) encountered along the particular path in
  question --- it is called the Maslov-Morse index.
\end{enumerate}

Now, we are ready to employ this machinery in order to draw several important
conclusions. Here are them:

\begin{enumerate}
\item The partition function, seen as a function of the parameters of the
  potential, $\mathcal{Z} = \mathcal{Z}[\tau] = \mathcal{Z}[\text{mass},
  \text{coupling constants}]$, is a \emph{meromorphic} function: it is
  holomorphic on a subset of $\mathbb{C}$ except for the set of isolated points
  given by the values of the parameters along Stokes' lines (resp. critical
  lines of phase transition). We have to keep in mind that the parameter space
  had to be complexified in order to yield all possible solutions to the theory
  given; another way to think about this is in terms of the contours that render
  the partition function finite: changing these contours (looking for all
  possible ones that make the partition function converge) will affect the
  allowed values for the parameters, i.e., these contours ultimately determine
  the parameter space, as done in \cite{garciaguralnik} (see also
  \cite{mollifier}).
\item Under the [full] elliptic modular group, $\Gamma =
  \mathrm{SL}(2,\mathbb{Z}) = \bigl\{ \bigl(
  \begin{smallmatrix}
    a & b\\
    c & d
  \end{smallmatrix}
  \bigr) \;|\; a=1,\, b=0,\, c=0,\, d=1\bigr\}$ (see \cite{mfmf}), the partition
  function is a \emph{modular function}, i.e., for any $\mathds{M} \in \Gamma$
  we have that $\mathcal{Z}[\mathds{M}\, \tau] = \mathcal{Z}[\tau]$, where
  $\mathds{M}\, \tau = \tfrac{a\, \tau + b}{c\, \tau + d}$. (For a fuller
  appreciation of the importance of this fact, see, e.g., \cite{3dgravity}.)
\item The Action, written in terms of Jacobi's
  metric, $S[\phi] = \int_{\mathpzc{M}} \gE(\tilde{\pi},\tilde{\pi})$, can be
  thought of as the Morse-theoretic Energy functional, $E[\gamma] =
  \int\g(\gamma',\gamma')$, for the path $\gamma$, where $\gamma' =
  d\gamma/ds$. Therefore, we can readily say that all of the critical points of
  $S[\phi]$ are given by minimal [Lagrangian] manifolds, and, just like $E''$,
  $S''=\delta^2 S/\delta\phi(x)\delta\phi(y)$ is a well defined symmetric bilinear
  functional. This implies that $S'' = 0$ if, and only if, $\tilde{\pi}$ is a
  Jacobi field, which, in turn, implies that $\tilde{\pi}|_{\partial\mathpzc{M}} =
  0$. Therefore, using the split of the partition function
  \eqref{eq:multiclfldecomp1} in terms of a classical part and its quantum
  fluctuations, we see that the vacuum manifold (moduli space) of the quantum
  theory is given by its classical minimal Lagrangian manifold with the quantum
  corrections being given by extensions of it via Jacobi fields. That is, the
  classical minimal manifold is extended via the gluing of the quantum
  fluctuation manifold obtained as a solution to the equations of motion coming
  from $S_{\text{fl}} = \int_{\mathpzc{M}}
  \gE^{\tau}(\delta\tilde{\pi},\delta\tilde{\pi})$. Thus, the quantum
  corrections are handles attached to the classical solution, without changing
  the classical topology.
\item Continuing the reasoning above, it is not difficult to see that a phase
  transition will happen when the quantum corrections (in terms of handle
  attachments) actually do change the topology of the particular solution in
  question. In fact, if $V_{\tau}[\phi]$ crosses a critical point of index
  $\nu$, the handle to be attached is a $\nu$-cell (i.e., a $\nu$-simplex) ---
  the connection with what has been said before should be
  straightforward. Analogously, we can say the following: let $\Sigma_{\tau}^E = 
  V^{-1}_{\tau}[E] = \{ \phi \;|\; V_{\tau}[\phi] = E\}$, i.e.,
  $\Sigma_{\tau}^E$ is the set of all field configurations which have the same
  potential energy --- this manifold is the configuration space (resp. moduli
  space), an equipotential surface. Thus, the family of equipotentials
  $\{\Sigma_{\tau}^E\}_{E\in \mathbb{R}}$ foliates the configuration space
  (moduli space) in such a way that if $\Sigma_{\tau}^E$ is
  $\mathscr{C}^{\infty}$-diffeomorphic to $\Sigma_{\tau}^{\bar{E}}$ (for two
  different values of the energy, $E$ and $\bar{E}$) then there is no phase
  transition. Conversely, a phase transition will be characterized by the
  existence of a certain critical value $E_c$ such that
  $\{\Sigma_{\tau}^{E}\}_{E < E_c}$ is \emph{not}
  $\mathscr{C}^{\infty}$-diffeomorphic to
  $\{\Sigma_{\bar{\tau}}^{\bar{E}}\}_{\bar{E} > E_c}$ (note that, upon a phase
  transition, the parameters of the potential change from $\tau$ to
  $\bar{\tau}$). Further, the topological difference between these two leaves is
  a $\nu$-cell, where $\nu$ is the index of the critical point $E_c$. Loosely
  speaking, it can be said that the origin of phase transitions is this change
  in topology. Therefore, different phases of the theory are topologically
  inequivalent.
\item Finally, let us note that all of these conclusions we have drawn so far
  have one last implication, a topological constraint: the Euler Characteristic,
  $\chi$, computed from \gE gives the quantization rules [for the
  energy]. Therefore, our theory can be topologically quantized; in fact, each
  topologically inequivalent leaf $\{\Sigma_{\tau}^{E}\}$ has its own
  $\chi_{\tau}^{E}$ and, thus, its own quantization rules.
\end{enumerate}

With all of these facts in hand, let us see what is the framework they imply:
given a theory, $L = \g(\pi,\pi) + V_{\tau}[\phi]$, we can readily construct its
Jacobi metric, $\gE^{\tau} = 2\, (E - V_{\tau})\, \g$: from this, we can do two
things, either compute the geodesics of $\gE^{\tau}$ and classify their Jacobi
fields in terms of $\tau$, or calculate its Euler Characteristics,
$\chi_{\tau}^{E}$, and find the quantization rules (bearing in mind that they
will vary with $\tau$). The different geodesics (corresponding to different
Jacobi fields) will label diffeomorphically equivalent foliations
$\{\Sigma_{\tau}^{E}\}$, while the order of the zeros of
$F_{\tau}^{\nu}[\partial\mathpzc{M}]$ will correspond to the $\nu$-cells associated
with a particular phase transition --- furthermore, they will be responsible for
the Lee-Yang zeros and the Stokes phenomena of the theory. In turn, this gives
the partition function a meromorphic and a modular character when seen with
respect to $\tau$.

All of this was possible because we extended the parameter space, allowing
$\tau$ to be complex. This is completely analogous to complexifying the solution
space (moduli space) of the theory given, a fact which is made clear in
\cite{garciaguralnik} (and employed in lattice calculations in
\cite{mollifier}).

Lastly, using $\tau$, it will be possible to construct ``dualities'' between
different phases. However, in general, these dualities will be non-trivial
combinations of the parameters of the theory (as opposed to what we found in the
simple examples above, where $E/\mu \mapsto -E/\mu$ or $E \mapsto -E$ did the
job).
\section{The $\lambda\, \phi^{\mathbf{4}}$ Potential}\label{subsec:lp4}
This theory is defined for scalar-valued fields, $\phi$, by the Lagrangian $L =
\tfrac{1}{2}\, \bigl(\g(\pi,\pi) - \mu\, \phi^2 - \tfrac{\lambda}{2}\, 
\phi^4\bigr)$, i.e., the potential is given by $V(\phi) = \tfrac{1}{2}\,
\bigl(\mu\, \phi^2 + \tfrac{\lambda}{2}\, \phi^4\bigr)$; which is invariant by
$\mathbb{Z}_2$-reflection: $\phi \mapsto -\phi$.

The Jacobi metric for this theory is given by $\gE^{\tau} = 2\, \bigr(E -
\tfrac{\mu}{2}\, \phi^2 - \tfrac{\lambda}{4}\, \phi^4\bigr)\, \g$. And
solving the normalization condition $\gE^{\tau}(\gamma',\gamma') = 1$ --- where
$\gamma(s)$ is the geodesic we want to compute and $\gamma' =
\tfrac{d\gamma}{ds}$, where $s$ is the arc-length parameter --- should give us
the geodesic structure of the vacuum manifold (resp. moduli space):

\begin{align}
  \nonumber
  \gE^{\tau}(\gamma',\gamma') &= 1 \; ;\\
  \label{eq:lp4geq}
  2\, \bigr(E - \tfrac{\mu}{2}\, \gamma^2 - \tfrac{\lambda}{4}\,
    \gamma^4\bigr)\, (\gamma')^2 &= 1 \; ; \\
  \nonumber
  \hksqrt{E - \tfrac{\mu}{2}\, \gamma^2 - \tfrac{\lambda}{4}\,
    \gamma^4}\, d\gamma &= \tfrac{1}{2}\, ds \; .
\end{align}

The full picture presents itself upon a more detailed analysis of
$\int\bigl(E - \mu\, \gamma^{2}/2 - \lambda\, \gamma^{4}/4\bigr)^{1/2}\, d\gamma$, which
is the [elliptic] integral that needs to be solved in order to find
$\gamma(s)$. Therefore, it is useful to consider the polynomial $P(\gamma) =
(\gamma^{2} - r_1)\, (\gamma^{2} - r_2)$, where $(-\lambda/4)\, P(\gamma) = E -
\mu\, \gamma^{2}/2 - \lambda\, \gamma^{4}/4$ and $r_{1,2} = -\bigl(\mu \pm
\hksqrt{4\, E\, \lambda + \mu^{2}}\bigr)/\lambda$. It is the \emph{discriminant}
of this polynomial $P(\gamma)$ that will, ultimately, determine the different
solutions of the theory: $\Delta = (r_1 - r_2)^{2} = \lambda\, E + \mu^{2}/4
\lesseqgtr 0$.

As mentioned above, the ``dualities'' between different phases involves
non-trivial combinations of the parameters of the theory. In this case, the
dualities are given by the possible values of the discriminant above: $\Delta >
0$, $\Delta = 0$ or $\Delta < 0$.

The two inequalities, $\Delta > 0$ and $\Delta < 0$, are related by the
reflection of the $\tau$ parameter of this theory, where $\tau = \mu^2/\lambda$,
i.e., by the analytic continuation of the mass parameter such that $\mu^2
\mapsto -\mu^2$, which implies that $\tau \mapsto -\tau$ --- note that this can
be obtained by a modular transformation, $\mathds{M}\, \tau = \tfrac{a\, \tau +
  b}{c\, \tau + d} = -\tau$, such that $b=0=c$, $d=1$ and $a=-1$; further, the
analytic continuation of $\mu$ is such that $\mu \mapsto \pm i\, \mu$, or more
generically as $\mu \mapsto e^{\pm i\, \pi\, \nu/2}\, \mu$, for odd $\nu$ (this is
related to the Stokes phenomena discussed previously, where $\nu$ selects a
certain Riemann sheet for this analytic continuation, which depends on the
Maslov-Morse index $\nu$) ---; meanwhile, the case $\Delta = 0$ has to be
computed separately, once it establishes a fixed value $\tau = -4\, E$ --- note
that when $\Delta = 0$ the \emph{resultant} of $P(\gamma)$ and its derivative,
$P'(\gamma)$, also vanishes, this resultant being given by $\Res(P,P') =
-\tfrac{\lambda}{4}\, \Delta$, which says that $P$ and $P'$ have a common root
(which happens when $\tau = \tfrac{\mu^2}{\lambda} = -4\, E$).

This whole framework is potentially a very interesting result, once
discriminants are closely related and contain information about
\emph{ramifications} (``branching out'' or ``branches coming together'') in
Number Theory. As we just saw, these ramifications are related to the
topological structure of each solution, where the Maslov-Morse index, $\nu$,
selects the appropriate Riemann sheet (for the analytic continuation) and the
topology of the problem (determined by the attachment of a
$\nu$-cell). Therefore, this duality relating the two different solutions, for
$\tau$ and $-\tau$, is given by a [particular] modular transformation of $\tau$
(resp. analytic continuation of $\mu$), which turns out to be a measure of the
ramifications of the problem.

The analytical answers for the geodesic $\gamma(s)$ are found to be given by the
following implicit equations:

\begin{description}
\item[$\boldsymbol{\Delta = 0 \, ,\; \Res(P,P') = 0:}$]
  \begin{equation}
    \label{eq:deltaEQ0}
    \frac{1}{3}\, \gamma^{3} + \frac{\mu}{\lambda}\, \gamma + s = 0\; ;
  \end{equation}
\item[$\boldsymbol{\Delta > 0 \, ,\; \Res(P,P') < 0:}$]
  \begin{equation}
    \label{eq:deltaGT0}
    \begin{split}
      3\, s &+ \gamma\, \hksqrt{(\gamma^{2} - r_1)\,(\gamma^{2} - r_2)} + 2\, r_1\,
      \hksqrt{r_2}\,
      F\biggl(\frac{\gamma}{\hksqrt{r_1}}\boldsymbol{;}
      \hksqrt{\tfrac{r_1}{r_2}}\biggr) -\\
      &- \frac{2}{3}\, \frac{\mu}{\lambda}\,
      \hksqrt{r_2}\, \Biggl[F\biggl(\frac{\gamma}{\hksqrt{r_1}}\boldsymbol{;} 
      \hksqrt{\tfrac{r_1}{r_2}}\biggr) - E
      \biggl(\frac{\gamma}{\hksqrt{r_1}}\boldsymbol{;}
      \hksqrt{\tfrac{r_1}{r_2}}\biggr)\Biggr] = 0 \; ;
    \end{split}
  \end{equation}
\item[$\boldsymbol{\Delta < 0 \, ,\; \Res(P,P') > 0:}$]
  \begin{equation}
    \label{eq:deltaLT0}
    \begin{split}
      3\, s &+ \gamma\, \hksqrt{(\gamma^{2} - r_1)\,(\gamma^{2} - r_2)} + 2\, r_1\,
      \hksqrt{r_2}\,
      F\biggl(\frac{\gamma}{\hksqrt{r_1}}\boldsymbol{;}
      \hksqrt{\tfrac{r_1}{r_2}}\biggr) -\\
      &- \frac{2}{3}\, \frac{\mu}{\lambda}\,
      \hksqrt{r_2}\, \Biggl[F\biggl(\frac{\gamma}{\hksqrt{r_1}}\boldsymbol{;} 
      \hksqrt{\tfrac{r_1}{r_2}}\biggr) - E
      \biggl(\frac{\gamma}{\hksqrt{r_1}}\boldsymbol{;}
      \hksqrt{\tfrac{r_1}{r_2}}\biggr)\Biggr] = 0 \; .
    \end{split}
  \end{equation}
\end{description}

Note that, in the equations above, $F(z;k)$ is the incomplete elliptic integral
of the first kind, while $E(z;k)$ is the incomplete elliptic integral of the
second kind. Moreover, $r_{1,2}$ are defined (as shown above) as the solutions to
the polynomial equation $P(\gamma) = 0$: $r_{1,2} = -\bigl(\mu \pm \hksqrt{\mu^{2} +
  4\, \lambda \, E}\bigr)/\lambda$. On top of this, although \eqref{eq:deltaGT0}
and \eqref{eq:deltaLT0} have the same form, they will yield distinct solutions,
once the relation given by $\Delta = \lambda\, E + \mu^{2}/4$ will either be
positive or negative, which, in turn, affects the outcome of $r_{1,2}$ --- as
already discussed above for $\tau \mapsto -\tau$.

The graphical results are shown below, where $\tfrac{E}{\lambda} = 1,
\gamma(0) = 0$ and $\Delta$, respectively, assumes positive
($\Delta > 0$), null ($\Delta = 0$) and negative ($\Delta < 0$) values:

\newpage

\begin{center}
  \includegraphics[scale=0.45]{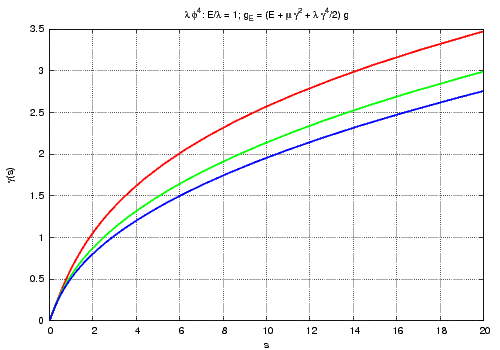} \hfill
  \includegraphics[scale=0.45]{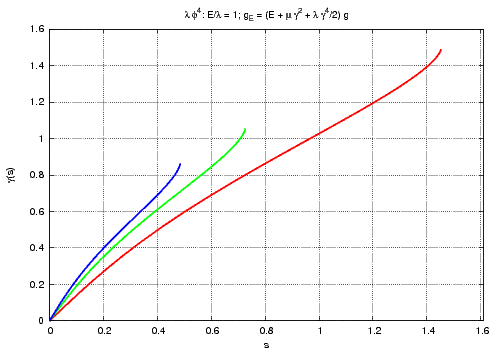}

  \includegraphics[scale=0.45]{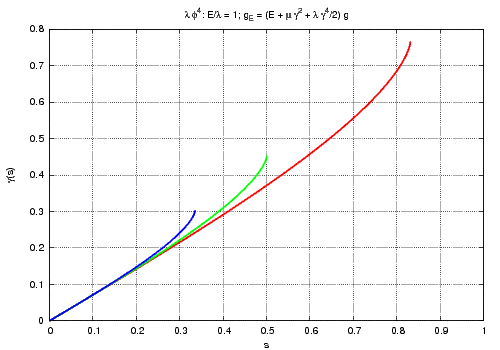}

  \bigskip
  {\footnotesize \noindent\textbf{Figure 3:} The top-left-corner plot shows the
    geodesic for $E/\lambda = 1$ and $\Delta > 0$, while the top-right-corner
    one depicts it for $E/\lambda = 1$ and $\Delta = 0$. The bottom-center plot
    shows $E/\lambda = 1$ and $\Delta < 0$.}
\end{center}
\bigskip \bigskip

We can clearly see that, for positive $\Delta$ (leftmost graph), we have a
smooth geodesic $\gamma(s)$ representing the symmetric phase. Then, when
$\Delta$ vanishes (middle graph), there is a clear change which delimits
the 2 different phases of the theory. Moreover, when $\Delta$ is negative
(rightmost graph), we have the third [broken-symmetric] phase of the theory
(which has a finite geodesic).
\chapter{Gauge Theory Examples}\label{sec:examples}
Now, we will consider examples of theories with gauge symmetry (rather than the
discrete $\mathbb{Z}_2$-symmetry of above). As seen in \eqref{eq:lp4geq},
taking $\gamma \mapsto -\gamma$ did not change the geodesic structure, i.e., the
procedure was covariant (resp. equivariant). This should come as no surprise,
once the potential, $V_{\tau}(\phi)$, is an equivariant function of the fields
and the spacetime (base manifold) metric does not depend on the gauge symmetry
involved either.

As mentioned before in section \ref{sec:appsqft}, we have assumed thus far that
the quantum fluctuations vanish at the boundary,
$\delta\phi|_{\partial\mathpzc{M}} = 0$. This, in turn, implies that the Jacobi
fields $\tilde{\pi}$ will satisfy $\tilde{\pi}|_{\partial\mathpzc{M}} =
0$. While the fields considered have no internal structure (i.e., gauge
symmetry), this is a fairly straightforward constraint, in the sense that its
solution is trivial (albeit labeled by $\tau$). However, when gauge symmetry is
present, there may be \emph{non-trivial} solutions to these constraints. This is
the phenomenon of spontaneous symmetry breaking.

Therefore, following the discussion done in section \ref{sec:appsqft}, the
following may happen: the quantum corrections, which are handles attached to the
classical solution, may now be able to change the topology of the classical
solution. So, when a symmetry is reduced from $G \rightarrow H$ (where $H$ is a
subgroup of $G$; see, e.g., \cite{ladies, kerbrat, fulpnorris, mayer,
  sardanashvily, neeman}), we start with a Jacobi field which satisfies
$\tilde{\pi}_{G}|_{\partial\mathpzc{M}} = 0$ and end up with a Jacobi field who only has
$H$ as symmetry, which implies that $\tilde{\pi}_{H}|_{\partial\mathpzc{M}} = 0$ ---
this means that the initial degrees-of-freedom that combined (respecting the $G$
symmetry) to yield the [initial] constraint $\tilde{\pi}_{G}|_{\partial\mathpzc{M}} =
0$, are not all available now, such that only part of the original symmetry is
still respected, yielding $\tilde{\pi}_{H}|_{\partial\mathpzc{M}} = 0$ (the remaining
degrees-of-freedom having recombined in non-trivial ways); thus, we end up with
a source (or sink) of Jacobi fields that only have $H$ as symmetry.

In this sense, the Atiyah-Singer Index Theorem can be used to measure this
variation (i.e., to measure the inequivalent representations of the algebra of
observables):

\begin{enumerate}
\item If all of the quantum corrections preserve the topology of the classical
  solution, the topological index does not change, which implies that the
  analytical index of the differential operator in question also does not
  change, which, in turn, leaves the vacuum state unchanged.
\item On the other hand, if the quantum fluctuations change the topology of the
  classical solution (as described above), the topological index will change
  (following the attachment of the appropriate $\nu$-cell), implying that the
  analytical index changes as well, which means that the vacuum state changes.
\end{enumerate}

From a different viewpoint, the question can be posed in the following way:
Given a certain symmetry breaking connection, how can the topology of the moduli
space of the associated Higgs Bundle be studied?

There are some studies in this direction, but no general answers: this is
because these types of characterizations are highly model-dependent, i.e., they
depend on the particular properties of the connection chosen for the Higgs
Bundle in question. For instance, following the discussion above, if the
boundary $\partial\mathpzc{M}$ is non-existent (i.e., $\mathpzc{M}$ is compact),
then $\tilde{\pi}|_{\partial\mathpzc{M}} = 0$ is trivialy satisfied; however, if
$\mathpzc{M}$ has, e.g., punctures (i.e., a finite set of points
omitted), then the connection in question will have a certain ramification
structure and the constraint $\tilde{\pi}|_{\partial\mathpzc{M}} = 0$ will have
non-trivial solutions (see, e.g., \cite{msheld}, and references therein);
therefore, the boundary conditions (in the sense of \cite{garciaguralnik,
  mollifier}) on the Jacobi fields ultimately determine the structure of the
connection (i.e., gauge field).

The examples below are twofold: the first one (Landau-Ginzburg Functional,
section \ref{subsec:lgf}) serves the purpose of showing the equivariance of
the method discussed in this work, once the connection (which is
$\mathfrak{u}(1)$-valued) does not undergo symmetry breaking; while the second
one (Seiberg-Witten Functional, section \ref{subsec:swf}) can be understood
as defined for pairs $(A, \phi)$, where $A$ is a Hermitian connection
(compatible with the holomorphic structure on the bundle in question) and $\phi$
is a section (of the bundle at hand) --- in this sense, one can consider the
space of solutions to Hitchin's equations (i.e., the moduli space of the
associated Higgs Bundle) given by,

\begin{align}
  \label{eq:hitchineq1}
  F_A + [\phi, \phi^{*}] &= 0 \; ; \\
  \label{eq:hitchineq2}
  d_A'' \phi &= 0 \; ;
\end{align}
where $F_A$ is the curvature of $A$ and $d_A'' \phi$ is the anti-holomorphic
part of the covariant derivative of $\phi$ (compare these with
equations \eqref{eq:sw3} and \eqref{eq:sw4}). In this sense, the different
solutions found via \eqref{eq:swjacobi} are a direct statement about the
topology of the moduli space of the Higgs Bundle under study.
\section{The Landau-Ginzburg Functional}\label{subsec:lgf}
To start off, let us consider the case where the base manifold is a compact Riemann
surface $\Sigma$ equipped with a conformal metric and the vector bundle is a Hermitian
line bundle $\mathpzc{L}$ (i.e., with fiber $\mathbb{C}$ and a Hermitian metric
$\hm{\cdot}{\cdot}$ on the fibers).

The Landau-Ginzburg functional is defined for a section $\varphi$ and a unitary connection
$D_A = \ed\, + A$ of $\mathpzc{L}$ as ($\sigma\in\mathbb{R}$ is a real scalar),

\begin{equation*}
  L(\varphi,A) = \int_{\Sigma} |F_A|^2 + |D_A\, \varphi|^2 + \frac{1}{4}\, \bigl(\sigma -
    |\varphi|^2\bigr)^2 \; .
\end{equation*}

Thus, its Euler-Lagrange equations [of motion] are given by:
\begin{align*}
  D_A^{*}\, D_A\, \varphi &= \frac{1}{2}\, \bigl(\sigma - |\varphi|^2\bigr)\, \varphi\;;\\
  D_A^{*}\, F_A &=  -\re\hm{D_{A}\, \varphi}{\varphi} \; ;
\end{align*}
where, $D_A^{*}$ is the dual of $D_A$, i.e., $D_A^{*} = -*\, D_A\, * = -*\, (\ed\,
+ A)\, *$. Note that the equation for $F_A$ (the second one above) is linear 
in $A$. Since $D_A$ is a unitary connection, $A$ is a $\mathfrak{u}(1)$-valued
1-form. This Lie algebra (of the group $U(1)$) will sometimes be identified with
$i\, \mathbb{R}$ --- in other words, our $A$ corresponds to $-i\, A$ in the
standard physics literature (where $A$ is real-valued).

Before we go any further, some notational remarks are in order. We decompose the
space of 1-forms, $\Omega^1$, on $\Sigma$ as $\Omega^1 = \Omega^{1,0} \oplus
\Omega^{0,1}$, with $\Omega^{1,0}$ spanned by 1-forms of the type $dz$ and
$\Omega^{0,1}$ by 1-forms of the type $d\bar{z}$. Here $z = x + i\, y$ is a
local conformal parameter on $\Sigma$ and $\bar{z} = x - i\, y$. Therefore, $dz
= dx + i\, dy$, $d\bar{z} = dx - i\, dy$, $\partial_z = \tfrac{1}{2}\,
(\partial_x - i\, \partial_y)$ and $\partial_{\bar{z}} = \tfrac{1}{2}\,
(\partial_x + i\, \partial_y)$. Furthermore, if $\partial_x$ and $\partial_y$
are an orthonormal basis of the tangent space of $\Sigma$ at a given point, we
have that $\hm{dz}{dz} = 2$, $\hm{d\bar{z}}{d\bar{z}} = 2$ and
$\hm{dz}{d\bar{z}} = 0$. Given that the decomposition $\Omega^1 = \Omega^{1,0}
\oplus \Omega^{0,1}$ is orthogonal, we may also decompose $D_A$ accordingly:
$D_A = \partial_A + \bar{\partial}_A$, where $\partial_A\varphi \in
\Omega^{1,0}(\mathpzc{L})$ and $\bar{\partial}_A\varphi \in
\Omega^{0,1}(\mathpzc{L})$ for all sections $\varphi$ of $\mathpzc{L}$
(holomorphic, $\bar{\partial}_A\, f(z,\bar{z}) = 0 \Leftrightarrow
f(z,\bar{z}) = f(z)$, and anti-holomorphic, $\partial_A f(z,\bar{z}) = 0
\Leftrightarrow f(z,\bar{z}) = f(\bar{z})$, parts; i.e., the space of 1-forms
and the space of connections is decomposable into a direct sum of its
holomorphic and anti-holomorphic parts: $\partial_A = \partial + A^{1,0}$ and
$\bar{\partial}_A = \bar{\partial} + A^{0,1}$; while the exterior derivative is
given by $\ed\, = \partial + \bar{\partial}$). As expected, we have that
$\partial_A\, \partial_A = 0 = \bar{\partial}_A\, \bar{\partial}_A$ and $F_A =
-\bigl(\partial_A\, \bar{\partial}_A + \bar{\partial}_A\, \partial_A\bigr)$.

It is not difficult to show \cite{geomana} that:

\begin{align*}
    L(\varphi,A) &= \int_{\Sigma} |F_A|^2 + |D_A\, \varphi|^2 + \frac{1}{4}\, \bigl(\sigma -
      |\varphi|^2\bigr)^2 \; ; \\
    &= 2\, \pi\, \deg(\mathpzc{L}) + \int_{\Sigma} 2\, \bigl|\bar{\partial}_A\,
      \varphi\bigr|^2 + \Bigl(*(-i\, F_A) - \frac{1}{2}\, \bigl(\sigma -
      |\varphi|^2\bigr)\Bigr)^2 \; ; \\
    \text{where}\quad \deg(\mathpzc{L}) &= c_1(\mathpzc{L}) = \frac{i}{2\, \pi}\, \tr(F_A)\; ;
\end{align*}
i.e., the degree of the line bundle is given by the 1st Chern class.

Therefore, a very useful consequence from the above is that if $\deg(\mathpzc{L})
\geqslant 0$ the lowest possible value for $L(\varphi,A)$ is realized if $\varphi$ and $A$
satisfy

\begin{align}
  \label{eq:lg1}
  \bar{\partial}_A \, \varphi &= 0 \; ;\\
  \label{eq:lg2}
  *(i\, F_A) &= \frac{1}{2}\, \bigl(\sigma - |\varphi|^2\bigr) \; .
\end{align}

These equations are just the expression of the self-duality of the
Landau-Ginzburg functional. If $\deg(\mathpzc{L}) < 0$, then these equations
cannot have any solution, thus one has to consider the self-duality equations
arising from the Landau-Ginzburg functional where the term $+2\, \pi\,
\deg(\mathpzc{L})$ is substituted by $\boldsymbol{-}2\, \pi\,
\deg(\mathpzc{L})$. Therefore, without loss of generality, we shall assume
$\deg(\mathpzc{L}) \geqslant 0$. A necessary condition for the
solvability of \eqref{eq:lg2} is that,

\begin{align}
  \nonumber
  2\, \pi\, \deg(\mathpzc{L}) = \int i\, F_A &= \frac{1}{2}\, \int_{\Sigma} \bigl(\sigma -
    |\varphi|^2\bigr) \leqslant \frac{\sigma}{2}\, \text{Area}(\Sigma)\; ; \\
  \label{eq:lgentropy}
  \therefore\; \sigma &\geqslant \frac{4\, \pi \deg(\mathpzc{L})}{\text{Area}(\Sigma)} \; ;
\end{align}
and the equality only occurs if, and only if, $\varphi \equiv 0$.

The following result is useful when studying the solutions of the above
functionals \cite{geomana}: \emph{Let $\Sigma$ be a compact Riemann surface with
  a conformal   metric and $\mathpzc{L}$ as before. For any solution of
  \eqref{eq:lg1}, we have that $|\varphi| \leqslant \sigma$ on $\Sigma$.} That
is, this maximum principle states that the amplitude of the field cannot exceed
the height of the potential (see the first plot below).

Let us now construct the Jacobi metric for this potential and find its possible
geodesics:
\begin{align*}
  \gE &= 2\, \big(E - V_{\sigma}(\gamma)\big)\, \g \; ; \\
  &= 2\, \bigg(E + \frac{1}{4}\,\big(\sigma - |\gamma|^2\big)^2\bigg)\, \g \; .
\end{align*}

In a complete analogy to what was previously done, we consider the $P(\gamma) =
E - V(\gamma)$ polynomial. However, in order to make this analysis clearer, let
us, in fact, use a slightly different polynomial given by $P^{\prime}(\gamma) =
4\, \bigl(P(\gamma) - E\bigr)$. It is straightforward to see that $P'(\gamma) =
\bigl(|\gamma|^2 - \sigma\bigr)\, \bigl(|\gamma|^2 - \sigma\bigr)$, which means
that $\sigma$ is the only root of $P'(\gamma)$, with double multiplicity (the
two roots coalesce into one).

This situation implies that the discriminant of $P'(\gamma)$ vanishes, i.e.,
$\Delta = 0$, and the different phases of the theory are labelled by $\sigma =
0$ and $\sigma > 0$.

The plots below have, respectively, the following
values for $\sigma$: $0.0$, $3.0$, $5.0$, $7.0$, $11.0$ and $31.0$. As they
show, when $\sigma = 0$ we have $c_1(\mathpzc{L}) = \deg(\mathpzc{L}) =
\frac{i}{2\, \pi}\, \tr(F_A) = 0$ and its geodesic has a clear character which
is quite different otherwise:

\begin{center}
  \includegraphics[scale=0.38]{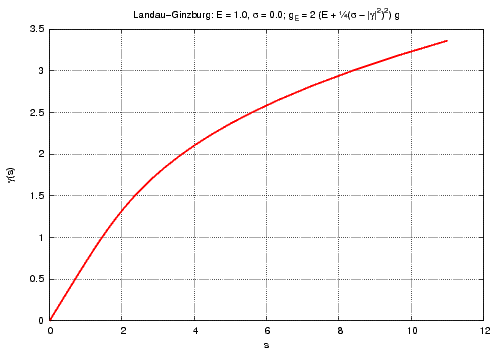}
  \includegraphics[scale=0.38]{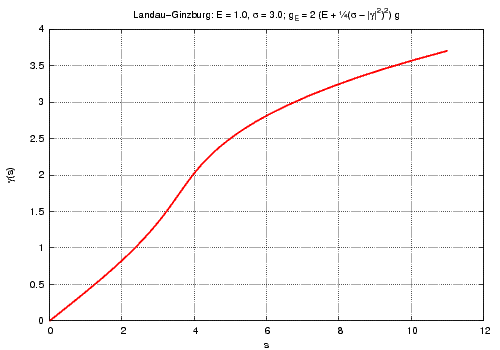}
  \includegraphics[scale=0.38]{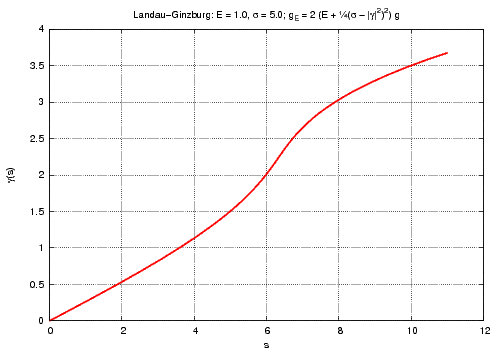}

  \includegraphics[scale=0.38]{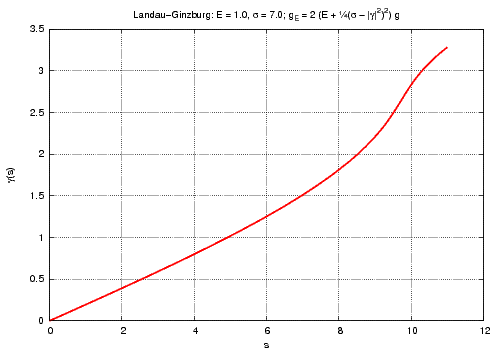}
  \includegraphics[scale=0.38]{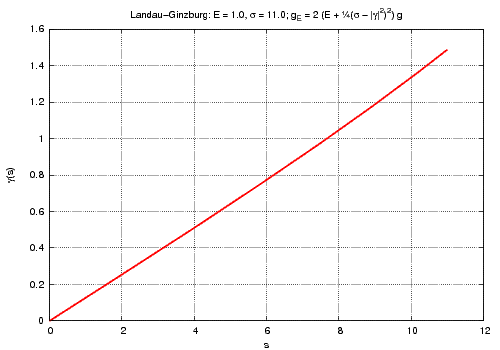}
  \includegraphics[scale=0.38]{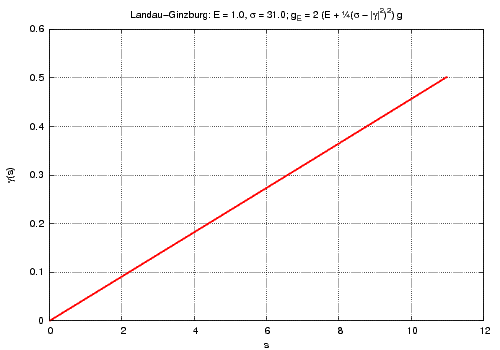}

  \bigskip
  {\footnotesize \noindent\textbf{Figure 4:} The plots above show the clear
    distinction between the $\sigma = 0$ solution and the $\sigma > 0$ ones; the
    first plot, on the upper left corner, has $\sigma = 0$, while the other ones
    have differing positive values for $\sigma$.}
\end{center}

As a last remark on this model, note it can be thought as defined by a
pair $(A, \varphi)$, where $A$ is a unitary connection compatible with the
holomorphic structure of the bundle in question, and $\varphi$ is a [global]
section of this bundle. In this sense, Hitchin's self-duality equations become,
\begin{align}
  \label{eq:hitchinlg1}
  \bar{\partial}_A \varphi &= 0 \; ; \\
  \label{eq:hitchinlg2}
  F_A + [\varphi, \varphi^{*}] &= 0 \; .
\end{align}
Compare these with \eqref{eq:lg1} and \eqref{eq:lg2}.

In this sense, the solutions found above are a direct statement about the
topology of the Higgs Bundle in question.
\section{The Seiberg-Witten Functional}\label{subsec:swf}
In this case, the base manifold, $\mathpzc{M}$, is a compact, oriented,
4-dimensional Riemannian manifold endowed with a spin${}^c$ structure, \ie, a
spin${}^c$ manifold. The determinant line of this spin${}^c$ structure will be
denoted by $\mathpzc{L}$ and the Dirac operator determined by a unitary
connection $A$ on $\mathpzc{L}$ will be denoted by $\mathcal{D}_A$. Recalling
the half spin bundle $\mathcal{S}^{\pm}$ defined by the spin${}^c$ structure, we
see that $\mathcal{D}_A$ maps sections of $\mathcal{S}^{\pm}$ into sections of
$\mathcal{S}^{\mp}$ \cite{seibergwitten,donaldson,liviu,geomana}.

In this fashion the Seiberg-Witten functional for a unitary connection $A$ on
$\mathpzc{L}$ and a section $\varphi$ of $\mathcal{S}^+$ is given by,

\begin{equation}
  \label{eq:swf}
  SW[\varphi,A] = \int_{\mathpzc{M}} |\nabla_A \varphi|^2 + |F^+_A|^2 + \frac{R}{4}\,
    |\varphi|^2 + \frac{1}{8}\, |\varphi|^4 \; ;
\end{equation}
where $\nabla_A$ is the spin${}^c$ connection induced by $A$ and the Levi-Civita
connection of $\mathpzc{M}$, $F_A^+$ is the self-dual part of the curvature of $A$ and $R$
is the scalar curvature of $\mathpzc{M}$. Its Euler-Lagrange equations are given by,

\begin{align}
  \label{eq:sw1}
  \nabla_A^*\, \nabla_A\, \varphi &= -\biggl(\frac{R}{4} + \frac{1}{4}\, |\varphi|^2
    \biggr)\, \varphi \; ; \\
  \label{eq:sw2}
  \ed{}^*\, F_A^+ &= -\re\hm{\nabla_{A}\, \varphi}{\varphi} \; .
\end{align}

Using a spin frame, it is not difficult \cite{geomana} to show that the Seiberg-Witten
functional can be written in the following form:

\begin{equation*}
  SW[\varphi,A] = \int_{\mathpzc{M}} |\mathcal{D}_A \varphi|^2 + \biggl|F_A^+ -
    \frac{1}{4}\, \hm{e_j\cdot e_k\cdot \varphi}{\varphi}\, e^j \wedge e^k\biggr|^2 \; ;
\end{equation*}
where $e^j$ are 1-forms dual to the tangent vectors $e_j$, $e^j(e_k) = \delta^j_k$;
$j,k=1,\dotsc,4$. As a corollary of the above, the lowest possible value of the
Seiberg-Witten functional is achieved if $\varphi$ and $A$ are solutions of the
\emph{Seiberg-Witten equations}:

\begin{align}
  \label{eq:sw3}
  \mathcal{D}_A \varphi &= 0 \; ;\\
  \label{eq:sw4}
  F_A^+ &= \frac{1}{4}\, \hm{e_j\cdot e_k\cdot \varphi}{\varphi}\, e^j \wedge e^k \; .
\end{align}

Thus, self-duality is at work yet again: the absolute minima of the Seiberg-Witten
functional satisfy not only the second order equations \eqref{eq:sw1} and \eqref{eq:sw2},
but also the first order Seiberg-Witten equations \eqref{eq:sw3} and \eqref{eq:sw4}.

Although our discussion of the Seiberg-Witten functional, so far, has mirrored our
discussion of the Landau-Ginzburg one, the parameter $\sigma$ on the latter has had no
analogue in the former. This can be accomplished with the introduction of a 2-form $\mu$
and the consideration of the perturbed functional,

\begin{align*}
  SW_{\mu}[\varphi,A] &= \int_{\mathpzc{M}} |\mathcal{D}_A \varphi|^2 + \biggl|F_A^+ -
    \frac{1}{4}\, \hm{e_j\cdot e_k\cdot \varphi}{\varphi}\, e^j \wedge e^k +
    \mu\biggr|^2 \; ; \\
  &= \int_{\mathpzc{M}} |\nabla_A \varphi|^2 + |F^+_A|^2 + \frac{R}{4}\,
    |\varphi|^2 + \biggl|\mu - \frac{1}{4}\, \hm{e_j\cdot e_k\cdot \varphi}{\varphi}\, e^j
    \wedge e^k\biggr|^2 + 2\, \hm{F_A^+}{\mu} \; .
\end{align*}
Their corresponding first order equations of motion are,

\begin{align*}
  \mathcal{D}_A \varphi &= 0 \; \\
  F^+_A &= \frac{1}{4}\, \hm{e_j\cdot e_k\cdot \varphi}{\varphi}\, e^j\wedge e^k - \mu\;.
\end{align*}

If we assume that $\mu$ is closed and self-dual, then we see that $\hm{F_A}{\mu} =
\hm{F_A^+}{\mu}$, once $\hm{F_A^-}{\mu} = 0$ due to the orthogonality between
anti-self-dual and self-dual forms. Thus, given that $F_A$ represents the first Chern
class $c_1(\mathpzc{L})$ of the line bundle $\mathpzc{L}$, and we assumed $\mu$ to be
closed (hence it represents a cohomology class $[\mu]$), the integral

\begin{equation*}
  \int_{\mathpzc{M}} \hm{F_A}{\mu} \; ,
\end{equation*}
does not depend on the connection $A$, thus representing a topological invariant, denoted
by $\bigl(c_1(\mathpzc{L})\wedge[\mu]\bigr)[\mathpzc{M}]$.

Just as before, we also have a maximum principle: \emph{For any solution $\varphi$ of
  \eqref{eq:sw1} --- in particular, for any solution of \eqref{eq:sw3} --- on a compact
  4-dimensional Riemannian manifold, we have that,}

\begin{equation*}
  \max_{\mathpzc{M}} |\varphi|^2 \leqslant \max_{x\in \mathpzc{M}} (-R(x), 0)\; .
\end{equation*}

As a direct consequence of this, if the compact, oriented, Riemannian Spin${}^c$ manifold
$\mathpzc{M}$ has nonnegative scalar curvature, the only possible solution of the
Seiberg-Witten equations is,

\begin{equation*}
  \varphi\equiv 0\; ; \quad F_A^+ \equiv 0 \; .
\end{equation*}

The Jacobi metric for this Seiberg-Witten model is given by

\begin{equation}
  \label{eq:swjacobi}
  \gE = 2\, \biggl(E + \frac{R}{4}\, |\gamma|^2 + \frac{1}{8}\, |\gamma|^4
    \biggr)\, \g \; ;
\end{equation}
for a geodesic $\gamma$; and, just like before, although $P(\gamma) = E -
V(\gamma) = E + R\, |\gamma|^2/4 + |\gamma|^4/8$, let us consider $P'(\gamma) =
8\, \bigl(P(\gamma) - E\bigr) = \bigl(|\gamma|^2 - 0\bigr)\, \bigl(|\gamma|^2 +
2\, R\bigr)$. It is then clear that when $R = 0$ the discriminant vanishes
$(\Delta = 0)$ and the 2 roots merge into 1, what constitutes one of the phases
of the theory. When $R \neq 0$, the discriminant is either $\Delta > 0$ $(R >
0)$ or $\Delta < 0$ $(R < 0)$, which accounts for the other 2 phases of the
theory.

The plots below were obtained with the choice of $E = 1.0$ and $R$ respectively
equal to $0.0$, $3.0$, $7.0$, $-3.0$, $-5.0$, $-7.0$.

\begin{center}
  \includegraphics[scale=0.38]{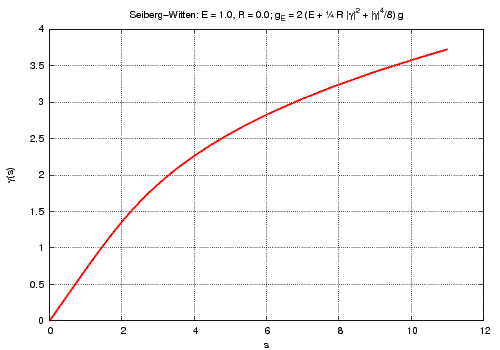}
  \includegraphics[scale=0.38]{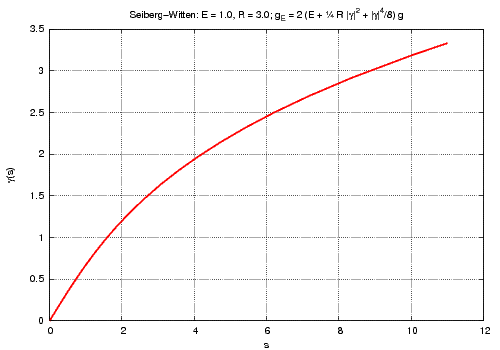}
  \includegraphics[scale=0.38]{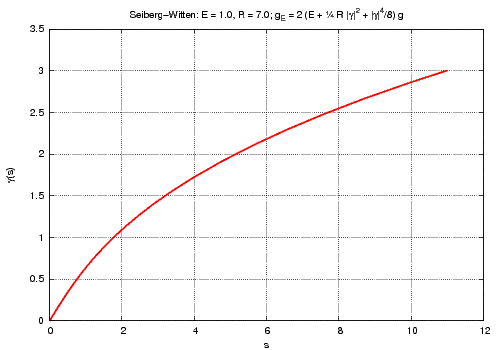}

  \includegraphics[scale=0.38]{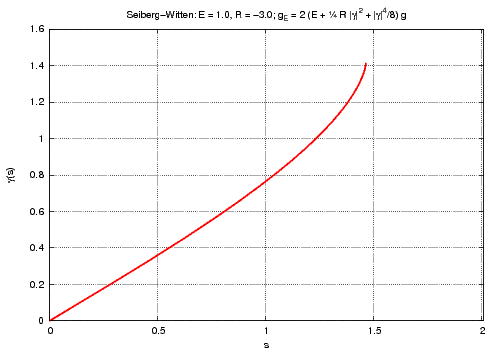}
  \includegraphics[scale=0.38]{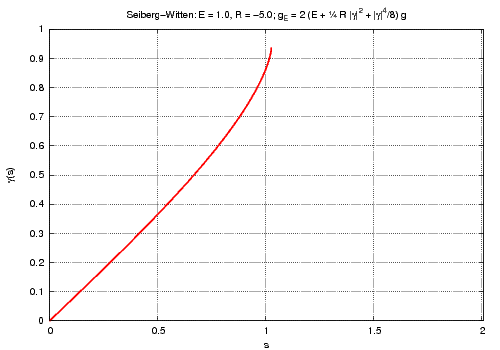}
  \includegraphics[scale=0.38]{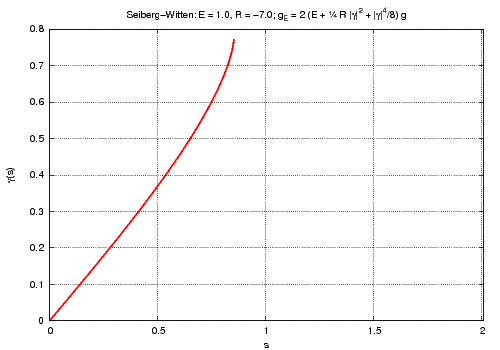}

  \bigskip
  {\footnotesize \noindent\textbf{Figure 5:} The plot in the upper left corner
    depicts the solution for $R=0$, while the plots in the upper center and
    upper right corner show solutions for $R > 0$. The plots in the lower row
    picture the solutions for $R < 0$.}
\end{center}
\chapter{Conclusions} \label{sec:conclusion}
Using a conformal transformation that amounts to finding the arc-length
reparameterization of the given problem, we were able to construct a new metric,
called Jacobi metric, such that its geodesic equation is equivalent to the
original equations of motion.

Then, we separated the given problem into its classical part and quantum
fluctuations and, by realizing that the action written in terms of the Jacobi
metric is the so-called ``energy'' in Morse Theory (resp. Morse-Bott theory), we
see can reinterpret this as a ``topological expansion'', in the sense that the
quantum fluctuations are handles attached to the classical solution.

Further, when these handles change the original (classical) topology, via the
gluing of an appropriate $\nu$-cell, there is a phase transition, in the sense
that we move from one solution [of our QFT] to another. This can be measured via
the use of an Index Theorem: once it relates the analytical index of the
differential operator in question with the topological index of the manifold
under study, we clearly see that if the topological index changes (because of a
certain handle attachment), the analytical index must change as well; which
means that the zero modes, the solutions of the equations of motion, have to
change.

Therefore, by studying the geodesics obtained from the Jacobi metric with
respect to its parameters, we can compute the topological index (via
generalizations of the Gauss-Bonnet theorem, using the Jacobi metric we derived
before) and thus compute the zero-mode solutions with respect to the different
values of the parameters. In this way, we are able to classify all possible
solutions to the QFT in question, keeping in mind that the quantum corrections
are given by handle attachments that may or may not (depending on the values of
the parameters) change the topology of the classical solutions.

To this picture, we add Lee-Yang zeros and Stokes phenomena, in order to obtain
a more robust view of what is at stake: the Lee-Yang zeros accumulate along
Stokes lines and pinch the parameter space, creating different regions of
``allowed values'' for the parameters of the theory (mass, coupling constants,
etc). This represents different sectors of theory, different phases of the
theory. And we showed that these are topologically inequivalent.

In this fashion, the partition function is meromorphic, once it is
singular along the Stokes' lines. However, it has various degrees of modular
symmetry, depending on the particulars of the theory in question. And, in this
sense, different sectors of the theory are related to each other [by an
appropriate modular transformation].

In turn, this implies that different solutions of the theory are related to each
other, a fact that we dubbed ``duality''. Ultimately, these dualities are
determined by the actual values of the parameters (and how they compose in order
to create the modular symmetry in question), which are determined by the
boundary conditions of the Schwinger-Dyson equations (as explained earlier).

Future work will focus on D-modules and dimensional construction of
0-dimensional theories (which can be completely solved). We also intend to
generalize the cubic potential showed earlier to matrix- and Lie-algebra-valued
fields \cite{lieairy}, analytically solving the its 0-dimensional counterpart
and dimensionally constructing it (via D-modules): it seems plausible that such
an extension might be related to current developments in three-dimensional gravity,
\cite{3dgravity}.

\part{Three-Dimensional Gravity and Airy Functions}
\chapter{Introduction} \label{chap:3dimintro}
Recently, new developments have shone new light on the problem of three (1
temporal, 2 spatial) dimensional gravity, \cite{3dgravity}. Loosely speaking,
here is the gist of the matter.

Classically, 3-dimensional gravity (plus a cosmological constant) is given by
the action,

\begin{equation*}
  S = \frac{1}{16\, \pi\, G}\, \int \biggl(R + \frac{2}{\ell^2}\biggr)\,
    \hksqrt{g}d^3x \; .
\end{equation*}
It is not difficult to see that the solutions to the above are all locally
equivalent and there are no gravitational waves. Thus, na{\"\i}vely, one may
think that this problem is tractable quantum mechanically.

However, on second thought, power counting tells us that $G$ has dimensions of
length, rendering the theory unrenormalizable. This leads to the conclusion that
the quantum theory does not exist. But this is premature as well, once
divergences in perturbation theory can be removed by field redefinitions
$(g_{\mu\, \nu} \mapsto g_{\mu\, \nu} + a\, R_{\mu\, \nu} + \dotsb)$ and a
renormalization of $\ell^2$.

Therefore, a slightly different venue of attack must be drawn in order to
resolve the conflicts above. What is customarily done is to write the above
Action as a gauge theory, which will lead us towards a Chern-Simons theory, in
the following way: write the metric in terms of its Cartan connection, $g_{\mu\,
\nu} = \eta_{\mu\, \nu}\, e_{\mu}^{i}\, e_{\nu}^{j}$, known as the ``vielbein''
(triad). From this, we write the spin connection, which is an
$\mathfrak{so}(2,1)$-valued 1-form: $\omega_{\mu}^{a\, b} =
e^{a}_{\mu}\, \partial_{\mu} e^{\nu\, b} + e^{a}_{\nu}\, e^{\sigma\, b}\,
\Gamma^{\nu}_{\sigma\, \mu}$.

At this point, we are ready to combine the vielbein and the spin connection into
an $\mathfrak{so}(2,2)$-valued connection, $A$:

\begin{equation*}
  A =
  \begin{pmatrix}
    \omega & \tfrac{e}{\ell} \\
    -\tfrac{e}{\ell} & 0
  \end{pmatrix} \; .
\end{equation*}

The Chern-Simons action, which is the gauge theoretical version of this problem,
is thus given by,

\begin{equation*}
  S = \frac{k}{4\, \pi}\, \int \tr\biggl(A\wedge\ed A + \frac{2}{3}\, A\wedge
    A\wedge A\biggr) \; .
\end{equation*}

Now, we can try and tackle this problem perturbatively. What we see is that,
once there is no appearance of the metric tensor in the action above, there are
no local densities in this problem, so there are no counterterm contributions in
this sense.

Further, we realize that the cosmological constant plays the role of the
structure constant of the gauge group. Therefore, it is not renormalizable.

This Chern-Simons description is valid only if the vielbein is invertible, which
is true for classical solutions; perturbation theory will not change this.

Finally, this theory is finite and renormalizable by power counting, once there
are no local counterterms. Therefore, the quantum theory seems to exist.

However, let us make some non-perturbative considerations:

\begin{itemize}
\item There is some unclear relationship between non-invertible vielbeins and
  the above picture for 3-dim gravity: classical solutions analogous to $A =
  \omega = e = 0$, which are clearly non-geometrical, must be included in a
  quantized theory of 3-dim gravity.
\item The equivalence between diffeomorphisms and gauge transformations is not
  straightforward, in the sense that gauge transformations are continuously
  connected to the identity, while more general diffeomorphisms also play a role
  in 3-dim gravity. There is no ``natural'' way to address this situation.
\item Further, the above picture in terms of a Chern-Simons theory does not
  require a sum over topologies, a basic ingredient of a theory of quantum
  gravity. There is no \emph{a priori} reason to add this to the picture
  described so far.
\item Lastly, for negative values of the cosmological constant, there is a whole
  class of classical solutions known as BTZ black holes (which make their
  appearance in the AdS/CFT correspondence). This clearly shows that the theory
  cannot be ``trivial''.
\end{itemize}

Therefore, we can summarize the above results in terms of the cosmological
constant, $\Lambda$, in the following fashion:

\begin{description}
\item[$\boldsymbol{\Lambda > 0}:$] it is not possible to define any set of
  observables that can be intrinsically measured.
\item[$\boldsymbol{\Lambda = 0}:$] this theory has no gravitons and no black
  holes, thus it has no S-matrix (no degrees-of-freedom).
\item[$\boldsymbol{\Lambda < 0}:$] this theory contains BTZ black holes as
  degrees-of-freedom; therefore its dual 2-dim CFT can be computed via the
  AdS/CFT duality.
\end{description}

So, at this point, the plan shifts to being that of finding the 2-dim CFT dual
(via AdS/CFT) to the Chern-Simons theory above, with negative cosmological
constant; for we know that there are degrees-of-freedom in this theory, given by
the BTZ black holes.

In this sense, the Euclidean partition function can be written as,

\begin{equation*}
  \mathcal{Z}[\beta,\theta] = \tr \exp\{-\beta\, H - i\, \theta\, J\} \; ;
\end{equation*}
where $H$ is the Hamiltonian, $J$ is the angular momentum (rotation of the
asymptotic AdS${}_{3}$), and $\beta$ is the imaginary time. This integral is
taken over 3-geometries conformal at infinity to a 2-torus with modular
parameter $\tau = \theta/2\pi + i\, \beta$.

However, there are two problems with this picture:

\begin{enumerate}
\item The sum of known contributions is not physically sensible because it
  cannot be written as $\tr\exp\{-\beta\, H - i\, \theta\, J\}$ such that $[H,J]
  = 0$.
\item This partition function is not convergent, once the action is not bounded
  from below.
\end{enumerate}

To remedy these, we note the following \cite{3dgravity}:

\begin{itemize}
\item Real saddle-points do not account for the whole theory.
\item Complex saddle-points are needed, which lead us towards a holomorphic
  factorized partition function consistent with an interpretation as
  $\tr\exp\{-\beta\, H - i\, \theta\, J\}$.
\item Classical geometry: in the semiclassical limit, where $G \rightarrow 0$
  with fixed AdS radius $\ell$, we obtain \emph{complex} $\beta$ and $\theta$
  --- justifying the complex saddle-points above.
\item Finally, the Hawking-Page phase transition with respect to $\beta$,
  between a thermal gas and a black hole, amounts to being a condensation, on the
  phase boundary, of the Lee-Yang zeros of the partition function.
\end{itemize}

It is at this point that we make contact with what was said in Part II of this
thesis, and also in \cite{rtnnm,mmc,mollifier,ferrante}: the parameter space
(i.e., the $(\beta,\theta)$-space) must be complexified in order to account for
all possible solutions of the theory --- this implies that \emph{complex
  geometries} must be considered in order to make sense of this theory.

Furthermore, all ingredients already described in Part II make their appearance
here as well: Lee-Yang zeros accumulate along Stokes' lines in order to account
for the Hawking-Page phase transition mentioned, and this corresponds to a
topological transition between the two phases. Also, in this case, the partition
function has modular symmetry, $\mathcal{Z}[\tau] = \mathcal{Z}[\mathds{M}\,
\tau]$, where $\mathds{M} \in \mathrm{SL}(2,\mathbb{C})/\mathbb{Z}_2$.

So, let us see what we can say about this in the next chapters.
\chapter{Extended Airy Functions} \label{chap:extairy}
The first thing we notice is that the Airy functions ($\Ai$ and $\Bi$) can be
extended from scalar-valued to matrix- and lie-algebra-valued functions, as done
in \cite{lieairy}.

In this sense, we end up with an integral representation of the form,

\begin{equation}
  \label{eq:lieairy}
  \Ai(\mathbf{J}) = \int_{\mathcal{V}} e^{i\,
    \tr\bigl(\tfrac{\boldsymbol{\Phi}^3}{3} - \mathbf{J}\,
    \boldsymbol{\Phi}\bigr)}\, d\boldsymbol{\Phi} \; ;
\end{equation}
where $\mathcal{V}$ can be either the space of $n\times n$ Hermitian matrices
(for the case of a matrix-valued Airy function) or, in general, a vector space
that contains the representation of the Lie algebra in question (for the case of
a Lie-algebra-valued Airy function); and $\mathbf{J}\in \mathcal{V}$.

This integral does not exist in the sense of Lebesgue, but it is well-defined as
a distribution. The formal similarity of these extensions of the Airy function
to the scalar-valued Airy function accounts for its name.

In this sense, just as remarked before in chapter \ref{par:airy} of Part II, an
integral that, \emph{a priori}, looks ill-defined (for it is unbounded from
below), does indeed have a finite solution and non-trivial content, given by the
Airy function (and its respective extensions). However, this extensions work for
more general polynomials of $\boldsymbol{\Phi}$, as long as these have the
so-called \emph{Airy property}, \cite{lieairy}. In particular, the differential
equations associated with this kind of integral representations is treated by
Harish-Chandra's study of invariant differential operators and integrals on
semisimple Lie algebras, \cite{hc}. Unfortunately, the study of the connections
between Harish-Chandra theory, \cite{hc}, and the Schwinger-Dyson equations of
these models is beyond the scope of this work.

It is worth remarking that these extensions of $\boldsymbol{\Phi}$ and
$\mathbf{J}$ are completely analogous to what we did before when we complexified
the $\phi$-space. In fact, this is a generalization of that idea: you can extend
from $\mathbb{R}$-valued to $\mathbb{C}$-valued scalars, but also to more
complicated fields, as matrix- or Lie-algebra-valued ones.

Once again, as observed before, the allowed values of $\mathbf{J}$ are
determined by the particular contours that render the integral representation
above finite, and there will be a certain ``duality'' between the different
solutions determined by the distinct set of allowed values (which is, in turn,
determined by the particular contour in question).

Therefore, summarizing, we have managed to generalize the original cubic
potential in different ways: from $\mathbb{R}$-valued to $\mathbb{C}$-valued
scalars, and from scalars to matrix- and Lie-algebra-valued fields, via equation
\eqref{eq:lieairy}. Now, two points are important to note:

\begin{enumerate}
\item Equation \eqref{eq:lieairy} is just a particular case of a more general
  one, given by \cite{lieairy}:
  \begin{equation}
    \label{eq:generallieairy}
    \mathcal{Z}_p(\mathbf{J}) = \int_{\mathcal{V}} e^{i\, (
      p(\boldsymbol{\Phi}) - \tr(\mathbf{J}\,
      \boldsymbol{\Phi}))}\, d\boldsymbol{\Phi} \; ;    
  \end{equation}
  where $p(\boldsymbol{\Phi})$ is a $G$-invariant polynomial on $\mathcal{V}$,
  and $G$ is the Lie group in question. Thus, the Airy case is recovered when we
  specialize to the case of $p(\boldsymbol{\Phi}) = \boldsymbol{\Phi}^3/3$; but
  other generalizations are possible, e.g., $p(\boldsymbol{\Phi}) =
  \tr\boldsymbol{\Phi}^4/4$, or $p(\boldsymbol{\Phi}) = \tr\bigl(\mathbf{A}\,
  \boldsymbol{\Phi}^2/2 + \mathbf{B}\, \boldsymbol{\Phi}^4/4\bigr)$, where
  $\mathbf{A}, \mathbf{B}\in \mathcal{V}$.
\item The equation above can be understood as defining a matrix model, where the
  general connected matrix-model partition function is given by,
  \begin{equation*}
    e^{\mathcal{Z}} = \int e^{-\tr(V(\mathbf{M}))}\, d\mathbf{M}\; ;
  \end{equation*}
  where $\mathbf{M}$ is an $n\times n$-matrix, and $V(\mathbf{M})$ is the
  Potential for the model; the integral is taken over the space called
  $\mathcal{V}$ above.
\end{enumerate}

At this point, we note that in spite of all of these generalizations we have
done so far, we are still talking about a 0-dimensional problem. This is no
surprise, for we started with a 0-dim problem and changed it in many different
ways, but we never added any dimension(s) to it. This is the next step, to
construct the dimensions.
\chapter{Dimensional Construction Sketch} \label{chap:sketch}
At this stage, the question poses that itself is: \emph{``Given the
  generalization \eqref{eq:generallieairy}, how does one go about engineering
  the dimensions in order to extend this problem from 0-dimensions to
  $d$-dimensions?''}

This is a very hard problem and, although it is outside of the scope of this
work to tackle it, we will give a very rough sketch of what should be done.

The whole problem of using \eqref{eq:generallieairy} in order to engineer the
dimensions from first principles is that two limiting processes are involved:
the limit where the size of each dimension tends to infinite (dubbed ``infinite
volume limit''), and the limit where the number of points in each dimension
tends to infinite (called ``thermodynamical limit''). What we want to do is to
``grow'' a space whose natural differential operator is given, in some sense,
by a combination of a derivative(s) and $p'(\boldsymbol{\Phi})$.

In our case (Chern-Simons), the Schwinger-Dyson equations of motion are
symbolically given by,

\begin{align*}
  \bigl(-i\, \ed\mathbf{A} - \mathbf{A}\wedge \mathbf{A}\bigr)\,
    \mathcal{Z}[\mathbf{J}] &= \mathbf{J} \; ;\\
  \bigl(-i\, \mathrm{d}(\delta_{\mathbf{J}}) - \delta_{\mathbf{J}}\wedge
    \delta_{\mathbf{J}}\bigr)\, \mathcal{Z}[\mathbf{J}] &= \mathbf{J} \; ;
\end{align*}
where $\mathbf{A}\mapsto -i\, \delta_{\mathbf{J}}$, and $\mathbf{A},
\mathbf{J}\in \mathfrak{so}(2,2)$. In this case, the operator given by $(-i\,
\mathrm{d} - \mathbf{A}\wedge \mathbf{A})$ is the natural differential operator
under study, and it is constructed out of the polynomial $p'(\boldsymbol{\Phi})
= \tr\mathbf{A}^2 = \mathbf{A}\wedge \mathbf{A}$, with the derivative operator
being given by $\ed\mathbf{A} = \tr(\ed\mathbf{A})$.

In some sense, the tool that allows us to engineer however many dimensions as we
want is called \emph{$D$-Module}. By definition, a $D$-module is a
module\footnote{A module is a generalization of a vector space over a field
  $\mathbb{K}$ when $\mathbb{K}$ is replaced by a ring.} over a ring $D$ of
differential operators. The methods of $D$-module theory have always 
been drawn from sheaf theory and other techniques with inspiration from the work
of Alexander Grothendieck in algebraic geometry. The approach is global in
character, and differs from the functional analysis techniques traditionally
used to study differential operators. The strongest results are obtained for
over-determined systems (holonomic systems), and on the characteristic variety
cut out by the symbols, in the good case for which it is a Lagrangian
submanifold of the cotangent bundle of maximal dimension (involutive
systems).

For more about the use of $D$-modules as partition functions, we refer the
reader to \cite{cgmm} (and references therein). In particular, for our problem
we want to compute the so-called ``Airy $D$-Module'' --- to learn more about
such problems, please refer to \cite{airydmod} (and references therein).

As a final note, we remark on the properties that will be determined by the two
contours that define \eqref{eq:lieairy}. As we already know, these contours
(resp. the boundary conditions for the Schwinger-Dyson equations) ultimately
define the ranges that the parameters (mass, coupling constants, etc) of the
theory can attain. More generally, when talking about gauge theories, we should
realize that, mathematically, the coupling constants are the parameters
describing the scalar product on the particular Lie algebra in question, while,
physically, the coupling constants appear in the structure constants [defining
the Lie algebra under study] and is interpreted as measuring the ``intensity''
of the interaction between fields, \cite{tlp}.

This means that different contours will determine equivalence classes of
structure constants: coupling constants in the same equivalence class determine
the same Lie algebra, while coupling constants in different equivalence classes
will determine distinct Lie algebras.

Therefore, the two contours that yield the $\Ai$ and $\Bi$ functions and its
respective extensions (as defined previously) will ultimately determine
different Lie algebras, i.e., two distinct theories.

This is not much of a surprise when we think in terms of the results already
derived in Part II, where the geodesics of the cubic potential (chapter
\ref{par:airy} of Part II) where shown to be significantly different (thus, the
group manifolds in question have to be just as distinct).

In particular, in this context of 3-dimensional gravity, one of the solutions is
particularly interesting, for it vanishes for a certain time, after which it
suddenly and abruptly starts to grow. It begs the question: does it really mean
that, in this particular solution, (2+1)-spacetime is non-existent for some
time, after which it expands quite aggressively?

These questions are, as of now, still unresolved.
%

\part{Final Conclusions}
\chapter{Conclusions and Future Work} \label{chap:conclusion}
To sum up this work, these are a the ``milestone results'' that should be clear:

\begin{itemize}
\item In general, physically interesting QFTs has multiple solutions, which are
  associated and determined by the contours of integration that render the
  partition function finite, which is absolutely analogous to finding all
  solutions to the Schwinger-Dyson equations of the problem (and its boundary
  conditions).
\item In order to be able to find all possible solutions, we are forced to
  extend the field configurations to complex-valued fields, or, more generally,
  to matrix-valued or Lie-algebra-valued fields. In turn, this implies that the
  parameters of the theory (mass, coupling constants, etc) are also extended in
  an analogous fashion.
\item The Lee-Yang zeros of these multiple solutions condense over Stokes'
  Lines, coalescing the solutions and creating ``sectors of analyticity'' for
  the partition function. Thus, we end up having a ``canonical'' set of
  superselection rules in order to determine the particular vacuum under study.
\item In this sense, the partition function should be understood as a
  meromorphic function of the parameters of the theory. This is in complete
  accord with the interpretation of the partition function as a particular
  $D$-module, constructed from the 0-dimensional version of the model at hand,
  in order to engineer a $d$-dimensional theory. Moreover, once the partition
  function is being understood as a function of the parameters of theory, which
  have been ``extended'', we can focus on its transformation properties, in
  particular, whether or not it possess modular symmetry
  (SL$(2,\mathbb{C})/\mathbb{Z}_2$), which is the automorphism group of the
  complexified parameter space (understood as a Riemann sphere).
\item On a slightly different token, when gauge symmetry is present, the
  boundary conditions of the Schwinger-Dyson equations (resp. contours that
  render the partition function finite) and the gauge symmetry completely
  determine the local system (in the sense of \cite{simpson}) in question and,
  in this fashion, they determine some properties of the Higgs bundle associated
  with this theory. In particular, it is possible to infer the topology of this
  Higgs bundles, based on the Jacobi metric of the problem. This information is
  important for the Geometric Langlands Duality, where the singularity structure
  of the gauge connection is of importance.
\item If we are given a pure gauge theory, $S = \tfrac{1}{g}\,
  \tr(\mathbf{F}^2)$, the contours of the partition function
  ultimately determine the coupling constant, which makes its appearance in the
  structure constants determining the Lie algebra of the gauge group at hand
  (analogously, mathematicians understand this as determining the scalar product
  of the Lie algebra), thus measuring the ``intensity'' of the field
  interactions. In this sense, the structure constants can be dynamically
  determined by the theory, and its different equivalence classes will determine
  distinct theories.
\end{itemize}

Furthermore, in the sense of \cite{gravgaugediffgeom,tqft}, we can understand
the process of starting from a ``bare'' (initial) theory and dynamically
determine its parameters (via the contours of the partition function or the
boundary condition of the Schwinger-Dyson equations) giving rise to multiple
(final) theories as a \emph{quantization} scheme. This is intimately related to
the fact that the quantum corrections, being handles attached to the classical
solution, deform the classical geometry into a quantum one --- a fact which is
related to the more common Geometric and Deformation Quantization schemes.

In terms of cobordisms, this can be picture as shown below, where arrows (which
represent the particular propagator in question) going from the initial theory
to each one of the final theories represents the quantization of that particular
solution:

\begin{center}
\begin{xy}
  0;/r.53pc/:
  (0,3.5)*+{\txt\footnotesize{Initial Theory}}="INlabel";
  %
  (0,0)*\ellipse(3,1){-};
  (-6,-8)*\ellipse(3,1){.}; 
  (6,-8)*\ellipse(3,1){.}; 
  (0,-8)*\ellipse(3,1){.}; 
  (-6,-8)*\ellipse(3,1)__,=:a(-180){-};
  (6,-8)*\ellipse(3,1)__,=:a(-180){-};
  (0,-8)*\ellipse(3,1)__,=:a(-180){-};
  (-3,12)*{}="1"; 
  (3,12)*{}="2"; 
  (-9,12)*{}="A2"; 
  (9,12)*{}="B2"; 
  (-3,0)*{}="A"; 
  (3,0)*{}="B"; 
  (-3,1)*{}="A1"; 
  (3,1)*{}="B1"; 
  (3,-16)*{}="1"; 
  (9,-16)*{}="2"; 
  "1";"2" **\crv{(3,-10) & (9,-10)}; 
  (-3,-16)*{}="1"; 
  (-9,-16)*{}="2"; 
  "1";"2" **\crv{(-3,-10) & (-9,-10)}; 
  (-15,-16)*{}="A2"; 
  (15,-16)*{}="B2"; 
  (-3,0)*{}="A"; 
  (3,0)*{}="B"; 
  (-3,-1)*{}="A1"; 
  (3,-1)*{}="B1"; 
  "A";"A1" **\dir{-}; 
  "B";"B1" **\dir{-}; 
  "B2";"B1" **\crv{(13,-6) & (2,-8)}; 
  "A2";"A1" **\crv{(-13,-6) & (-2,-8)};
  (18,-25)*+{\txt\footnotesize{Topologically Inequivalent Final Theories \\
      Equivalence Classes of points in the Moduli Space}}="OUTlabel";
  (-9,-18)*+{}="OUT1";
  (3,-18)*+{}="OUT2";
  (14,-18)*+{}="OUT3";
  {\ar@{->} "OUTlabel";"OUT1"};
  {\ar@{->} "OUTlabel";"OUT2"};
  {\ar@{->} "OUTlabel";"OUT3"};
\end{xy}
\end{center}
\bigskip

As for future work, some options are clear, e.g., a better understanding of how
the boundary conditions of the Schwinger-Dyson equations determine the structure
constants in a pure gauge theory, and how this affects the different multiple
solutions of the theory; a deeper realization of which role the Geometric
Langlands Duality plays in this framework; and, of course, a more robust
comprehension of the 3-dimensional gravity problem. On the mollifier front, we
hope to be able to tackle more physically interesting problems as soon as we
have better hardware and software available.

%

\part{Appendices}
\appendix
\chapter{Convolutions and Smoothing} \label{sec:background}
With the aid of the tools developed below, the smooth approximations of functions can be
done in a mathematically rigorous fashion. This is useful to justify the
statements made in this work, where this technology is applied to the generating
functional of an arbitrary QFT. For proofs of the theorems shown below, see
\cite{pde} \cite{mmmp2}.
\section{Mollifiers}\label{subsubsec:mollifiers}
\begin{notat}
  If $U\subset \mathbb{R}^n$ is open, $\partial U$ is its boundary and $\varepsilon > 0$, let
  $U_{\varepsilon} = \{x\in   U\; |\; \mathrm{dist}(x,\partial U) >
  \varepsilon\}$. Further, let $B(0,\varepsilon)$ be the ball centered on $0$ and with
  radius $\varepsilon$.
\end{notat}

\begin{deff}\label{deff:mollifier}
  (Standard mollifier.)

  A \emph{mollifier}, $\eta$, also called an \emph{approximate identity}, is a positive
  $\cont{\infty}\/(\mathbb{R}^n)$ function. The \emph{standard mollifier} is defined in
  the following way:

  \begin{itemize}
    \item Define $\eta\in \cont{\infty}(\mathbb{R}^n)$ to be,
      \begin{equation*}
        \eta(x) =
        \begin{cases}
          C\, \exp\Bigr(\frac{1}{\abs{x}^2 - 1}\Bigr), & \text{if } \abs{x} < 1; \\
          0, & \text{if } \abs{x} \geqslant 1.
        \end{cases}
      \end{equation*}
      The constant $C>0$ selected so that $\int_{\mathbb{R}^n} \eta(x)\, dx = 1$.
    \item $\forall\, \varepsilon >0$, set
      \begin{equation*}
        \eta_{\varepsilon}(x) = \frac{1}{\varepsilon^n}\, \eta(x/\varepsilon) \; .
      \end{equation*}
  \end{itemize}
  $\eta$ is called the \emph{standard mollifier}. The functions $\eta_{\varepsilon}\in
  \cont{\infty}$ satisfy $\int_{\mathbb{R}^n} \eta_{\varepsilon}(x)\, dx = 1$ and
  $\mathrm{supp}(\eta_{\varepsilon}) \subset B(0,\varepsilon)$.
\end{deff}

\begin{deff}\label{deff:mollification}
  (Mollification.) If $f:\, U\to\mathbb{R}$ is locally integrable, define its
  mollification to be,
  \begin{equation*}
    U_{\varepsilon}\ni\; f^{\varepsilon} = \conv{\eta_{\varepsilon}}{f}\; .
  \end{equation*}
  That is, $\forall \; x\in U_{\varepsilon}$,
  \begin{equation}
    \label{eq:moll-conv}
    f^{\varepsilon}(x) = \int_U \eta_{\varepsilon}(x-y)\, f(y)\, dy =
      \int_{B(0,\varepsilon)} \eta_{\varepsilon}(y)\, f(x-y)\, dy \; .
  \end{equation}
\end{deff}
\section{Properties of Mollifiers}\label{subsubsec:propmollifiers}
\begin{thm}
  (Properties of mollifiers.)

  \begin{enumerate}
    \item $f^{\varepsilon} \in \cont{\infty}(U_{\varepsilon})$;
    \item $f^{\varepsilon} \to f$, almost everywhere, as $\varepsilon \to
      0$;
    \item If $f\in \cont{}(U)$, then $f^{\varepsilon} \to f$ uniformly on compact
      subsets of $U$; \&
    \item If $1 \leqslant p < \infty$ and $f\in L^p_{\mathrm{loc}}(U)$, then
      $f^{\varepsilon} \to f$ in $L^p_{\mathrm{loc}}(U)$.
  \end{enumerate}
\end{thm}
\chapter{Entropy Calculations} \label{sec:entropy}
In the cases considered in \ref{subsubsec:ld}, the information theoretic entropy can be
computed analytically. There are 2 types of functions to be considered:

\begin{align*}
  f^1_{\epsilon}(x) &= \exp\biggl\{-\frac{1}{2}\, \biggl(\frac{x +
    c}{\epsilon}\biggr)^{2}\biggr\} \; ;\\
  f^2_{\epsilon}(x) &= \exp\biggl\{-\frac{1}{2}\, \biggl(\frac{x^{2} +
    c}{\epsilon^{2}}\biggr)\biggr\} \; ;
\end{align*}
where $c$ is a constant scalar. The first type above deals with the Airy Function action
for $J < 0$ ($c = \pm \hksqrt{-J}$) and the $\phi^4$ action with $\mu < 0$ ($c = \pm\hksqrt{-\mu/2\,
  g}$), while the second type above handles the Airy Function action for $J \geqslant 0$
($c = \pm J$) and the $\phi^4$ action for $\mu \geqslant 0$ ($c = \pm i\,
\hksqrt{\mu/2\, g}$).

The information theoretic entropy has the general form,

\begin{align*}
  \mathcal{S}^i(\epsilon) &= -\frac{\int f^i_{\epsilon}(x)\,
    \log\bigl(f^i_{\epsilon}(x)\bigr)\, dx}{\int f^i_{\epsilon}(x)\, dx +
    \log\biggl(\int f^i_{\epsilon}(x)\, dx\biggr)} \; .\\
  \intertext{Therefore, we see that the analytical results are given by:}
  \mathcal{S}^1(\epsilon) &= \frac{(\epsilon/2)\, \hksqrt{2\, \pi}}{\hksqrt{2\, \pi}\,
    \epsilon + \log\bigl(\hksqrt{2\, \pi}\, \epsilon\bigr)} \; ;\\
  \mathcal{S}^2(\epsilon,c) &= \frac{e^{-c/2\, \epsilon^2}\,\hksqrt{2\, \pi}\,
    \bigl(\epsilon + c/2\,\epsilon\bigr)}{e^{-c/2\, \epsilon^2}\, \hksqrt{2\, \pi}\,
    \epsilon + \log\bigl(e^{-c/2\, \epsilon^2}\, \hksqrt{2\, \pi}\, \epsilon\bigr)} \; .
\end{align*}

\newpage

And, thus, we have closed forms for the entropies of the examples given. The graphs below
show these entropies: the leftmost one shows $\mathcal{S}^1(\epsilon)$ while the rightmost
one shows $\mathcal{S}^2(\epsilon,c)$ for a range of $c$ values.

\begin{center}
  \hspace{\fill}
  \scalebox{0.5}{\includegraphics{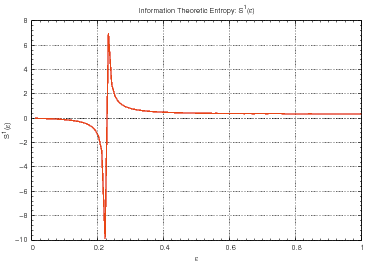}}
  \hspace{\fill}
  \scalebox{0.6}{\includegraphics{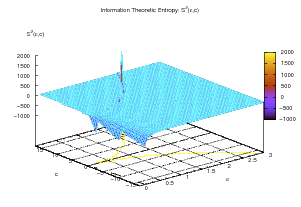}}
  \hspace{\fill}
\end{center}
\chapter{Parabolic Cylinder Functions} \label{sec:pcf}
The differential equation for the Parabolic Cylinder functions is usually
written as,

\begin{equation*}
  \frac{d^2 f}{dz^2} + (a\, z^2 + b\, z + c)\, f = 0 \; ;
\end{equation*}
where $z \in \mathbb{C}$.

The 0-dimensional $\phi^4$ action is given by $S[\phi] = \mu\, \phi^2/2 + g\,
\phi^4/4$, which yields two Schwinger-Dyson equations:

\begin{align*}
  \text{Perturbative solution:}\qquad & -i\, \delta_{J}\mathcal{Z} - J\,
    \mathcal{Z} = 0 \; ; \\
  \text{Non-perturbative solutions:}\qquad & \delta^{2}_{J}\mathcal{Z} + J\,
    \mathcal{Z} - \beta\, \mathcal{Z} = 0 \; ;
\end{align*}
where $\beta = \mu/g$. The match to the previous form is obtained when $a = 0,\,
b = +1,\, c = -\beta$. Note that there are two non-perturbative solutions, one
being solitonic and the other one being called simply broken-symmetric.
%

\backmatter
\part{Bibliography}


\printindex
\end{document}